\documentclass[3p,twocolumn]{elsarticle}
\usepackage{amsmath}
\usepackage{graphicx}
\usepackage{lineno}
\usepackage{listings}
\usepackage{xcolor}
\usepackage{subfig}
\usepackage{hyperref}
\usepackage{float}
\usepackage{booktabs}
\usepackage{tikz}
\usepackage{natbib}
\usepackage{multirow}
\usepackage{appendix}
\usepackage[absolute,overlay]{textpos}
\setcitestyle{open={(},authoryear,close={)}}
\journal{Astroparticle Physics}
\begin{document}
    \def\figureautorefname{Fig.}
    \def\tableautorefname{Tab.}
    \def\sectionautorefname{Sec.}
    \def\subsectionautorefname{Sec.}
    \def\subsubsectionautorefname{Sec.}
    \def\equationautorefname{Eq.}
    \def\appendixautorefname{} 
    \def\appendixname{App.}
    \begin{textblock*}{16cm}(2.75cm,26.65cm) 
        \begin{footnotesize}
            \noindent \textit{Astroparticle Pysics's internal reference: 102933, volume: 158} \newline
            \textit{International Standard Serial Number (ISSN): 0927-6505} \newline
            \textit{Digital Object Identifier (DOI): \url{https://doi.org/10.1016/j.astropartphys.2024.102933}}
        \end{footnotesize}
    \end{textblock*}
    \begin{frontmatter}
        \title{
            Exploring one giga electronvolt cosmic gamma~rays with a Cherenkov~plenoscope capable of recording atmospheric light~fields\\
            Part 1: Optics
        }
        \author[mpik,ethz_ipa]{Sebastian Achim Mueller\corref{corresponding_author_email}}
        \author[ethz_geo]{Spyridon Daglas\fnref{fnspyridon}}
        \author[ethz_ipa]{Axel Arbet Engels\fnref{fnengels}}
        \author[ethz_ipa]{Max Ludwig Ahnen\fnref{fnahnen}}
        \author[ethz_ipa]{Dominik Neise}
        \author[ethz_geo]{Adrian Egger}
        \author[ethz_geo]{Eleni Chatzi}
        \author[ethz_ipa]{Adrian Biland}
        \author[mpik]{Werner Hofmann}
        \cortext[corresponding_author_email]{sebastian-achim.mueller@mpi-hd.mpg.de}
        \affiliation[mpik]{
            organization={Max-Planck-Institute for Nuclear Physics},
            addressline={Saupfercheckweg\,1},
            city={Heidelberg},
            postcode={69117},
            state={Germany}
        }
        \affiliation[ethz_ipa]{
            organization={Institute for Particle Physics and Astrophysics, ETH-Zurich},
            addressline={Otto-Stern-Weg\,5},
            city={Zurich},
            postcode={8093},
            state={Switzerland}
        }
        \affiliation[ethz_geo]{
            organization={Dept. of Civil, Environmental and Geomatic Engineering, ETH-Zurich},
            addressline={Stefano-Franscini-Platz\,5},
            city={Zurich},
            postcode={8093},
            state={Switzerland}
        }
        \fntext[fnspyridon]{
            s.daglas@schnetzerpuskas.com,
            now at Schnetzer Puskas Ingenieure AG, Basel
        }
        \fntext[fnahnen]{
            max.ahnen@positrigo.com,
            ETH Zurich affiliated, now at Positrigo AG, Zurich
        }
        \fntext[fnengels]{
            now at Max-Planck-Institute for Physics, Boltzmannstr. 8, 85748 Garching, Germany
        }
        \begin{abstract}
            Detecting cosmic gamma~rays at high rates is the key to time-resolve the acceleration of particles within some of the most powerful events in the universe.
            Time-resolving the emission of gamma~rays from merging celestial bodies, apparently random bursts of gamma~rays, recurring novas in binary systems, flaring jets from active galactic nuclei, clocking pulsars, and many more became a critical contribution to astronomy.
            For good timing on account of high rates, we would ideally collect the naturally more abundant, low energetic gamma~rays in the domain of one giga electronvolt in large areas.
            Satellites detect low energetic gamma~rays but only in small collecting areas.
            Cherenkov~telescopes have large collecting areas but can only detect the rare, high energetic gamma~rays.
            To detect gamma~rays with lower energies, Cherenkov-telescopes need to increase in precision and size.
            But when we push the concept of the --far/tele-- seeing Cherenkov~telescope accordingly, the telescope's physical limits show more clearly.
            The narrower depth-of-field of larger mirrors, the aberrations of mirrors, and the deformations of mirrors and mechanics all blur the telescope's image.
            To overcome these limits, we propose to record the --full/plenum-- Cherenkov-light field of an atmospheric shower, i.e. recording the directions and impacts of each individual Cherenkov photon simultaneously, with a novel class of instrument.
            This novel Cherenkov~plenoscope can turn a narrow depth-of-field into the perception of depth, can compensate aberrations, and can tolerate deformations.
            We design a Cherenkov~plenoscope to explore timing by detecting low energetic gamma~rays in large areas.
        \end{abstract}
        \begin{keyword}
            timing\sep
            gamma~ray~astronomy\sep
            atmospheric Cherenkov~method\sep
            telescope\sep
            plenoscope\sep
            optics\sep
            light~field\sep
            tomography\sep
            stereo\sep
            cosmic~ray\sep
            burst\sep
            flare\sep
            transient\sep
            variability
        \end{keyword}
    \end{frontmatter}
    \newcommand{\NameAcp}{Portal}
    \newcommand{\NumPix}{8,443}
    \newcommand{\NumPax}{61}
    \newcommand{\NumLix}{515,023}
    \newcommand{\ReflectorFocalLenght}{106.5}
    \newcommand{\ReflectorDiameter}{71}
    \newcommand{\NumFacets}{1,842}
    \newcommand{\LightField}{\mathcal{L}}
    \newcommand{\Response}{\mathcal{R}}
    \newcommand{\ResponseListRepr}{\Response{}_\text{l}}
    \newcommand{\ResponseHistRepr}{\Response{}_\text{h}}
    \newcommand{\Image}{\mathcal{I}}
    \newcommand{\LightFieldGeometry}{\mathcal{G}}
    \newcommand{\Beam}{\mathcal{B}}
    \newcommand{\OpticalPath}{\mathcal{P}}
    \newcommand{\PerfectImageBeam}{\mathcal{B}_\text{img}}
    \newcommand{\ListOfArrivalTimes}{\mathcal{T}}
    \newcommand{\Eye}{eye}
    \newcommand{\std}{\text{std}}
    \newcommand{\median}{\text{median}}
    \newcommand{\BeamSolidAngle}{\Omega}
    \newcommand{\BeamArea}{A}
    \newcommand{\BeamTimeSpread}{T}
    \newcommand{\BeamEfficiency}{E}
    \newcommand{\StateMirrorCamera}[2]{(mirror: \textit{#1}, camera: \textit{#2})}
    \newcommand{\MirrorGood}{good}
    \newcommand{\MirrorBad}{deformed}
    \newcommand{\CameraGood}{good}
    \newcommand{\CameraBad}{misaligned}
    \newcommand{\Pone}{\mbox{P-1}}
    \newcommand{\Pseven}{\mbox{P-7}}
    \newcommand{\PsixtyOne}{\mbox{P-61}}
    \newcommand{\RotPara}[1][0.03]{\includegraphics[angle=0, width=#1\textwidth]{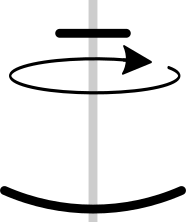}}
    \newcommand{\RotPerp}[1][0.03]{\includegraphics[angle=0, width=#1\textwidth]{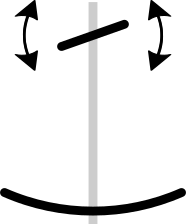}}
    \newcommand{\TransPara}[1][0.03]{\includegraphics[angle=0, width=#1\textwidth]{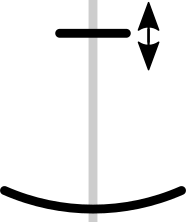}}
    \newcommand{\TransPerp}[1][0.03]{\includegraphics[angle=0, width=#1\textwidth]{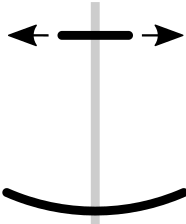}}
    \newcommand{\subfigureautorefname}{\figureautorefname}
    \newcommand{\FigPsfTextIntro}{Images of a star for three angles off the optical axis taken by the three tele- and plenoscopes \Pone{}, \Pseven{}, and \PsixtyOne{}}
    \newcommand{\FigPsfVsOffAxisTextIntro}{The spread of a star's image vs. the star's angle off the optical axis for the three tele- and plenoscopes \Pone{}, \Pseven{}, and \PsixtyOne{}}
    \newcommand{\FigPsfTextAutoRef}[1]{\autoref{#1} shows the corresponding spread averaged over many stars.}
    \newcommand{\FigPsfVsOffAxisTextAutoRef}[1]{\autoref{#1} shows corresponding example images of the star.}
    \newcommand{\FigPsfTextSeeLegend}[1]{Find legend in caption of \autoref{#1}.}
    \section{Introduction}
        \label{SecIntroduction}
        The observation of cosmic gamma~rays offers unique insights to the most violent and energetic events in the universe.
        Unlike the lower energetic light ranging from radio to X-rays, which mostly originates in thermal emission of cosmic remnants slowly transitioning towards a new equilibrium, the higher energetic gamma~rays are often a direct probe of sudden cosmic events.
        In many cosmic sources, we observe flares and bursts of gamma~rays.
        Because of this, the observation and interpretation of timing in the sky of gamma~rays became a vivid part of astronomy.
        Already with our current instruments, we observe occasional and apparently random bursts \citep{ackermann2014fermi} of gamma~rays so bright that they outshine an entire galaxy.
        We observe rapid changes in the flux of gamma~rays coming from active galactic nuclei which host massive black holes.
        Each change in flux is a probe to better constrain the inner processes in the vicinity of the black hole, complementary to and often better than probes based on angular resolution \citep{aleksic2014blackholelightning}.
        Pulsars, which were expected to have steady pulsed emissions, apparently also can emit powerful bursts of gamma~rays \citep{tavani2011discovery}.
        With the observations of gravitational waves from cosmic mergers, astronomy starts to look for counterparts in gamma~rays \citep{abbott2017gravitational} hoping that the variations in time of both gravitational waves and gamma~rays will allow to better understand the moments before, and during these cosmic catastrophes.

        However, the frequent use of timing in the astronomy of gamma~rays pushed our current instruments to their limits.
        Cosmic sources emit gamma~rays mostly with the energy distributed in steep power~laws where lower energetic gamma~rays in the 1\,GeV domain are much more abundant than higher energetic gamma~rays with 10\,GeV or above.
        Pulsars even have their emission cut off above $\sim 10\,$GeV \citep{magic2008pulsar,abdo2009population}.
        Also the universe is not transparent for higher energetic gamma~rays which are absorbed by the infrared light found in between galaxies \citep{magic2008distantQuasar}.
        To improve the exploration of timing, for not only nearby but also for distant sources, we want to detect many low energetic gamma~rays in large collecting areas in order to gather significant statistics in short time.

        Motivated by initial findings \citep{mueller2019phd}, we propose to combine the atmospheric Cherenkov~technique with plenoptics to build the Cherenkov~plenoscope, a novel class of instrument, which can detect gamma~rays with energies down to $\approx 1\,$GeV in large collecting areas exceeding $\approx 10^3\,$m$^{2}$.
        We present a specific Cherenkov~plenoscope, named \NameAcp{}, which can be built with existing technology and which might become the next generation's explorer for the timing in the sky of gamma~rays.
        \subsection{Current Methods}
            \label{SubSecCurrentMethods}
            Currently, gamma~rays in the $1\,$GeV domain are only accessible with detectors in space like Fermi-LAT \citep{acero2015fermi3fgl} or AGILE \citep{tavani2011discovery}.
            Detectors in space directly interact with the cosmic particle and have typically a tracker to measure the direction, a calorimeter to measure the energy, and a surrounding detector for electrically charged cosmic~rays to distinguish them from neutral gamma~rays.
            Despite their high cost, detectors in space provide only a modest $\sim 1\,$m$^2$ area to detect gamma~rays.
            For steady sources, the small area can be compensated by multi-year exposure \citep{acero2015fermi3fgl}, but it significantly limits the exploration of the highly-variable sky of gamma~rays with energies in the GeV domain.

            As the collecting area of detectors in space is not expected to rapidly grow in the foreseeable future, we want to shift our focus to indirect detection in earth's atmosphere using Cherenkov~light.
            Here cosmic particles induce showers in earth's atmosphere.
            Only a few products from the shower reach the ground, for example a flash of Cherenkov~photons.
            The Cherenkov~telescope's mirror and image~camera effectively bin the photons into a sequence of images.
            A trigger searches this sequence of images for flashes of Cherenkov~photons and redirects those images to permanent storage while it dismisses the rest.
            The trigger is a critical, but necessary compromise that reduces the extensive and continuous response from the photosensors down to an amount that one can process and store.
            The sequences of images selected by the trigger are then the basis~\citep{hillas1985cerenkov} to reconstruct the cosmic particle's energy, direction, and type.
            The effective collecting area of a Cherenkov~telescope to detect a gamma~ray can be as large as the area illuminated by the gamma~ray's shower on ground, reaching $\sim 10^5\,$m$^2$.
            The fact that a Cherenkov~telescope effectively detects gamma~rays in an area much larger than its own physical size makes it very efficient.
            Current \citep{magic2008pulsar} and upcoming \citep{bernlohr2013monte} Cherenkov~telescopes can detect gamma~rays with energies as low as 20\,GeV.
            The lowest detectable energy is governed by the area and efficiency to detect Cherenkov~photons, as their number is roughly proportional to the cosmic particle’s energy \citep{naurois2015ground}, and experience shows that a minimum of $\approx$100 photons has to be detected in order to reconstruct a cosmic particle from its shower's Cherenkov~light.
            Cherenkov~telescopes have put great effort into lowering their energetic thresholds, by enlarging their mirrors, by using more efficient photosensors, by improving the reconstruction, and identification of showers, and by moving onto mountains, closer to the shower, where the density of Cherenkov~photons is higher.
            For a proposed array of five 20\,m large Cherenkov~telescopes at 5\,km altitude, a threshold of 5\,GeV was targeted \citep{aharonian2001}.

            However, two physical limits reduce the maximum diameter for a mirror in a Cherenkov~telescope to $\sim 30\,$m.
            First, the narrower depth-of-field on larger mirrors blurs more parts of the shower's image and thus erodes the power to reconstruct the cosmic particle \citep{hofmann2001focus,mirzoyan1996optical}.
            Showers do not happen in a field far away but have their maximum usually in a depth not further than 30\,km in front of the mirror while they naturally extend over a depth in the range of 10\,km.
            The narrow depth-of-field is a central reason why the mirrors in the next generation of Cherenkov~telescopes will not exceed $\sim 23\,$m in diameter \citep{bernlohr2013monte}.
            Second, the square-cube-law complicates the construction of larger Cherenkov~telescopes rapidly.
            When enlarging a telescope, the forces scale cubic with the dimensions but the areas, which have to tolerate these forces, only scale quadratic.
            This limits the cost effective construction of a large Cherenkov~telescope with sufficient rigidity for the optics.

            Since large telescopes have their limits, one could also consider to solve the technological challenge of combining the images from many small telescopes.
            The idea \citep{jung2005star} is to lower the energetic threshold by building an array of small telescopes which feed their continuous sequences of images into a central trigger.
            The critical challenge is that all signals need to be transmitted to a central location, where they are appropriately delayed according to the telescopes pointing, and summed up into pixels of an image.
            However, only in the image seen by the central trigger the Cherenkov~light emerges out of the random light coming from the nightly sky.
            Thus the individual telescopes can not reduce their signals before they send them to the central trigger.
            Overall, the required rates of data that need to be transmitted over the array's size, delayed, and reorganized into an image, be it analog or digital, are daunting.
            Although impressive progress \citep{schroedter2009topological,lopez2016topo,dickinson2018image} was made, today the central trigger for an array of telescopes remains a challenge.
            Today, the first trigger, where most information is reduced, is bound to the individual telescope \citep{bulian1998characteristics,funk2004trigger,weinstein2007veritas,lopez2016topo}.
        \subsection{Proposing a Novel Class of Atmospheric Cherenkov~Instrument}
            \label{SubSecPlenopticForAtmosphericCherenkov}
            When one investigates the causes for the telescope's narrow depth-of-field, and its suffering from deformations and aberrations it turns out that all this is related to the telescope not knowing the position where a photon is reflected on its mirror.

            Assume there was a novel class of instrument just like the telescope but with a different camera.
            Instead of only measuring the position where a photon is absorbed on the camera's screen, the novel camera was further able to measure the direction of the photon relative to its screen.
            Given the alignment of the camera relative to the mirror, one can compute the trajectory of the photon on its path from the mirror towards the camera.
            Given further the shape of the mirror, one can compute the photon's trajectory before it gets reflected by the mirror.
            Given further the pointing of the mirror, one can compute the photon's trajectory in the atmosphere.
            Even if the novel instrument's mirror deformed and its alignment relative to the camera changed, one can still reconstruct the photon's trajectory correctly as long as one measures the instrument's alignment and shape.

            At this point, when the instrument is capable of reconstructing the trajectories of individual photons, one says: the instrument has plenoptic perception \citep{lippmann1908}, and is capable of recording light fields.

            From here on, one can mentally leave the instrument's optics behind and instead concentrate on the recorded trajectories of the photons.
            Now one can take the observed bundle of photons and simulate the image it would make when one guides it into the perfect model for a telescope with its focus set to an arbitrary depth of one's choice \citep{ng2005lightfieldphotography}.
            One could make a whole stack of images with the depth set to different foci highlighting different slices of the shower as it progresses downwards in the atmosphere.
            One could simulate the images that the bundle of photons would make when one guides it into an array of telescopes.
            Here one could make images seen from different perspectives and apply the established reconstructions used in arrays of Cherenkov~telescopes.
            Of course the images created by a perfect model for a telescope are free of aberrations and distortions \citep{hanrahan2006digital}.
            One could also take the bundle of photons and their trajectories to reconstruct the three-dimensional shower in the atmosphere using tomography~\citep{ng2006lightfieldmicroscopy, engels2017master}.
            The only remaining limitations would come from the finite resolution of the camera to measure the photon's position and direction, and the instrument's uncertainties about its alignment and shape.
            Clearly, the possibilities of such an instrument would be remarkable.
            We call this novel class of instrument the Cherenkov~plenoscope\footnote{\citet{bergholm2002plenoscope} uses `plensocope' for a related, hand held optics that allows a single human eye to perceive depth.}.
    \section{Cherenkov~Plenoscope}
        \label{SecCherenkovPlenoscope}
        Both the Cherenkov~telescope and -plenoscope use an imaging mirror.
        But the Cherenkov~plenoscope's camera is different and changes the optics fundamentally.
        \subsection{Optics}
            \label{SubSecCherenkovPlenoscopeConcept}
            \label{SubSecCherenkovPlenoscopeOptics}
\begin{figure}{}
                \centering
                \begin{minipage}{0.49\columnwidth}
                    \centering
                    \includegraphics[width=1\columnwidth, trim={25mm 10mm 25mm 10mm}]{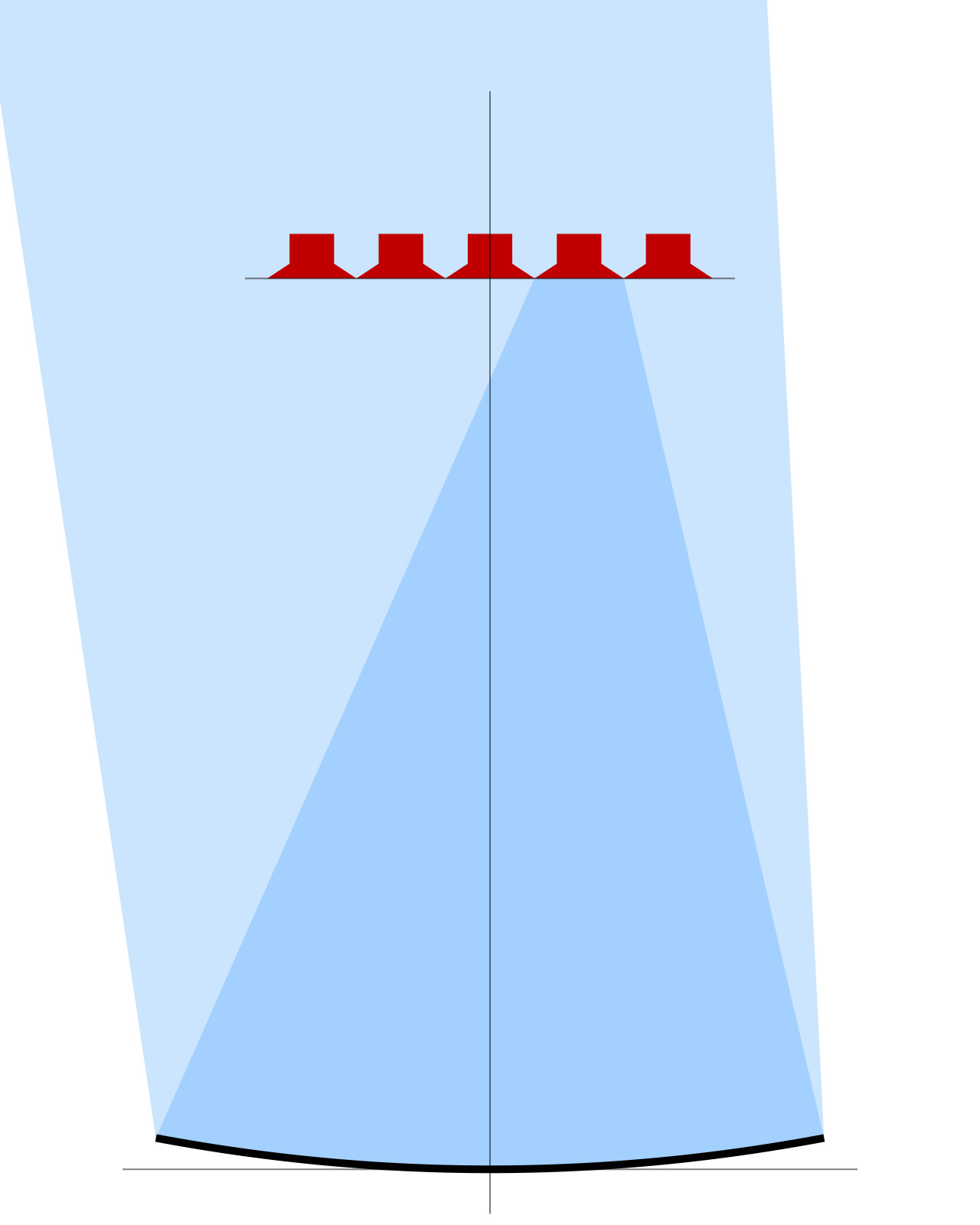}
                    \textcolor{gray}{
                        \tiny{
                            \StateMirrorCamera{\MirrorGood{}}{\CameraGood{}}
                        }
                    }
                \end{minipage}
                \begin{minipage}{0.49\columnwidth}
                    \centering
                    \includegraphics[width=1\columnwidth, trim={25mm 10mm 25mm 10mm}]{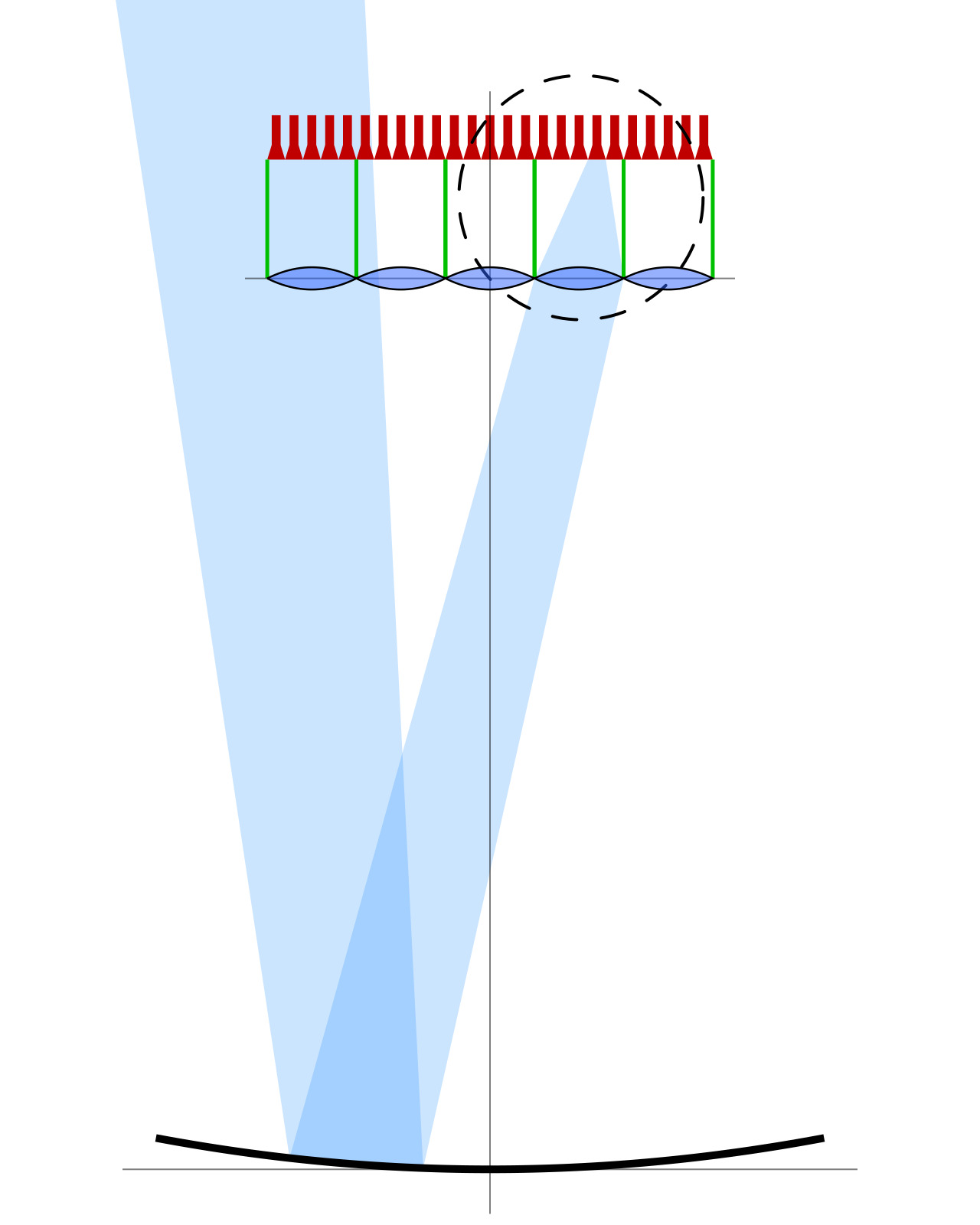}
                    \textcolor{gray}{
                        \tiny{
                            \StateMirrorCamera{\MirrorGood{}}{\CameraGood{}}
                        }
                    }
                \end{minipage}
                \caption{
                    A telescope on the left and a plenoscope on the right.
                    Both have identical mirrors.
                    photosensors are red.
                    Beams, which bundle all the photons trajectories reaching a given photosensor, are blue.
                    On each instrument only one beam is shown.
                    \autoref{FigOpticsConceptOverviewClose} shows the dashed circle up close.
                }
                \label{FigOpticsConceptOverview}
            \end{figure}

\begin{figure}{}
                \centering
                \includegraphics[width=0.7\columnwidth]{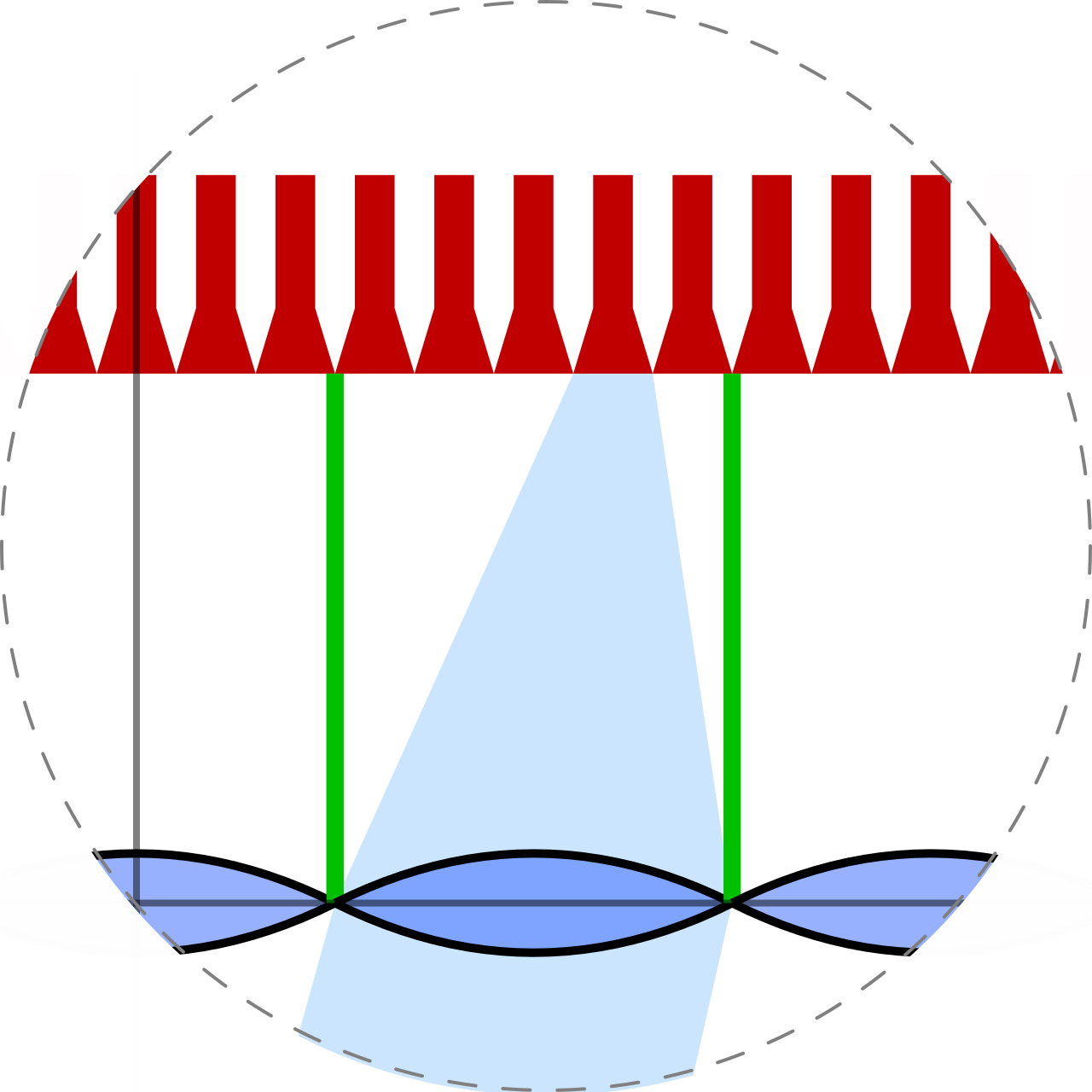}
                \caption{
                    A close-up on the plenoscope's light-field~camera shown in \autoref{FigOpticsConceptOverview}.
                    Green walls are the housings of \Eye{}s.
                    Curved black lines show the bi-convex lenses of \Eye{}s.
                }
                \label{FigOpticsConceptOverviewClose}
            \end{figure}

            \autoref{FigOpticsConceptOverview} shows the optics of a telescope and a plenoscope that use identical mirrors.
            To illustrate the difference in optics, we show on each tele- and plenoscope a single beam, defined as the set of all the optical paths a photon can take to reach a given photosensor.
            The plenoscope's camera needs to measure both the position and the direction of the same photon.
            Measuring the photon's position on the camera gives the photon's direction relative to the mirror, and measuring the photon's direction relative to the camera gives the photon's position of reflection on the mirror.
            To achieve this, consider an imaging lens that guides light into a closed box where a plane of photosensors opposes the lens on the back of the box, see \autoref{FigOpticsConceptOverviewClose}.
            For short, one might call this arrangement an \Eye{}\footnote{
                Outside of astronomy one would probably use the word `camera' to describe a box with a lens and a screen.
            }.
            Such an \Eye{} can measure the direction of a photon while it can further determine the same photon's position to have been within the limits of its lens.
            However, a single \Eye{} is not yet enough.
            For the plenoscope's camera to not only measure the photon's direction but also it's position, one packs multiple, small \Eye{}s into a dense array.
            This array of \Eye{}s is called a light-field~camera.
            When one knows the \Eye{} in which a photon was absorbed, one has measured the photon's position on the light-field~camera.
            And when one knows the specific photosensor inside the \Eye{} that has absorbed the photon, one has measured the photon's direction relative to the light-field~camera.
            A light-field~camera can measure the trajectories of all photons without blocking and masking them.
            While the telescope's beams spread across the entire mirror, the plenoscope's beams only cover parts of the mirror.
            \begin{figure}{}
                \centering
                \begin{minipage}{0.49\columnwidth}
                    \centering
                    \includegraphics[width=1\columnwidth, trim={25mm 10mm 25mm 10mm}]{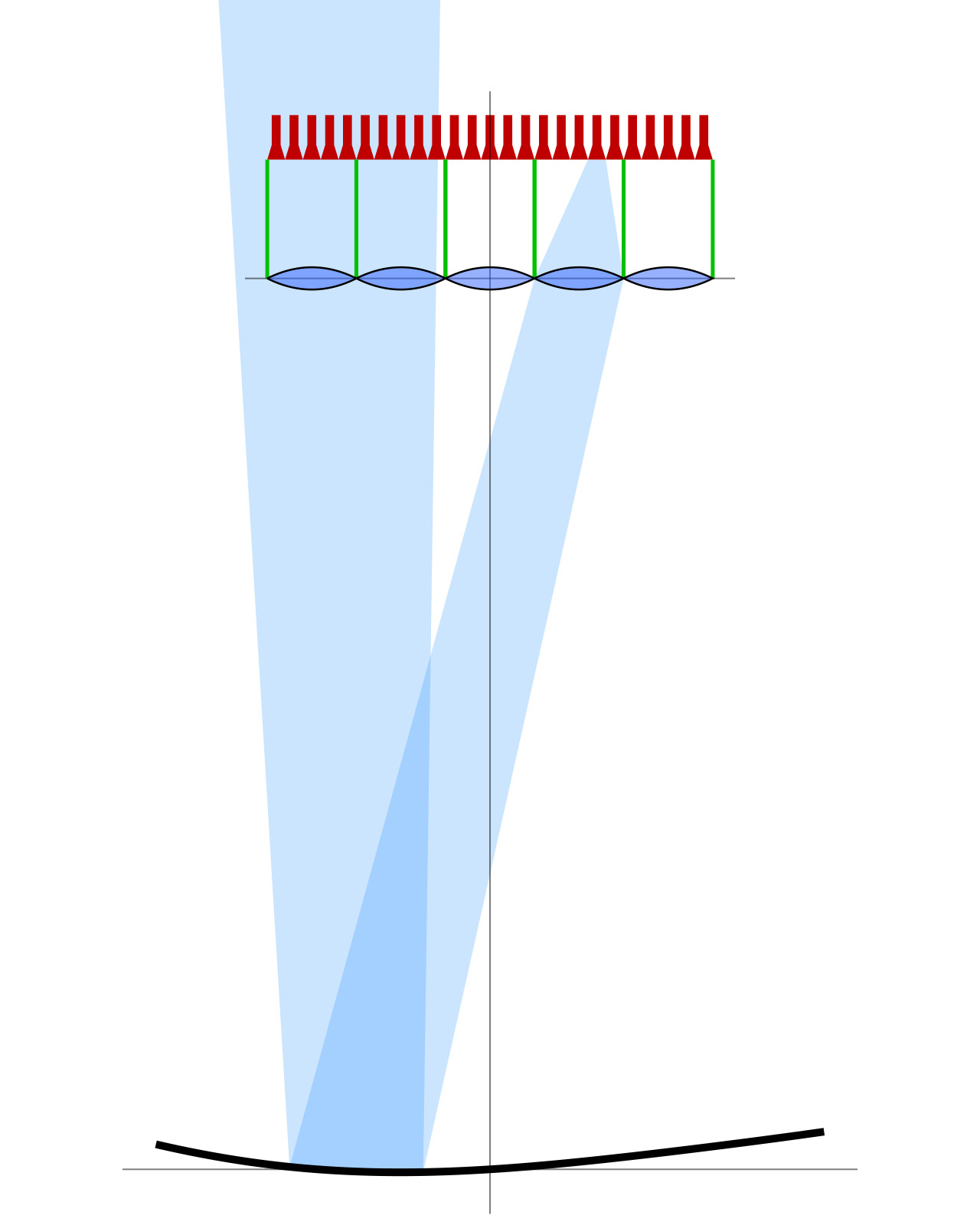}
                    \tiny{\StateMirrorCamera{\MirrorBad{}}{\CameraGood{}}}
                \end{minipage}
                \begin{minipage}{0.49\columnwidth}
                    \centering
                    \includegraphics[width=1\columnwidth, trim={25mm 10mm 25mm 10mm}]{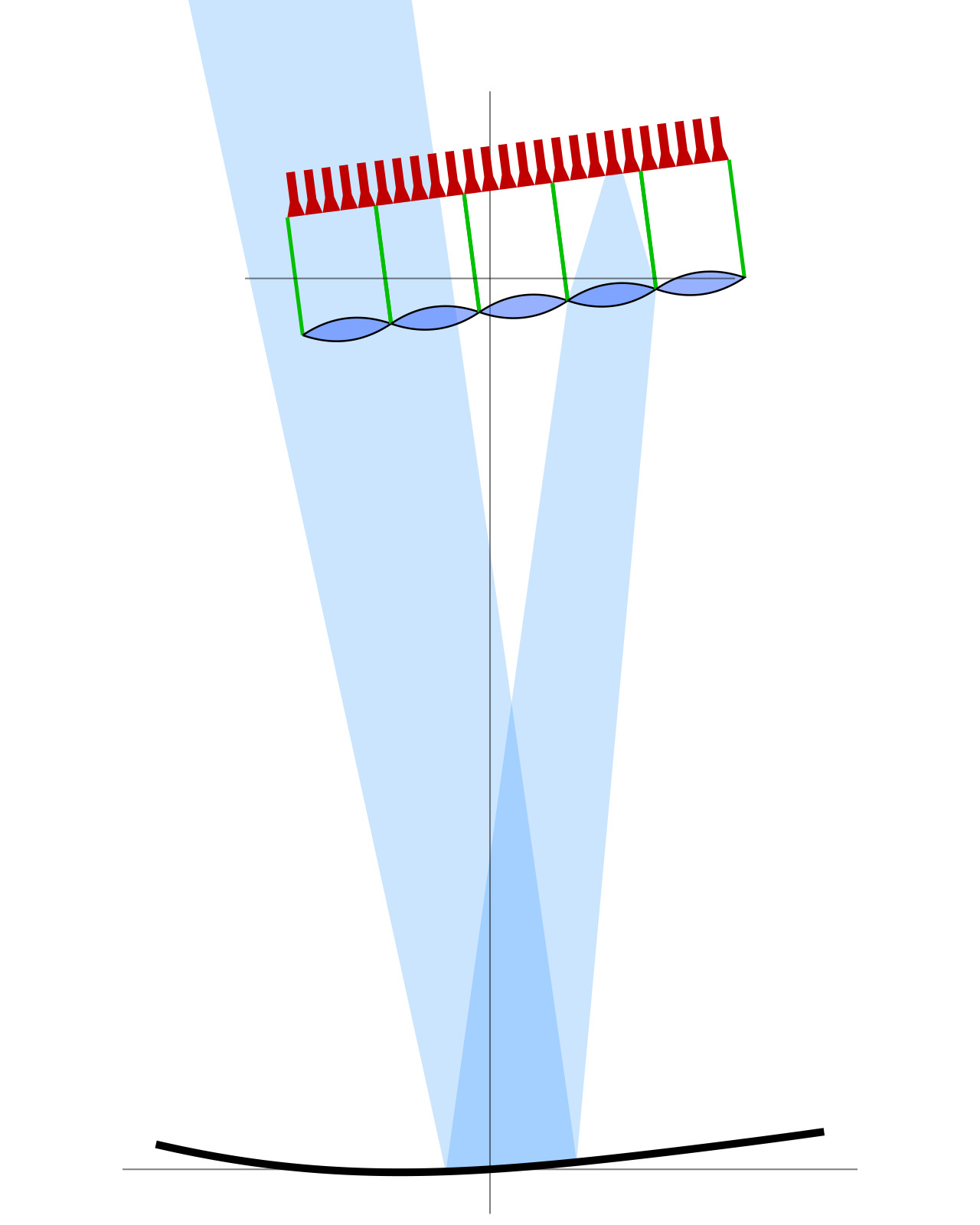}
                    \tiny{\StateMirrorCamera{\MirrorBad{}}{\CameraBad{}}}
                \end{minipage}
                \caption{
                    A plenoscope with a deformed mirror on the left and a plenoscope with both a deformed mirror and a misalignment on the right.
                    Compare with the plenoscope's default geometry in \autoref{FigOpticsConceptOverview}.
                }
                \label{FigOpticsConceptOverviewDeformedMisaligned}
            \end{figure}
            Both the telescope and plenoscope shown in \autoref{FigOpticsConceptOverview} have five directional bins to measure a photon's direction.
            The telescope has five photosensors which serve as its directional bins, and the plenoscope has five eyes which serve as its directional bins.
            While the telescope has only a single positional bin to measure a photon's point of reflection on its mirror, the plenoscope has five positional bins.
            In \autoref{FigOpticsConceptOverviewClose} one finds five photosensors inside each \Eye{} of the light-field~camera which serve as the plenoscope's five positional bins.

            To convert the telescope's image~camera into the plenoscope's light-field~camera, one has to replace each of the image~camera's photosensors with an \Eye{}.
            The other way around, if one adds the photosensors signals from within each \Eye{} into a single signal, the light-field~camera degenerates into an image~camera.
            Only if one records the individual signals, one can take advantage of the plenoscope's individual beams being only effected by small parts of the mirror.
            This can resolve aberrations.

            Aberrations occur when the photon's position on the camera's screen not only depends on the photon's direction relative to the mirror, but also on the photon's position of reflection on the mirror.
            So if one can measure the photon's position of reflection on the mirror, one can compensate aberrations.
            \autoref{FigOpticsConceptOverviewDeformedMisaligned} shows that the geometry of the plenoscope's beams change when its mirror is deformed, or its camera is misaligned.
            However, we will show that this change of the beams geometry does hardly effect the plenoscope's performance as long as one knows the geometry of the beams.
            The plenoscope's ability to compensate deformations is in stark contrast to the telescopes demand for rigidity.
            Postponing the square-cube-law by compensating deformations is one key in the plenoscope's quest to be large enough to detect gamma~rays with lower energies.
            But while one could postpone the square-cube-law also on larger telescopes by spending exponentially more resources and engineering, the telescope's physical limit of a narrowing depth-of-field remains.
        \subsection{Depth-of-Field}
            \label{SubSecCherenkovPlenoscopeDepthOfField}
            \label{ChDepthOfField}
\begin{figure}{}
                \centering
                \includegraphics[width=1\columnwidth]{
                    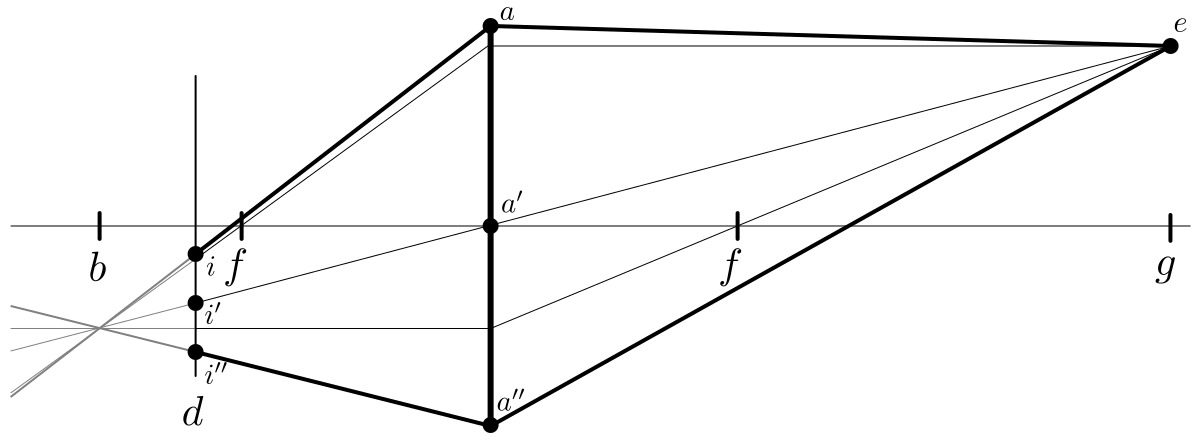}
                \caption[Extended apertures, focusing]{
                    The `thin lens' with focal-length $f$ and aperture $\overline{aa''}$.
                    Horizontal line is the optical axis, thick vertical line is the aperture's principal plane.
                    The three principal rays emitted by an object in depth $g$ converge in the image's distance $b$.
                    However, the principal rays do not converge on the camera's screen which is in distance $d \neq b$.
                    The image of the object is not sharp but stretches from $i$ to $i''$.
                    For readability, this figure shows a transmitting lens.
                    But \autoref{EqThinLens} is also valid when the figure was folded along the aperture's principal plane so that the two focal-points $f$ became one and $g$ and $b$ were on the same site of the aperture.
                    In that case the figure would show a reflective mirror as we find it on most Cherenkov~telescopes.
                    In the context of atmospheric Cherenkov emission: note that one can interpret the point $e$ as the location of a nearby charged particle emitting Cherenkov light in a cone along e.g. the lines $\overline{ea}$ and $\overline{ea'}$.
                }
                \label{FigThinLens}
            \end{figure}

            The bigger the mirror of a telescope becomes, the narrower becomes its depth-of-field.
            The telescope's depth-of-field is the `field' (range) along its depth (optical axis) where objects create sharp images.
            When this depth narrows we are forced to adjust the telescope's focus \citep{trichard2015enhanced} in order to have at least a sharp image of the most relevant depth while we have to accept blurring in all remaining depth.
            The thin-lens~equation
            \begin{eqnarray}
                \frac{1}{f} &=& \frac{1}{g} + \frac{1}{b}
                \label{EqThinLens}
            \end{eqnarray}
            and \autoref{FigThinLens} describe this blurring and show
            the distance $b$ where the image forms and where the camera's screen must be positioned in order to record a sharp image of an object in depth $g$ when the focal-length is $f$.
            \autoref{FigThinLens} shows that the image $i,i',i''$ is a scaled projection of the aperture $a,a',a''$.
            This projection is the reason for the blurring and is referred to
as Bokeh in some literature \citep{merklinger1997bokeh,ahnen2016bokeh}.

            In \citet{bernlohr2013monte}, one finds an estimate
            \begin{eqnarray}
                g_\pm &=& g(1 \pm pg/(2fD))
                \label{EqDepthOfField}
            \end{eqnarray}
            for the object's starting depth $g_{-}$ and ending depth $g_{+}$ in order to define the depth-of-field on a Cherenkov~telescope.
            Here $D$ is the diameter of the mirror, and $p$ is the diameter of an area\footnote{On some Cherenkov telescopes, this area is the opening of a light guide (called Winston cone) which couples to a photosensor. On some other telescopes, this area is the sensitive surface of a photosensor itself.} in the camera's screen from where the light will be further guided into a photosensor.
            This estimate is based on \citet{hofmann2001focus}, which also uses \autoref{EqThinLens}.
            For example, a Cherenkov~telescope with $D=71\,$m and $f=106.5\,$m would have a depth-of-field extending from $g_{-} = 9.2\,$km to $g_{+} = 10.8\,$km when its focus is set to $g = 10.0\,$km.
            This will blur large parts of a shower's image because a shower's depth exceeds this only $1.6\,$km narrow depth-of-field by about an order-of-magnitude.
            In the past, Cherenkov~telescopes lowered their energetic threshold by enlarging their mirrors (enlarging $D$ and $f$), by moving up in altitude closer to the shower (lowering $g$), and by increasing angular resolution and the number of photosensors (lowering $p$).
            But \autoref{EqDepthOfField} shows that these three measures all narrow the depth-of-field.
            Thus imaging itself becomes a physical limit that prevents Cherenkov~telescopes from being large enough to observe gamma~rays with energies below $\approx20\,$GeV.

            But if the camera in \autoref{FigThinLens} is a light-field~camera, it not only observes that three photons arrive in the points $i$, $i'$, and $i''$, but it observes that the photons travel on the trajectories $\overline{ia}$, $\overline{i'a'}$, and $\overline{i''a''}$.
            With this one knows that the photons approached on the trajectories $\overline{ae}$, $\overline{a'e}$, and $\overline{a''e}$.
            With a light-field~camera, one obtains a strong hint that the photons were produced in point $e$ in depth $g$ despite the camera being out of focus with $d \neq b$.
            On a Cherenkov~plenoscope, the narrowness of the depth-of-field expressed in \autoref{EqDepthOfField} becomes an estimate for the plenoscope's resolution of this depth.
            So the enlarging of the mirror, the increasing of angular resolution, the moving up in altitude onto mountains, all these measures to lower the Cherenkov~method's energetic threshold do not narrow, but actually extend and sharpen the Cherenkov~plenoscope's perception of depth.
            Therefore, at least from the optic's point-of-view, the Cherenkov~plenoscope is promising to be made large enough in order to detect low energetic gamma~rays.
    \section{Reconstruction and Interpretation}
        \label{SecLightFieldGeometry}
        To reconstruct the Cherenkov~photons of an atmospheric shower with a Cherenkov~plenoscope, while accounting for misalignments and deformations, we propose to describe the perception of the Cherenkov~plenoscope with the help of three-dimensional rays.
        Reconstructing a photon here means to measure its three-dimensional trajectory and its time of arrival.
        \subsection{Defining the Light-Field~Calibration}
            \label{SecDefiningTheLightFieldCalibration}
            We describe the perception of an atmospheric Cherenkov~instrument using the beams of light that the instrument's photosensors observe in the atmosphere.
            We describe a photosensor's beam using the many optical paths that a photon can travel in order to reach this photosensor.
            Each of these optical paths is a set
            \begin{eqnarray}
                \OpticalPath{} &=& \begin{Bmatrix}
                    {s_x}, {s_y}, {c_x}, {c_y}, \tau, \eta
                \end{Bmatrix}
                \label{EqOpticalPathDefinition}
            \end{eqnarray}
            which describes a ray
            \begin{eqnarray}
                \vec{r}(\chi) &=&
                \underbrace{
                    \begin{pmatrix}
                        s_x \\ s_y \\ 0
                    \end{pmatrix}
                }_{\vec{s}}
                + \chi
                \underbrace{
                    \begin{pmatrix}
                        c_x \\ c_y \\ \sqrt{1 - {c_x}^2 - {c_y}^2}
                    \end{pmatrix}
                }_{\vec{c}}
                \label{EqRay}
            \end{eqnarray}
            that the photon travels on, in opposing direction, in order to reach the instrument's aperture, and later a photosensor.
            In the same frame, the aperture's principal plane is the $xy$-plane,
            and the aperture's optical axis is the $z$-axis.
            This way, the ray's support $\vec{s}$ is within the aperture's principal plane, and the ray's direction $\vec{c}$ is pointing from the aperture up into the sky, anti parallel to the incoming Cherenkov photon from a shower.
            By defining\footnote{Definition and wording is inspired by \citet{heck1998corsika}.} the rays relative to the aperture's principal plane, the four-tuple ($s_x$, $s_y$, $c_x$, $c_y$), compiled from the $xy$-components of the ray's support $\vec{s}$ and direction $\vec{c}$, is sufficient to describe the geometry of an optical path $\OpticalPath{}$.
            \autoref{FigIdealImageBeams} shows a ray $\vec{r}(\chi)$ and its vectors $\vec{s}$ and $\vec{c}$ in a later use to construct image~rays.

            Because Cherenkov~instruments can measure the arriving time of photons in a regime of 1\,ns while having optics with sizes in the regime of several $1$\,m, one also adds for each path the time $\tau$ that it takes a photon to travel from the aperture's principal plane to the photosensor.
            Further, we add relative efficiencies $\eta$ for each path to respect differences in the instrument's reflection, and transmission.
            With this, we approximate a photosensor's beam by using a list of $P$ optical paths
            \begin{eqnarray}
                \Beam{} &=&
                \begin{bmatrix}
                    \OpticalPath{}_1, \OpticalPath{}_2, \dots \OpticalPath{}_P
                \end{bmatrix}.
                \label{EqBeamDefinition}
            \end{eqnarray}
            The optical paths are randomly drawn from a uniform distribution and the number $P$ of paths which reach the given photosensor is sufficiently large to represent the beam appropriately.
            Now we can describe the perception of an atmospheric Cherenkov~instrument with the list
            \begin{eqnarray}
                \LightFieldGeometry{} &=&
                \begin{bmatrix}
                    \Beam{}_1,
                    \Beam{}_2,
                    \,\dots\,
                    \Beam{}_K
                \end{bmatrix}
                \label{EqLightFieldGeometryDefinition}
            \end{eqnarray}
            of its $K$ beams, one beam for each photosensor.
            We name $\LightFieldGeometry{}$ the instrument's light-field~calibration.
        \subsection{Defining the Instrument's Response}
            A photosensor measures the arriving time $t'$ of a photon.
            photosensors have limitations, but one can express their response using a list
            \begin{eqnarray}
                \ListOfArrivalTimes{} &=& [t'_{1},\,t'_{2},\,\,\dots\,\,t'_{L}]
                \label{EqPhotoSensorResponseDefinition}
            \end{eqnarray}
            of arriving times of $L$ photons that is most probable to have caused the photosensor's recorded signal.
            In turn one can describe the instrument's total response using a list of lists
            \begin{eqnarray}
                \ResponseListRepr{} &=&
                \begin{bmatrix}
                    \ListOfArrivalTimes{}_1,
                    \ListOfArrivalTimes{}_2,
                    \dots
                    \ListOfArrivalTimes{}_K
                \end{bmatrix}
                \label{EqRecordDefinition}
            \end{eqnarray}
            of all the $K$ photosensors responses.
            Alternatively and equally valid, one can represent this as a histogram with the number of photons $i_k,t$ in the $k$-th beam and the $t$-slice in time
            \begin{eqnarray}
                \ResponseHistRepr{} &=&
                \begin{bmatrix}
                    i_{1,\,1} & i_{1,\,2} & \ldots & i_{1,\,T}\\
                    i_{2,\,1} & i_{2,\,2} & \ldots & i_{2\,T}\\
                    \vdots    & \vdots    & \ddots    & \vdots\\
                    i_{K,\,1} & i_{K,\,2} & \ldots & i_{K,\,T}\\
                \end{bmatrix}.
                \label{EqResponseHist}
            \end{eqnarray}
        \subsection{Reconstructing the Photons Observables}
            \label{SecReconstructingPhotons}
\begin{figure}
                \centering
                \includegraphics[width=1.0\columnwidth]{
                    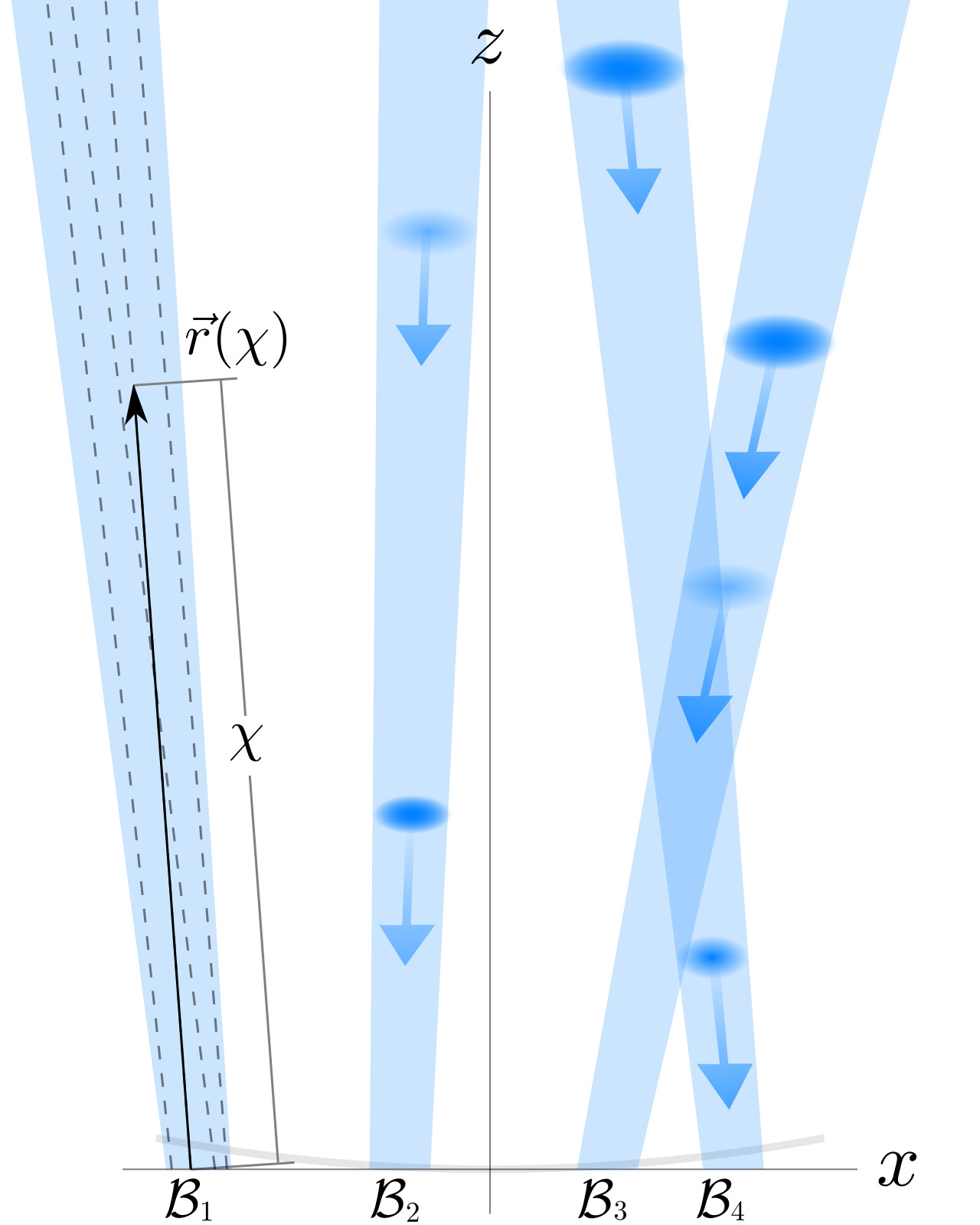
                }
                \caption[]{
                    Reconstructing photons on a Cherenkov~plenoscope using beams.
                    The plenoscope's physical mirror is a curve in faint gray.
                    See description on reconstruction in \autoref{SecReconstructingPhotons}.
                    Note that the blue beams $\Beam{}$ and dashed rays $\vec{r}(\chi)$ start on the aperture's principal plane ($\vec{r}(\chi=0)$), pass through the physical mirror, and extent up into the sky.
                    On the other hand, the blue-cloudy photons, which represent the physical Cherenkov photons coming down from an atmospheric shower, will not pass the physical mirror.
                    These photons will be reflected on the physical mirror.
                    The light-field calibration does not know on what paths the physical photons find their ways into their photosensors.
                    The light-field calibration does not need to know this.
                    All that the light-field calibration needs to knows is, that a physical photon, which approaches the aperture's principal plane on a specific ray $\vec{r}_k$ (that is a ray within $\Beam{}_k$), will be absorbed by a specific and corresponding photosensor $k$.
                }
                \label{FigBeamsRaysAndPhotons}
            \end{figure}

            When a photosensor $k$ detects a photon to be absorbed at time $t'$ one can use an optical path from within beam $\Beam{}_k$ to compute the time
            \begin{eqnarray}
                t_\text{aperture} &=& t' - \tau_k
                \label{EqTimeAperture}
            \end{eqnarray}
            at when the photon would have passed the aperture's principal plane.
            For any moment in time $t$, one can now compute the distance
            \begin{eqnarray}
                \chi(t) &=& - \frac{{c_0}}{n}(t - t_\text{aperture})
                \label{EqChi}
            \end{eqnarray}
            that the photon still has to travel before it will reach the aperture's principal plane.
            Here $c_0$ is the speed of light and $n$ is the refractive index of the instrument's surrounding medium.
            With this one can compute the photon's position $\vec{r}(\chi(t))$ for any moment in time $t$, while keeping in mind that from a single photon's observables $(x, y, c_x, c_y, t')$ we can not constrain the time of its creation.
            \autoref{FigBeamsRaysAndPhotons} shows the concepts we defined so far.
            In pale blue one sees some of the photosensors beams $\Beam{}_1$ to $\Beam{}_4$ reaching up from the aperture's principal plane into the atmosphere.
            But unlike \autoref{FigOpticsConceptOverview}, this does not show the beams inside the plenoscope which reach from the mirror to the photosensors.
            In the most left beam $\Beam{}_1$ one can see the rays $\vec{r}(\chi)$ of the optical paths which describe this beam.
            All the figure shows the same moment in time $t$.
            Along the beams, darker blue clouds indicate the reconstructed volumes of photons.
            The darker the blue, the more photons are reconstructed to be in this volume.
            The volume's spread depends both on how fuzzy the beam is and on how well the beam's photosensor can reconstruct the photon's time of absorption $t'$.
            When one applies the light-field~calibration $\LightFieldGeometry{}$ to the photosensors responses $\Response{}$, the result is a measurement of five observables ($x\,,y\,,c_x,\,c_y\,,t$) for each photon.
            One might call this measurement a sequence (in time $t$) of light~fields ($x\,,y\,,c_x,\,c_y$).
            There are many interchangeable ways to represent sequences of light~fields.
            For example, in combination with the light-field~calibration $\LightFieldGeometry{}$, the histogram in \autoref{EqResponseHist} or the list of lists in \autoref{EqRecordDefinition} are possible representations.
    \section{Imaging}
        \label{SecImageRays}
\begin{figure}
            \centering
            \includegraphics[width=1\columnwidth]{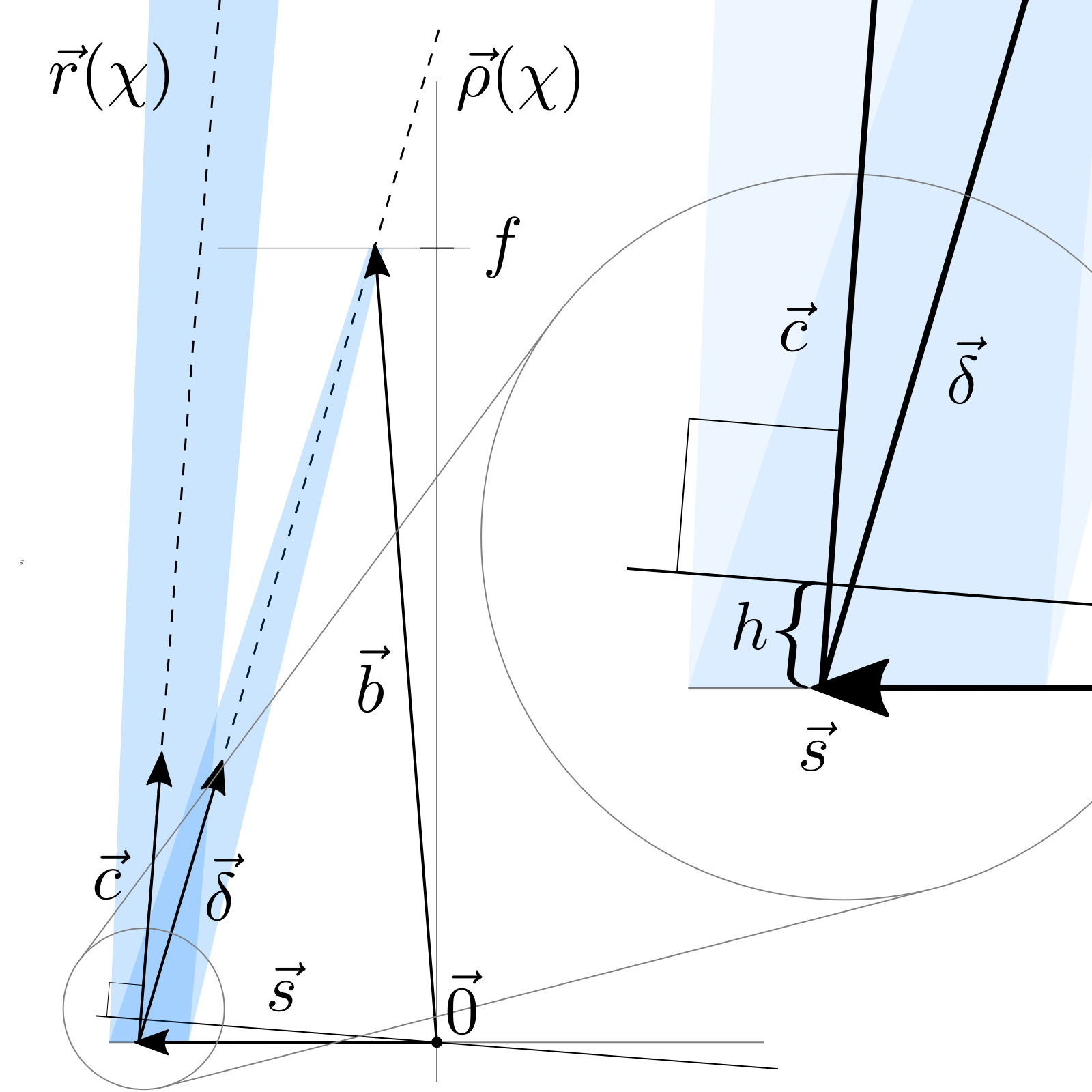}
            \caption[Image~rays]{
                A dashed ray $\vec{r}(\chi)$ in its blue beam $\Beam{}$ and its corresponding dashed image~ray $\vec{\rho}(\chi)$ in its own blue image~beam $\PerfectImageBeam{}$.
                In the close-up, one can see the distance $h$ to correct the length of the optical path for `ideal' timing.
                The construction of $\vec{b}$ is shown in \autoref{SecCalculatingImageRays}.
            }
            \label{FigIdealImageBeams}
        \end{figure}

        Since \citet{hillas1985cerenkov}, imaging is well established in the atmospheric Cherenkov~method to reconstruct cosmic gamma~rays.
        So projecting the Cherenkov~plenoscope's light~field onto an image is a good start to get familiar with plenoptic perception.
        The idea is to take the Cherenkov~plenoscope's reconstructed photons, defined in \autoref{SecReconstructingPhotons}, and simulate the propagation of these photons in an ideal imaging optics.
        The imaging optics being `ideal' depends on one's purpose and field-of-view as there is no general projection from a sphere onto a plane.
        However, the thin lens, see \autoref{EqThinLens}, is a good start.
        Before the reconstructed photons pass the aperture's principal plane, they travel along the beams $\Beam_1$ to $\Beam_K$ defined in the plenoscope's light-field~calibration $\LightFieldGeometry{}$.
        The rays in these beams correspond to the lines $\overline{ea}$, $\overline{ea'}$, and $\overline{ea''}$ in \autoref{FigThinLens}.
        Now one computes for each beam $\Beam{}_k$ the thin lens' corresponding image~beam $\PerfectImageBeam{}_k$ which, just like $\Beam{}_k$, is a set of optical paths.
        But the optical paths in the image~beam $\PerfectImageBeam{}_k$ are based on image~rays
        \begin{eqnarray}
            \vec{\rho}(\chi) &=& \vec{s} + \chi \vec{\delta}.
            \label{EqImageRay}
        \end{eqnarray}
        For every ray $\vec{r}(\chi)$ one can calculate the corresponding image~ray $\vec{\rho}(\chi)$.
        Both rays $\vec{r}(\chi)$, and $\vec{\rho}(\chi)$ share the same support $\vec{s} = (s_x, s_y, 0)^T$ on the aperture's principal plane.
        The image~ray's direction $\vec{\delta}$ is constructed based on the thin lens' model for imaging as \autoref{SecCalculatingImageRays} shows in more detail.
        Image~rays $\vec{\rho}(\chi)$ correspond to the lines $\overline{ai}$, $\overline{a'i'}$, and $\overline{a''i''}$ in \autoref{FigThinLens}.
        Figure \autoref{FigIdealImageBeams} shows the construction of an image~ray $\vec{\rho}(\chi)$ from a ray $\vec{r}(\chi)$.
    \subsection{Projecting the Plenoscope's Response onto Images}
        \label{SecProjectingTheLightFieldOntoImages}
        One can project the response of the plenoscope's $K$ photosensors $\ResponseHistRepr{}$ onto an image with $N$ pixels.
        Here, a pixel is just a picture-cell in a histogram over directions, and has no physical counterpart on the plenoscope.
        These pixels do neither need to have the shape of the plenoscope's eyes, nor does their number $N$ need to match the number of eyes in the plenoscope.
        One can define the plenoscope's image
        \begin{eqnarray}
            \Image{} &=&
            \begin{bmatrix}
                j_1, j_2, \dots j_N
            \end{bmatrix}.
            \label{EqDefineImage}
        \end{eqnarray}
        as a list of $N$ intensities $j$.

        For simplicity, we integrate over time here so that $\ResponseHistRepr{}$ lists only one intensity for each of the plenoscope's $K$ photosensors.

        Now, one can project the response $\ResponseHistRepr{}$ onto an image $\Image{}_g$ which has its focus set to an arbitrary depth $g$ of one's choice by multiplying
        \begin{eqnarray}
            \Image{}_g &=& U(g) \cdot \ResponseHistRepr{}
            \label{EqRefocusedImaging}
        \end{eqnarray}
        with the imaging matrix
        \begin{equation}
            \label{EqRefocusedImagingMatrix}
            \begin{aligned}
            U(g) &=&
                \begin{bmatrix}
                    u_{1,\,1}(g) & u_{1,\,2}(g) & \ldots & u_{1,\,K}(g)\\
                    u_{2,\,1}(g) & u_{2,\,2}(g) & \ldots & u_{2\,K}(g)\\
                    \vdots    & \vdots    & \ddots    & \vdots\\
                    u_{N,\,1}(g) & u_{N,\,2}(g) & \ldots & u_{N,\,K}(g)\\
                \end{bmatrix}
            \end{aligned}
        \end{equation}
        which elements depend on the desired depth $g$.
        Once one knows an instrument's light-field~calibration $\LightFieldGeometry{}$, the computation of its imaging matrix $U(g)$ is straight forward as shown in \autoref{SecCalculatingImagingMatrix}.
    \subsection{Timing in Images}
        \label{SecIsochronousImaging}
        Timing in the regime of $1\,$ns is important to the atmospheric Cherenkov~method to identify flashes of Cherenkov~photons in between the other photons from the nightly sky.
        Optics effect this timing, and can induce undesired spreads when optical paths differ in length.
        A segmented mirror with its facets placed on a parabola has good timing in the center of its image.
        Only when such a mirror is large, and angles off its axis increase, the spread in time becomes relevant.
        For example: A parabolic mirror with $D=71$\,m in diameter and a screen in $f=106.5\,$m focal-length induces a spread of $\approx 25\,$cm$\,{c_0}^{-1} \approx 0.8\,$ns in between the paths of two parallel running photons which are $3.25^{\circ}$ off the mirror's optical axis and are being reflected on opposite edges of the mirror.
        On the plenoscope, one can resolve this spread in time.
        For this, one needs a model for ideal timing in an image.
        We define ideal timing such that photons, which travel in the same direction $\vec{c}$ (i.e. before they enter the instrument) and within a plane perpendicular to their direction of motion, are assigned to the same bin in time.
        \autoref{FigIdealImageBeams} shows the distance
        \begin{eqnarray}
            h &=& \vec{s}^T \cdot{} \vec{c} = s_x c_x + s_y c_y
        \end{eqnarray}
        between the point of the photon being closest to the aperture's origin,
        and the point of the photon intersecting with the aperture's principal plane.
        With this distance, and the time of the photon's passing through the aperture's principal plane in \autoref{EqTimeAperture}, one can compute the photon's time of arrival
        \begin{eqnarray}
            t_\text{img} &=& t_\text{aperture} + h\frac{n}{c_0}
        \end{eqnarray}
        in a sequence of images.
        The speed of light $c_0$ and the medium's refractive index $n$ match \autoref{EqChi}.

        This means that when an instrument records a sequence of images over time (i.e. a movie), then the moment $t_\text{img}$ in this sequence to which the photon must be assigned to is defined as the moment when the photon would have been closest to the central point of the instrument's aperture, if nothing was ever stopping the photon on its initial trajectory coming down from the sky.

        Correcting the timing becomes relevant when the parabolic mirror deforms, or if one wants to use a different design for the segmented mirror in the first place.
        For example, the plenoscope can reduce the spread in time induced by a mirror with \citet{davies1957design_solar_furnace} design, which otherwise becomes significant for diameters beyond $D\approx15\,$m.
    \section{Introducing \NameAcp{}}
        \label{SecPortal}
        \NameAcp{} is a specific Cherenkov~plenoscope designed to detect cosmic gamma~rays with energies as low as 1\,GeV in an effective collecting area of several 10$^3$\,m$^{2}$.
        Here we motivate Portal's design from the optics point-of-view and demonstrate its powerful perception of depth and its ability to compensate aberrations, deformations, as well as misalignments.
        Together, we designed \NameAcp{}'s mechanical structure in an interdisciplinary collaboration of civil~engineering and particle~physics \citep{daglas2015master, zinas2016design}.
        \subsection{Overview}
            \label{SecPortalOverview}
\begin{figure}
                \centering
                \includegraphics[width=1.0\columnwidth]{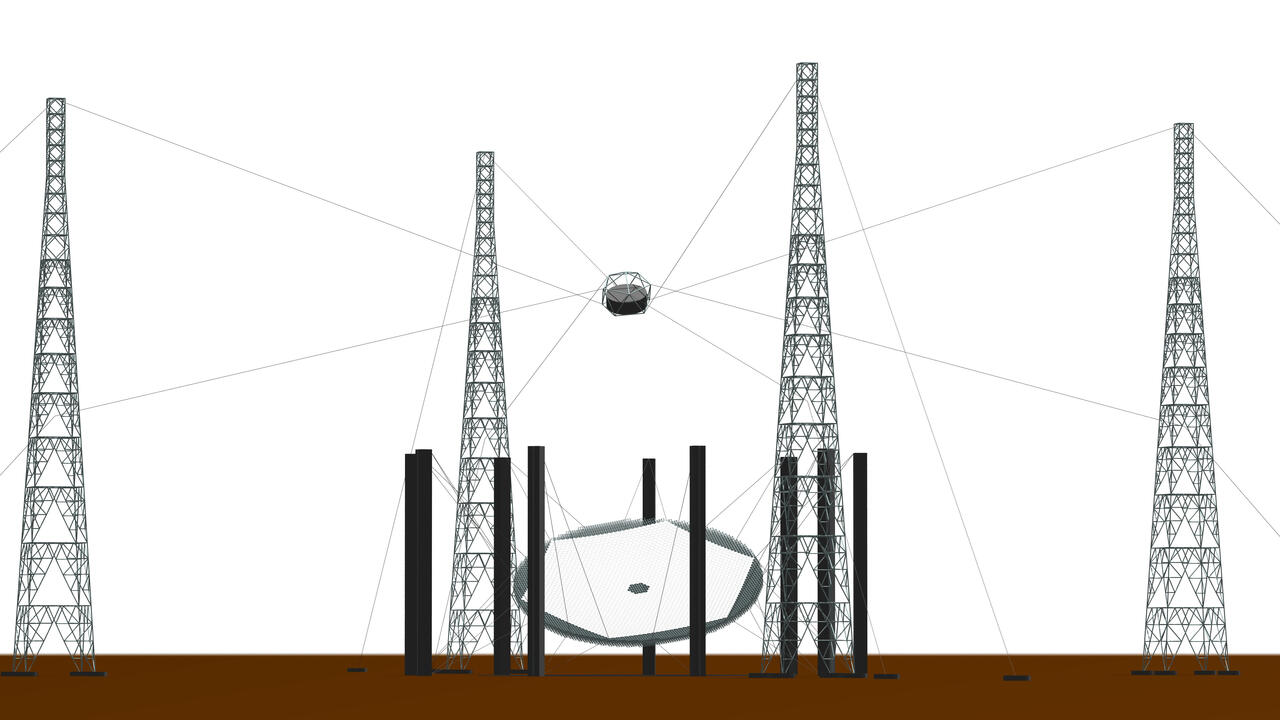}
                Side from a distance.
                \includegraphics[width=1.0\columnwidth]{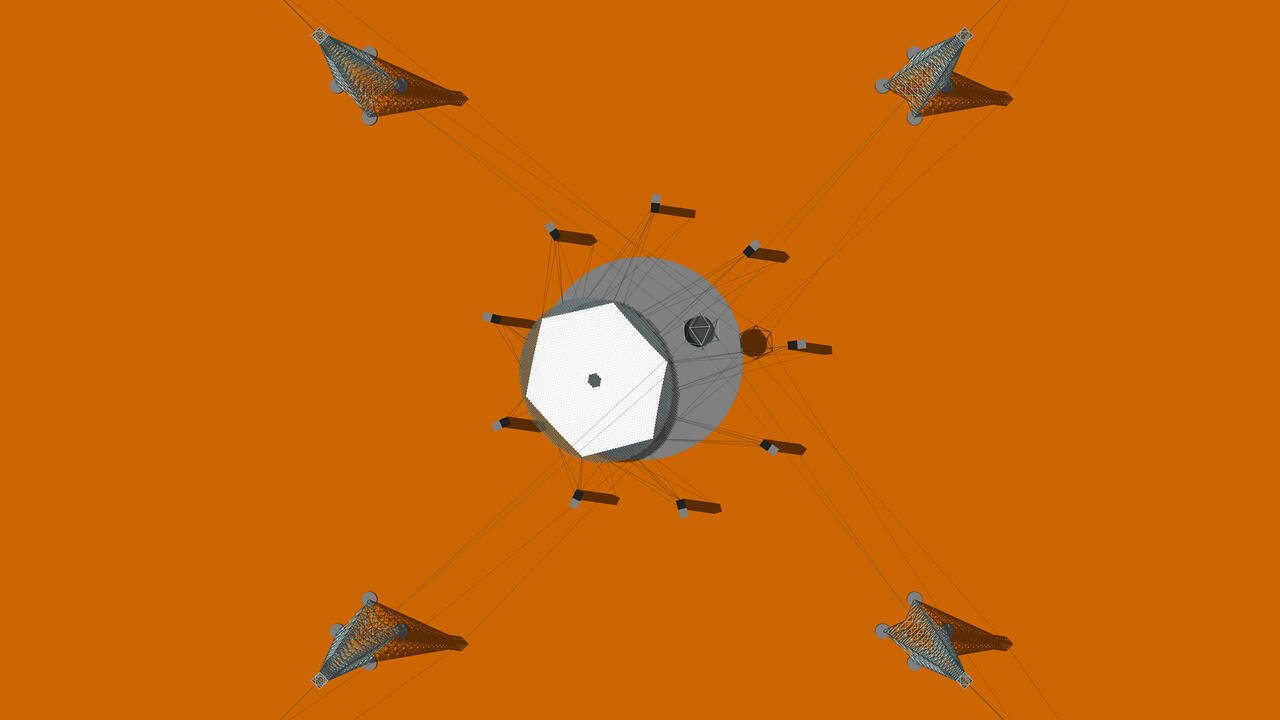}
                Top from a distance.
                \includegraphics[width=1.0\columnwidth]{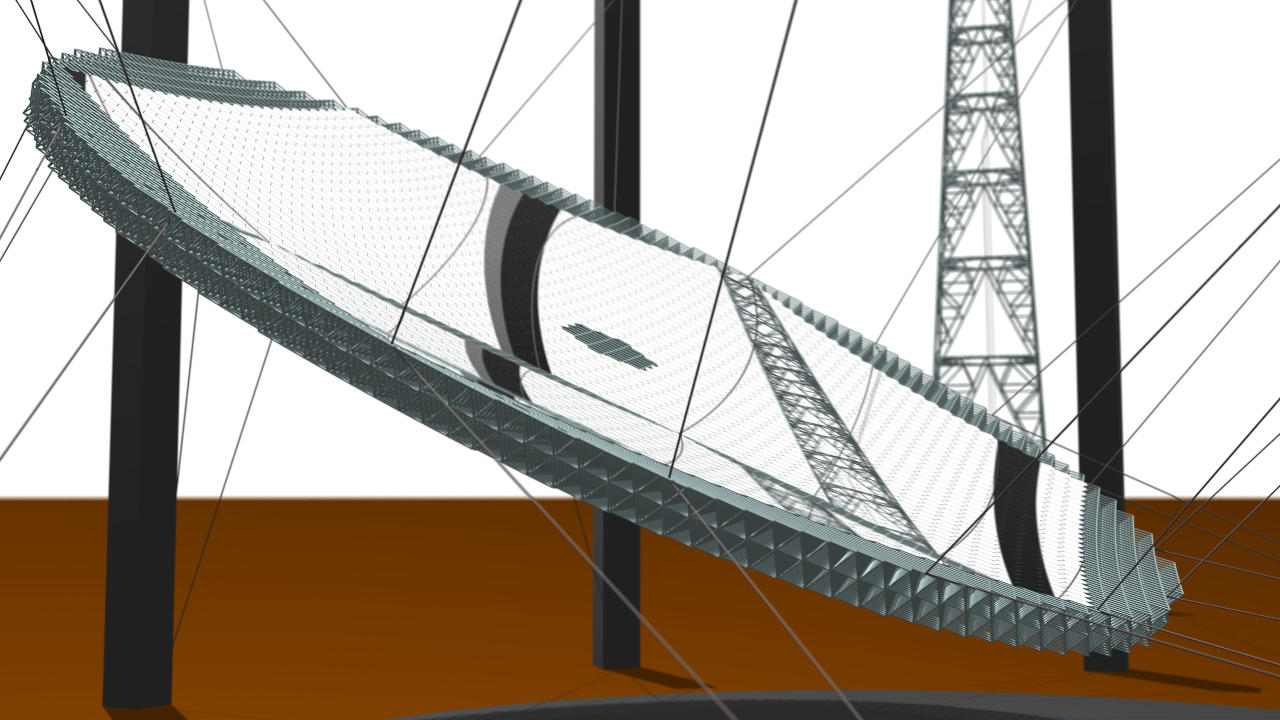}
                Mirror with own mount suspended from cables.
                \includegraphics[width=1.0\columnwidth]{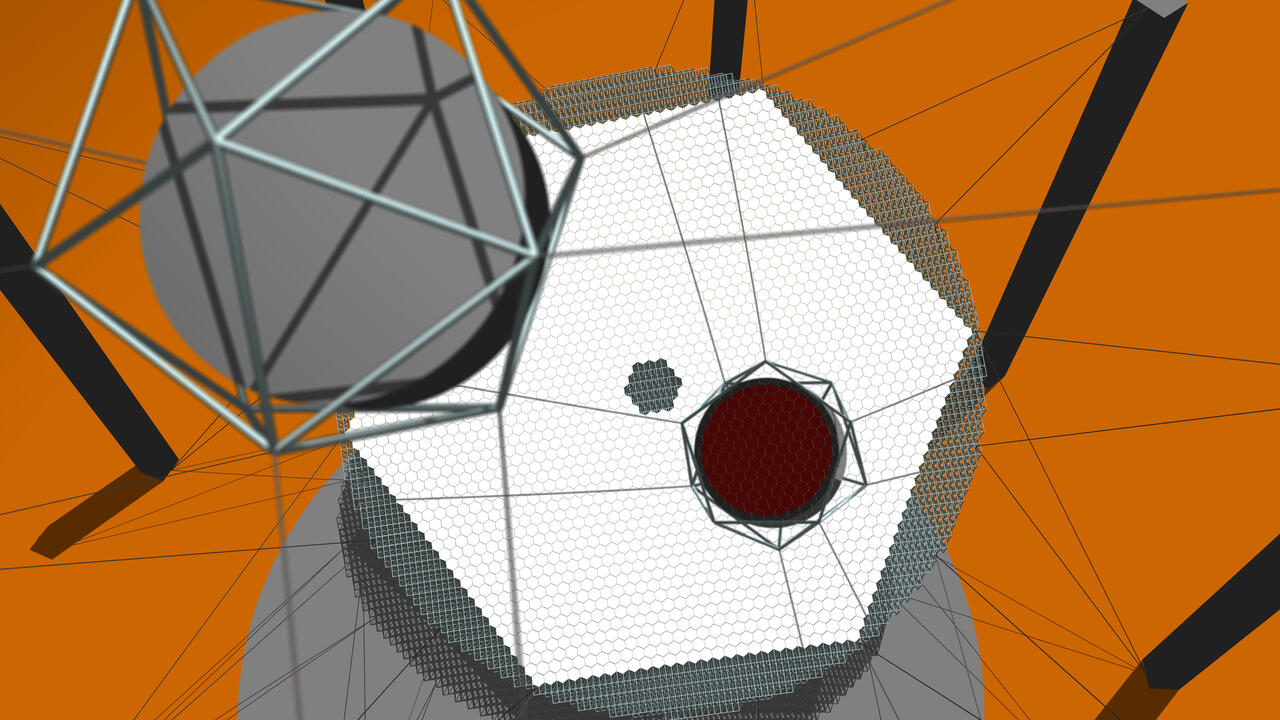}
                light-field~camera above mirror.
                \caption[]{
                    The \NameAcp{} Cherenkov~plenoscope with its two independent mounts for the mirror and the light-field~camera.
                    Renderings similar to \citet{mueller2019phd} (mirror is different).
                }
                \label{FigPortalMount}
            \end{figure}

            \autoref{FigPortalMount} shows \NameAcp{} with its two main components, first its imaging mirror, and second its light-field~camera.
            Plenoptic’s relaxed constrains for rigidity and alignment allow \NameAcp{} to mount its mirror and its camera independently on two mounts.
            The first mount moves the mirror which has a diameter of $D = 71\,$m, a focal-length of $f = 106.5\,$m, and a focal-ratio of $F = 1.5$.
            The second mount moves the light-field~camera which has a diameter of $12.1\,$m corresponding to a field-of-view of $6.5^\circ$.
            All optical surfaces are spherical to ease fabrication.
            All of \NameAcp{}'s optics are made from only two repeating primitives to ease mass production: One reflective facet for the mirror, and one transparent lens for the \Eye{}s in the light-field~camera.
            The primitives are truly identical so that any facet can replace any other facet, and any lens can replace any other lens.
            \NameAcp{}'s mounts are actuated by cable~robots \citep{miermeister2016cablerobot} which can point \NameAcp{} up to $45^{\circ}$ away from zenith and have only cables shadowing the mirror.
            When hunting flares and bursts of gamma~rays, \NameAcp{} points with an estimated $90^{\circ}$\,min$^{-1}$ and can always move along the shortest route along the great circle because unlike the altitude-azimuth~mount \citep{borkowski1987near} \NameAcp{}'s kinematic has no singularity near the zenith.
            To keep the mounts small, the mirror moves in one direction while the camera moves to the other what effectively rotates the unity of mirror and camera around its common center.
            During the day, the light-field~camera parks on a pedestal to the side to ease service, while the mirror parks flat on the ground.
            \begin{figure}
                \centering
                \includegraphics[width=1.0\columnwidth]{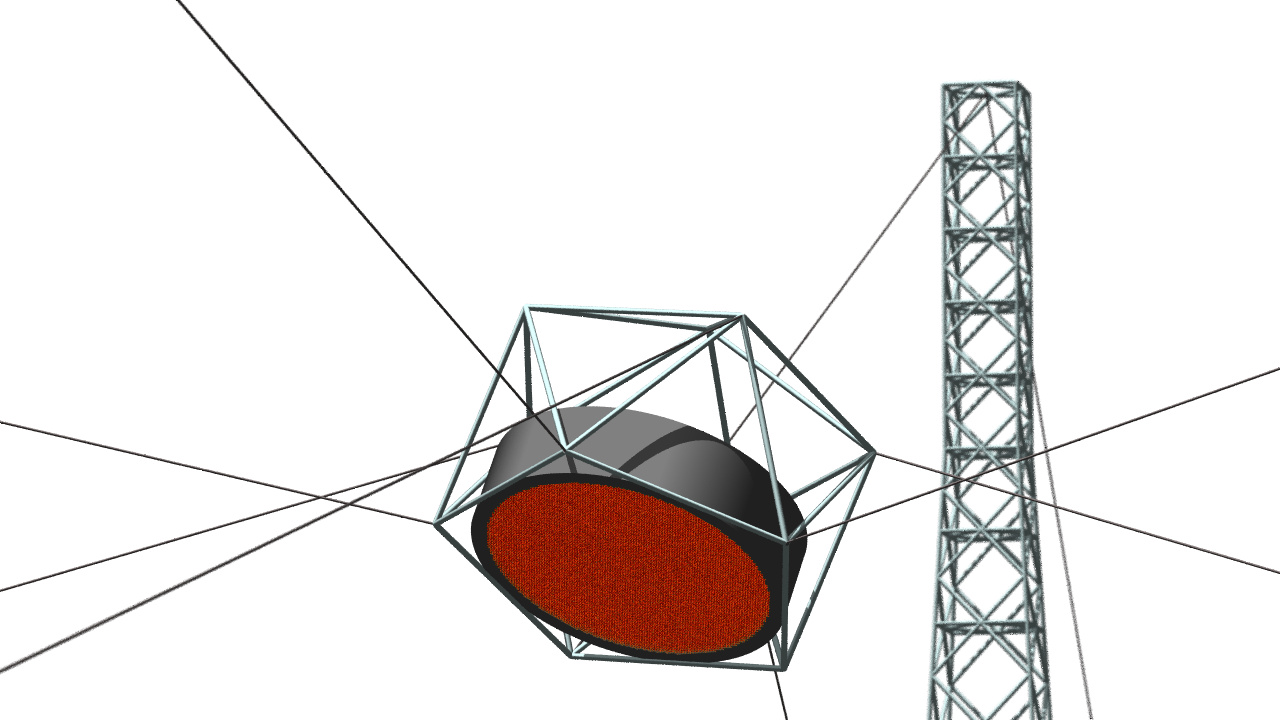}
                light-field~camera with own mount suspended from cables.
                \includegraphics[width=1.0\columnwidth]{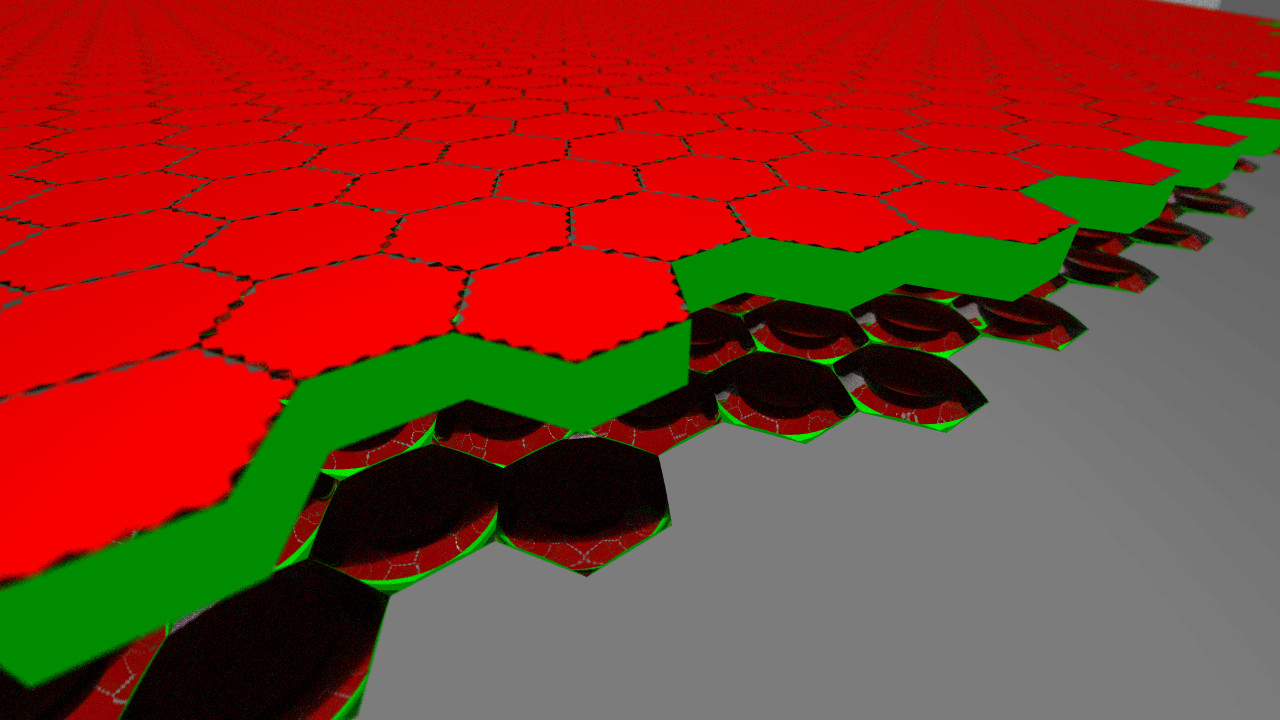}
                Inside light-field~camera, array of \Eye{}s.
                \includegraphics[width=1.0\columnwidth]{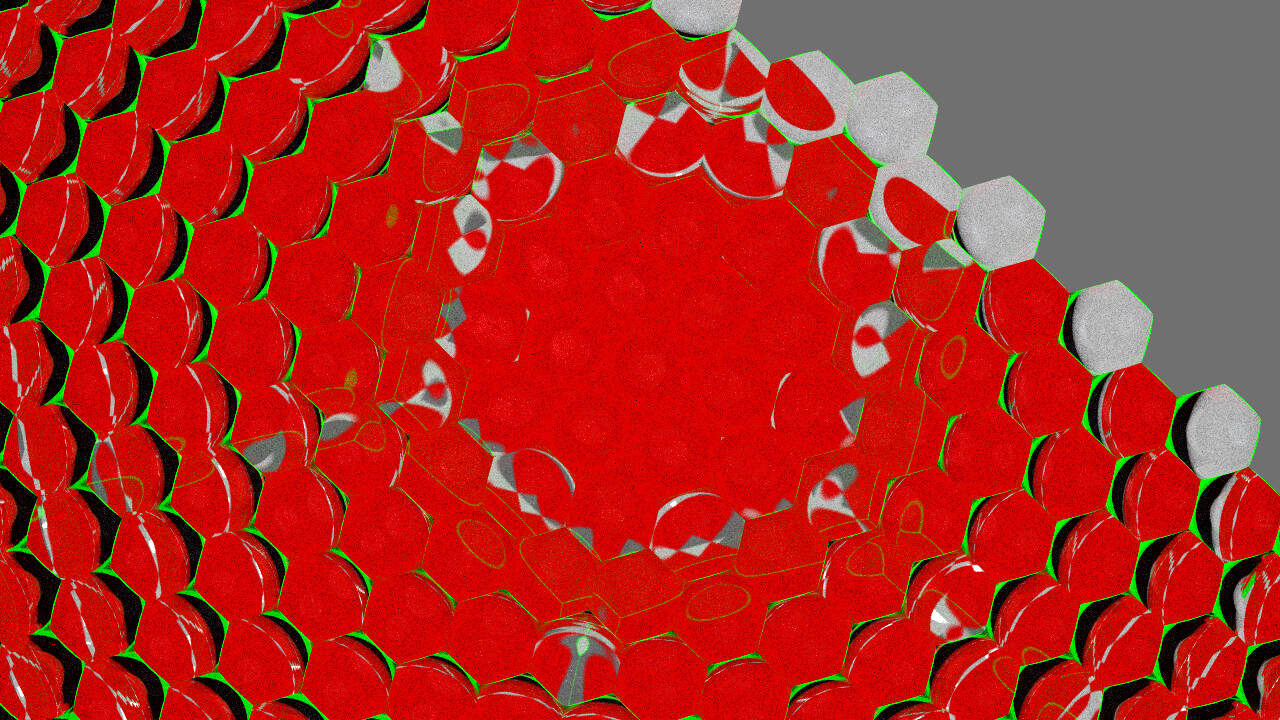}
                Below light-field~camera, looking up into the \Eye{}s lenses.
                \includegraphics[width=1.0\columnwidth]{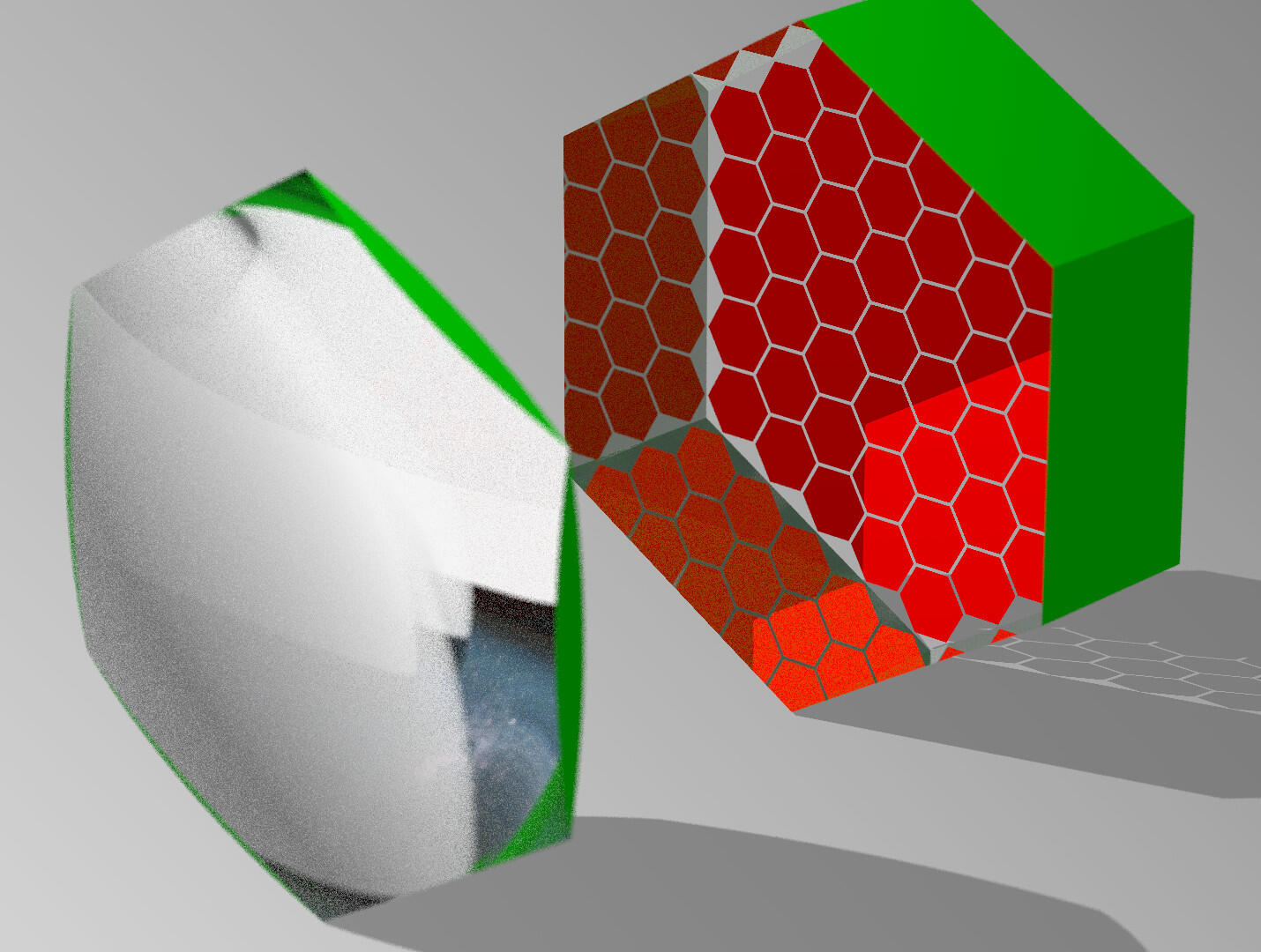}
                Single \Eye{} taken from light-field~camera.
                \caption[]{
                    \NameAcp{}'s light-field~camera in more detail.
                    Renderings similar to \citet{mueller2019phd}.
                }
                \label{FigPortalSensor}
            \end{figure}
            \begin{figure}
                \centering
                \includegraphics[width=1.0\columnwidth]{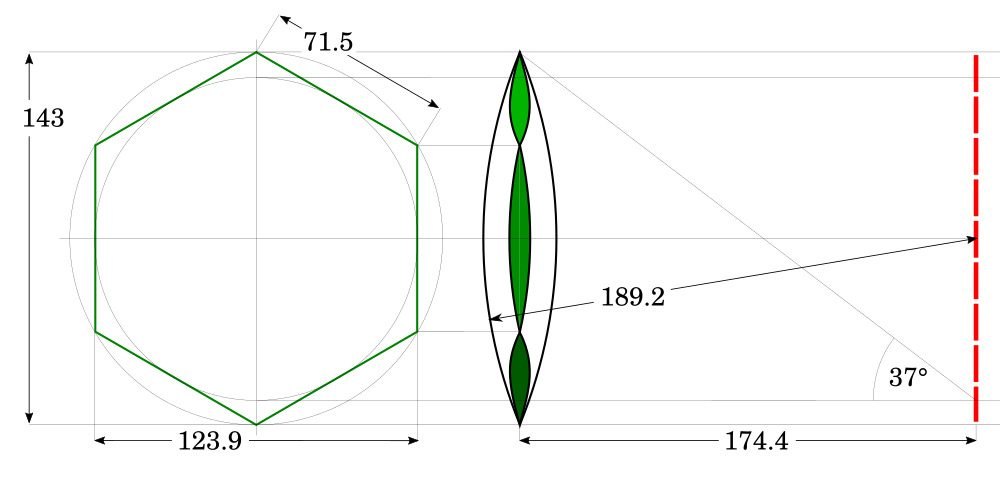}
                \caption[]{
                    An \Eye{} in the light-field~camera.
                    The lens' refractive index is $\approx 1.47$ for a wavelength of $400\,$nm.
                    Left: Front-view. Right: Side-view.
                    The photosensors are red and the opaque faces of the lens are green.
                    Dimensions\,/\,mm.
                    Adopted from \citet{mueller2019phd}.
                }
                \label{FigSmallCameraGeometry}
            \end{figure}
            \begin{figure}
                \centering
                \includegraphics[width=1\columnwidth]{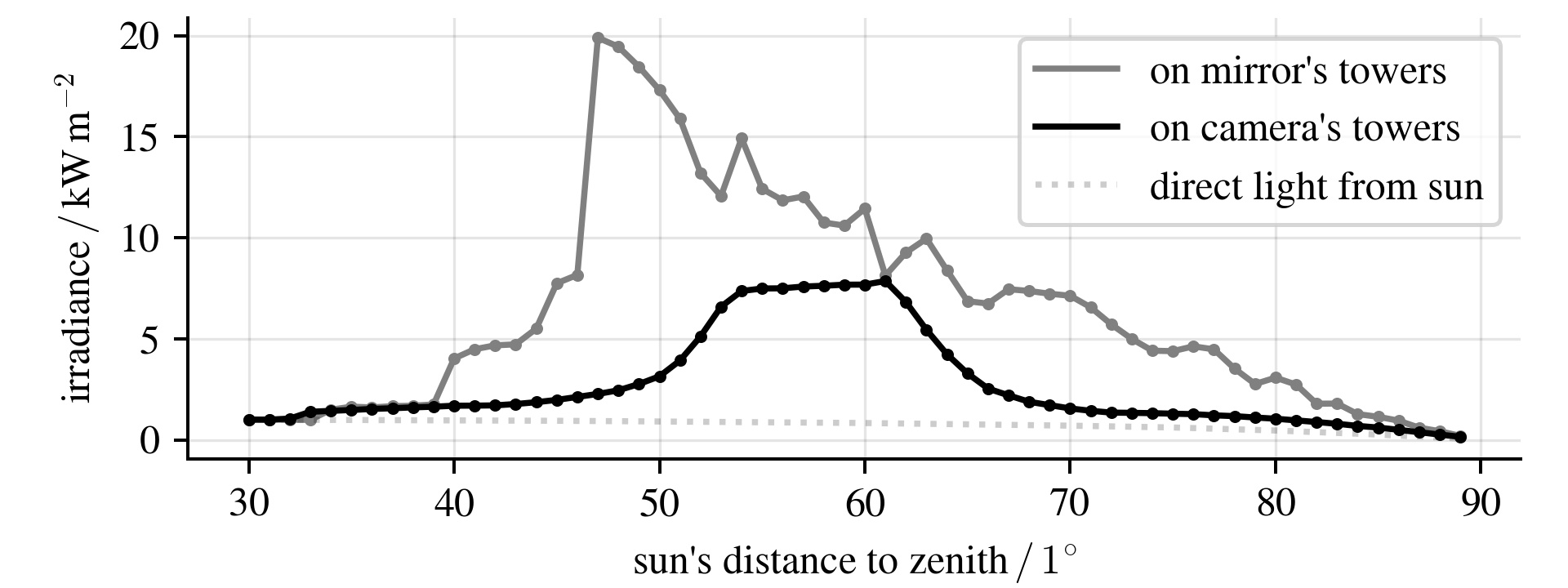}
                \caption[]{
                    The worst case irradiation from reflected sunlight found in hot spots on \NameAcp{}'s towers when the mirror has its facets perfectly aligned and is parking flat on the ground.
                }
                \label{FigSunHotSpots}
            \end{figure}
        \subsection{Light-Field~Camera}
            \label{SecPortalSensor}
            \autoref{FigPortalSensor} shows \NameAcp{}'s light-field~camera which is a dense array of $N=\NumPix{}$ identical \Eye{}s arranged in a hexagonal grid.
            The camera's field-of-view spans $6.5^{\circ}$ on the sky, what corresponds to a diameter of $\approx 12.1\,$m.
            A single eye's field-of-view spans $0.067^{\circ}$ on the sky, what is on a par with existing Cherenkov telescopes \citep{cornils2005optical}.
            Each \Eye{} is made out of a bi-convex, spherical lens and an array of $M=\NumPax{}$ photosensors encased in a hexagonal cylinder, see \autoref{FigSmallCameraGeometry}.
            Thus \NameAcp{}'s light-field~camera has a total of $K = M \times N = \NumLix{}$ beams.
            In general, the lens' focal-ratio must match the mirror $F \geq F_\text{lens}$,
            in order to have the lens' image of the mirror fully contained within the \Eye{}'s body.
            In \NameAcp{} the lens' focal-ratio is $F_\text{lens} \approx{} 1.4$ to ensure the lens' image of the mirror is a bit smaller than the \Eye{}'s body.
            This way the array of photosensors inside the \Eye{} does not need to cover the full cross-section of the \Eye{} and instead can leave space for the walls of the \Eye{}'s body.
            Since the \Eye{}'s body is hexagonal, the array of photosensors inside the body is also a hexagon.
            This in turn allows the lens' image of the mirror to be hexagonal which motivates the hexagonal outer shape of \NameAcp{}'s mirror.
            The inner walls of the \Eye{}'s hexagonal body reflect light, at least a part of the wall close to the photosensors.
            This is because the lens refracts parts of the light coming from the outer parts of the mirror too far off its optical axis.
            We found that making the flat inner walls reflective helps to redirect this light mostly back into the photosensors which are meant to receive it in the first place.

            At the time of writing, the area of photosensors, and to some lesser extent the number of read-out channels inside the light-field camera seem to determine \NameAcp{}'s costs.
            This assessment is motivated by our early listing of \NameAcp{}'s costs \citep{mueller2019phd}, which includes listings of the cable-robot mount and the support structure \citep{daglas2015master}.
        \subsection{Mirror}
            \label{SecPortalMirror}
            \NameAcp{}'s mirror is a dense array of $\NumFacets{}$ identical, reflective facets arranged in a hexagonal grid.
            Each facet's spherical surfaces has a radius of curvature of $2f$, and an area of $\approx 2\,$m$^{2}$, what is common on large Cherenkov~telescopes \citep{pareschi2013status}.
            The inner, flat-to-flat diameter of the hexagonal facets is $1,500\,$mm and the gap between the facets is $25\,$mm.
            The outer perimeter of the mirror is a hexagon to fill the photosensors in the \Eye{}s, see in \autoref{SecPortalSensor}.
            The facets centers are on a parabola with focal-length $f$.

            The quality of the facets surfaces does not need to exceed the quality of the surfaces used in existing Cherenkov telescopes.
            On most Cherenkov telescopes, the quality of the facets surfaces allows the telescopes to concentrate parallel light, from e.g. a star, into an area on their camera's screen which feeds only into a single photosensor (a single directional bin on the telescope).
            Similar on the plenoscope, the quality of the facets surfaces must only be good enough to concentrate parallel light into the lens of a single eye (a single directional bin on the plenoscope).
            On both the telescope and the plenoscope, the exact position where a photon enters a directional bin (e.g. a light-guide on the telescope or an eye's lens on the plenoscope) is not relevant.
            In case of the plenoscope's eye, this is because the eye's lens is an imaging optics itself which can be described with \autoref{FigThinLens} and \autoref{EqThinLens}.
            This means that a photosensor inside the eye receives light from only one specific direction relative to the lens, but from any possible impact position on the lens.
            So as the position where light passes through an eye's lens does not matter, the mirror facets for the Cherenkov plenoscope do not need to exceed the surface quality of existing mirrors on Cherenkov telescopes.

            Because of the many facets, \NameAcp{} adopts the established alignments of Cherenkov~telescopes which:
            First, measure all the facet's orientations in parallel \citep{mccann2010new, ahnen2016normalized}.
            And second, adjust all the facet's orientations in parallel with an active mirror~control \citep{biland2007active}.
        \subsection{Estimating the Light-Field~Calibration}
            \label{SecPortalLightFieldCalibration}
            Estimating \NameAcp{}'s light-field~calibration with simulations is straight forward, and simulations of optics are well established \citep{bernlohr2013monte, okumura2016robast} in the atmospheric Cherenkov~method.
            But simulations depend on how well one can measure the plenoscope's geometry.
            To ease this, \NameAcp{} takes two measures:

            First, \NameAcp{}'s mechanical structures are designed to deform steady and reproducible with respect to the pointing \citep{daglas2015master}.
            This makes it easier to estimate the plenoscope's geometry from fewer position~sensors, and allows to interpolate between estimates.
            Large Cherenkov~telescopes already use such designs to ease the alignment of their mirror's facets \citep{biland2007active}.

            Second, \NameAcp{} measures the shape of its mirror using the response of its photosensors to a known source of light.
            The idea is inspired by Cherenkov~telescopes which use CCD~cameras and the light of a star to measure the normal vectors on the surface of a mirror in great detail \citep{arqueros2003novel,mccann2010new,ahnen2016normalized}.
            When involving the \Eye{}s in \NameAcp{}'s light-field~camera directly, deformations of the mirror with low spatial frequencies can be kept track of during regular observations.
        \subsection{Estimating the Risk from Sunlight}
            \label{SecSunLight}
            An active mirror~control can also misalign the facets to prevent the mirror from focusing sunlight onto its surrounding.
            Depending on the location on earth, the mirror can also park inclined away from the sun to further minimize the risk from focused sunlight.
            In the worst case when the mirror parks flat on the ground and its facets are well aligned the ground is still safe but reflected sunlight will reach the surrounding towers, see \autoref{FigSunHotSpots}.
            This worst case irradiance of 20\,kWm$^{-2}$ needs to be addressed with passive protection and access limitations on some of the towers, but is still more than a factor of 200 below the irradiance found in solar concentrators \citep{dibowski2023solar}.
        \subsection{Imaging}
            \label{SecPortalImaging}
\begin{figure}{}
                \centering
                \includegraphics[width=0.5\textwidth]{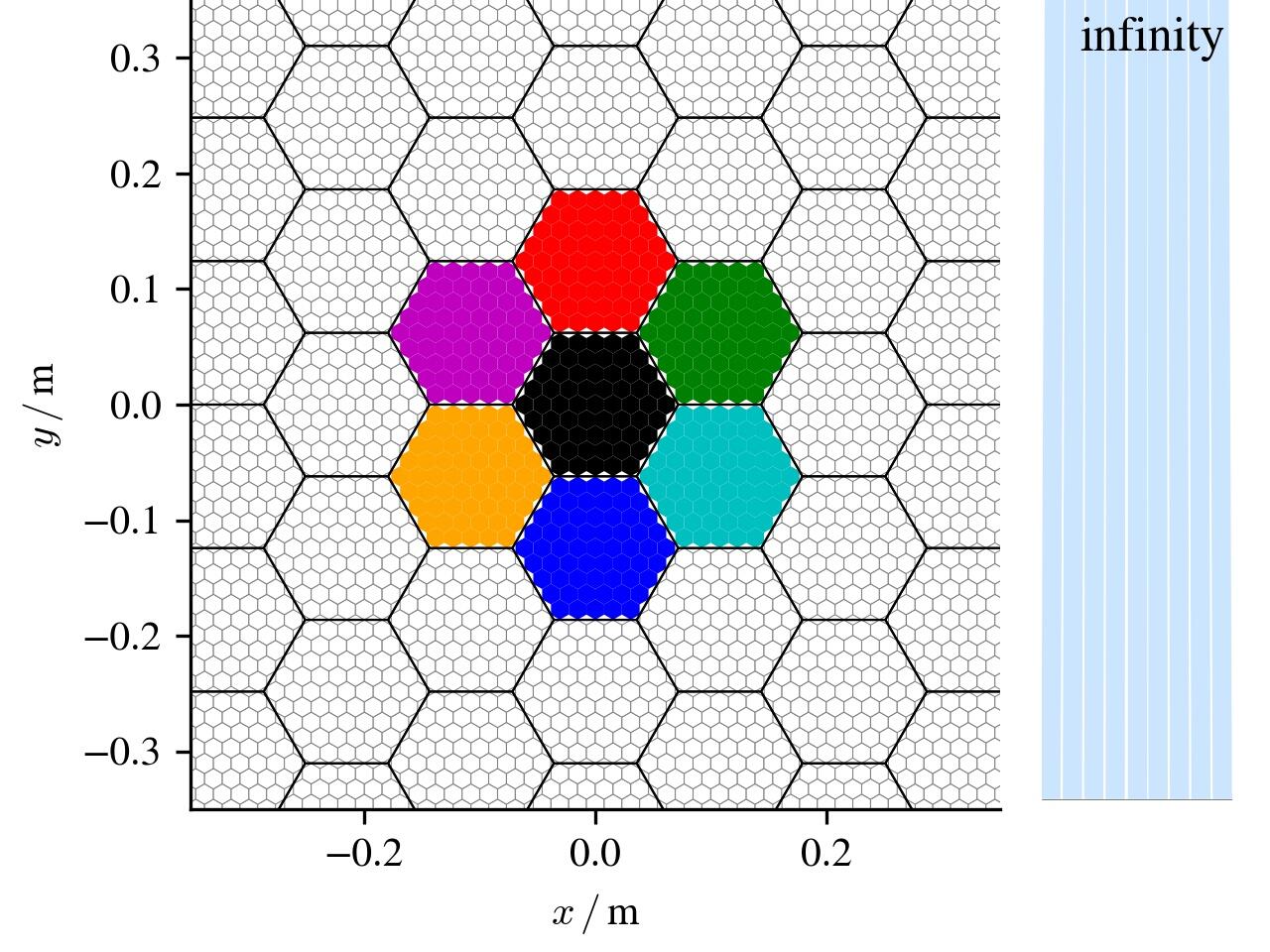}\\
                \vspace{-0.5cm}
                \includegraphics[width=0.5\textwidth]{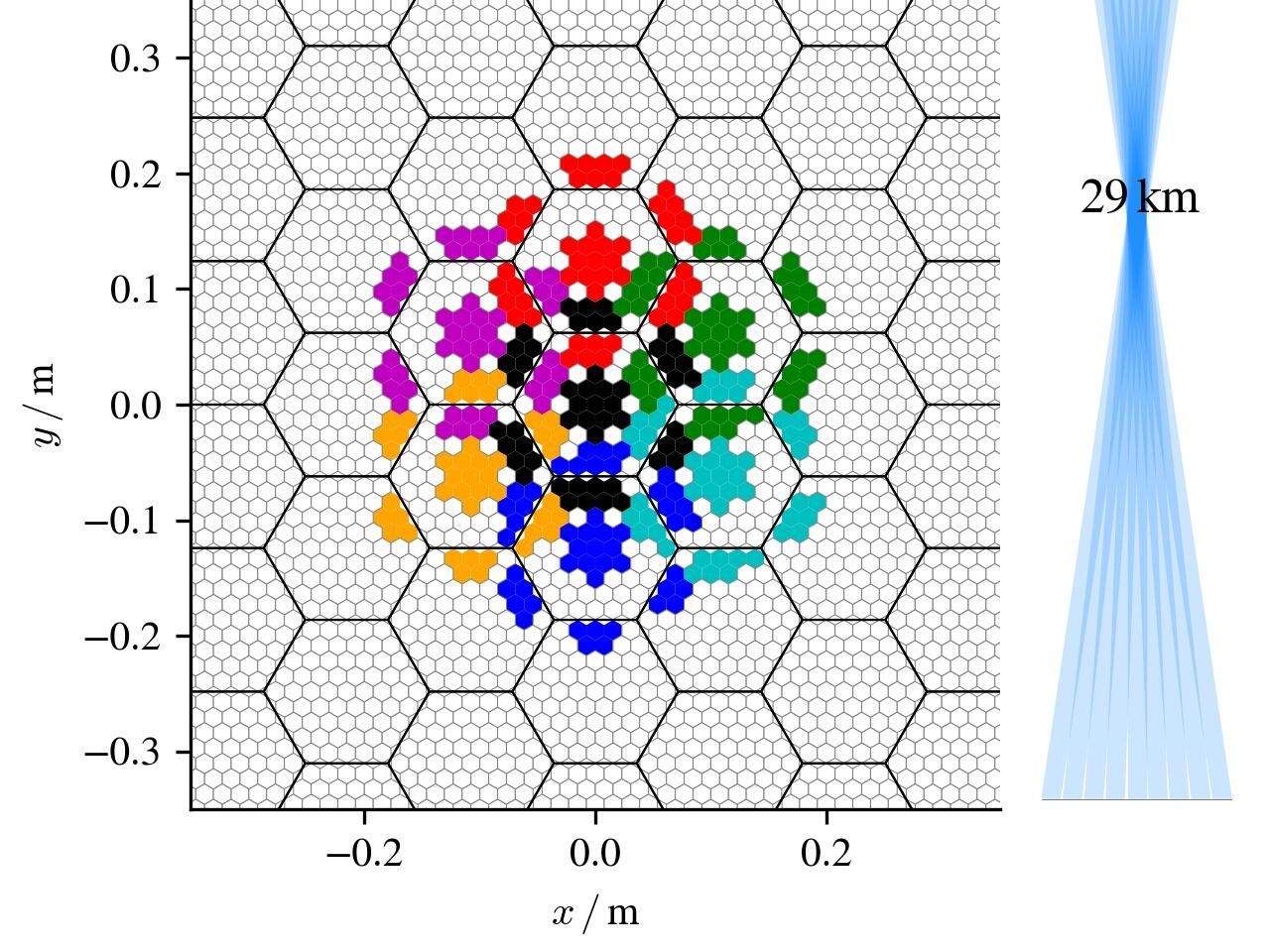}\\
                \vspace{-0.5cm}
                \includegraphics[width=0.5\textwidth]{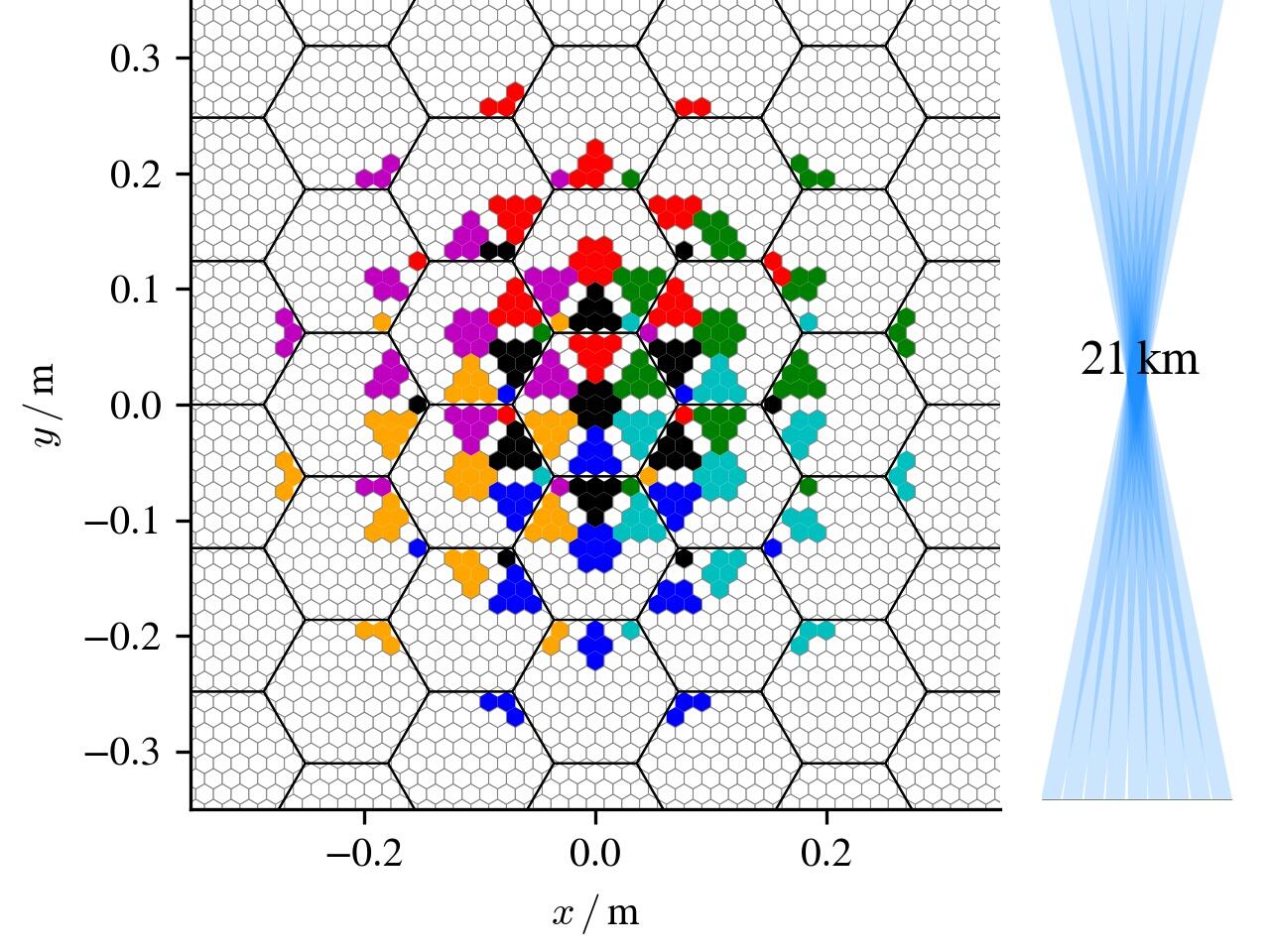}
                \caption[Refocusing, different linear combinations of lixels]{
                    A close-up on the photosensors in \NameAcp{}'s light-field~camera.
                    Large hexagons are \Eye{}s.
                    Small hexagons are photosensors.
                    Seven colors correspond to the seven central pixels of the final image.
                    Three panels show the final image's focus set to different depths in $g = (\infty{}$, 29\,km, 21\,km).
                    Compare \autoref{FigAberrationsNameAcpCloseUp} which also shows $U(g)$, but for larger angles off the mirror's optical axis.
                    State of \NameAcp{}: \StateMirrorCamera{\MirrorGood{}}{\CameraGood{}}.
                }
                \label{FigPixelRefocusedLinearCombinationSensorPlaneColored}
            \end{figure}

            \autoref{FigPixelRefocusedLinearCombinationSensorPlaneColored} visualizes the elements of matrix $U(g)$ from \autoref{EqRefocusedImagingMatrix} on the \NameAcp{} Cherenkov~plenoscope.
            Here in this demonstration, we set the binning of the pixels in the final image so that the central pixel falls together with the central \Eye{} in \NameAcp{}'s light-field~camera.
            The final image itself is not shown here, but one finds that in this special case all the photosensors of an \Eye{} contribute to the same pixel.
            At least when the pixel is close to the mirror's optical axis where aberrations are low, and when the image's focus corresponds to the camera's focus.
            This very special case is the upper panel in \autoref{FigPixelRefocusedLinearCombinationSensorPlaneColored}.
            When the final image changes its focus, one finds that photosensors from multiple \Eye{}s contribute to a pixel.
            This is what the lower panels in \autoref{FigPixelRefocusedLinearCombinationSensorPlaneColored} show.
            One finds that photosensors which contribute to the same pixel are in close, mechanical proximity.
            This proximity can be minimized when the camera's focus corresponds to the final image's focus.
            In the field, one would focus the light-field~camera to the average depth of the showers one wants to record to minimize this proximity.
            Minimizing this proximity eases the implementation of fast imaging close to the hardware.
            And fast imaging is relevant for the trigger.
        \subsection{Trigger}
            \label{SecPortalTrigger}
            \NameAcp{} adopts two powerful concepts from the Cherenkov~telescope's trigger.
            First, it acts on an image, and second it sums up the signals of multiple photosensors \citep{garcia2014status} to reduce accidental triggers.
            To provide the trigger with an image, \NameAcp{}'s light-field~camera continuously projects its light~field onto an image.
            \autoref{EqRefocusedImaging} shows how this projection is implemented in hardware as the matrix's elements $u_{n,\,k}(g)$ mask if a photosensor $k$ contributes its signal to the sum of pixel $n$ or not.
            Inside the light-field~camera are $N_\text{trigger}$ hubs which sum the signals of specific photosensors.
            These hubs are evenly spread in between the light-field~camera's \Eye{}s and represent the $N_\text{trigger}$ pixels of the trigger's image.
            Each hub runs signal~routes according to matrix \autoref{EqRefocusedImagingMatrix} to specific photosensors.
            Due to the plenoscope's design, these photosensors are already in the mechanical proximity of the hub, see \autoref{FigPixelRefocusedLinearCombinationSensorPlaneColored}, and do not require to be delayed in time for different pointings.
            This way from a signal-processing point-of-view, \NameAcp{}'s trigger is a sum~trigger that acts on an image exactly the way a Cherenkov~telescope's sum~trigger acts.

            In addition the plenoscope's trigger can be extended to perceive depth.
            To do so, \NameAcp{}'s trigger projects its light~field not only onto one, but onto two images.
            By installing the hubs for summation twice, but applying two different imaging matrices $U(g_\text{far})$, and $U(g_\text{close})$, \NameAcp{}'s trigger acts on two images.
            The far image focuses on the depth $g_\text{far}$, and the close image focuses on the depth $g_\text{close}$.
            With this perception of depth, \NameAcp{} can trigger faint showers exclusively when they emit their light high up in the atmosphere (far image), what is typical for low energetic gamma~rays.
            This way, and on a statistical basis, \NameAcp{}'s trigger can be made to favor low energetic gamma~rays over faint sources that emit their light further down in the atmosphere, such as parts of hadronic showers impacting at large distances.
        \subsection{Optical Performance}
\begin{figure}{}
                \centering
                \includegraphics[width=1.0\columnwidth]{
                    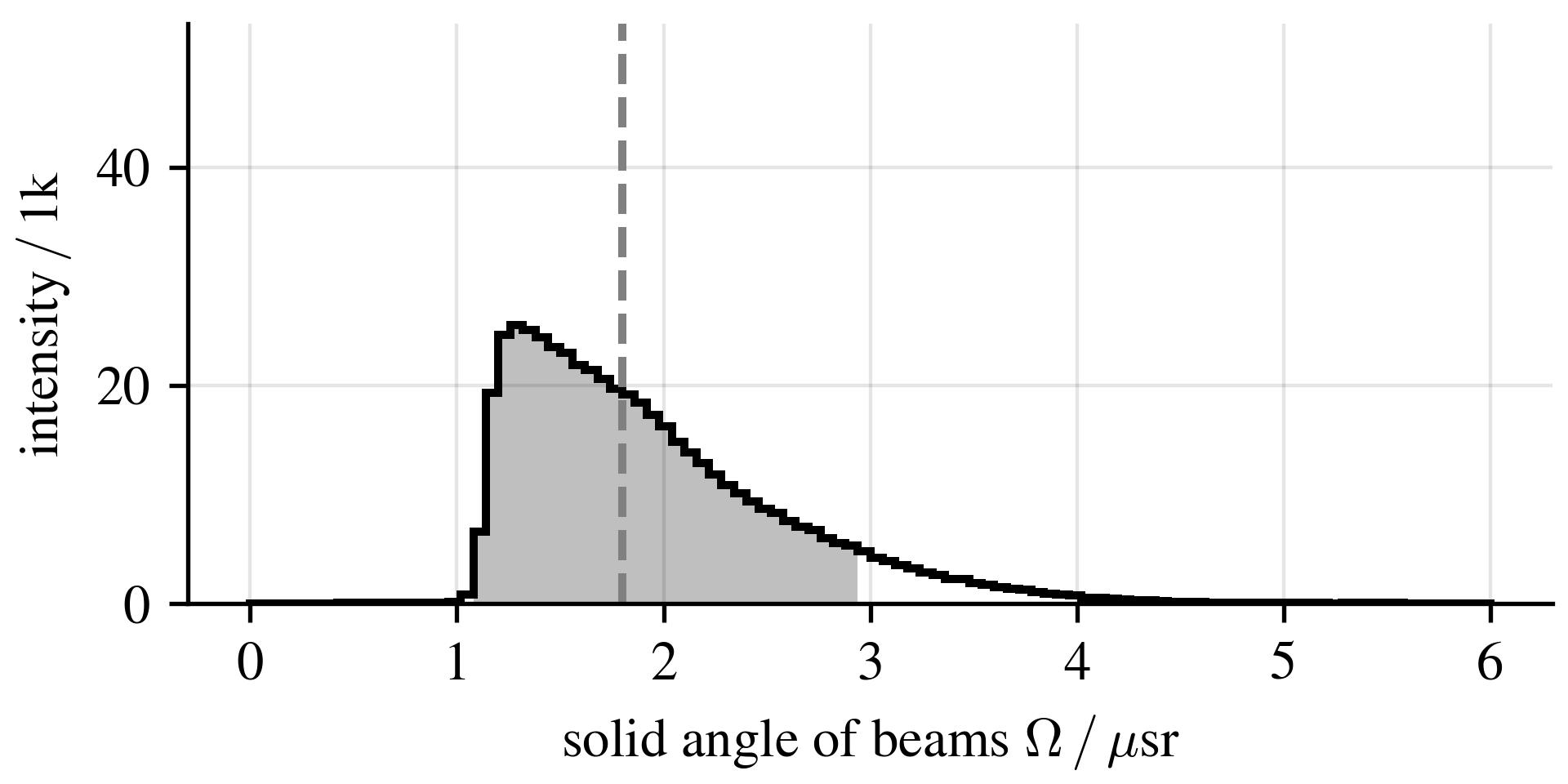
                }
                \includegraphics[width=1.0\columnwidth]{
                    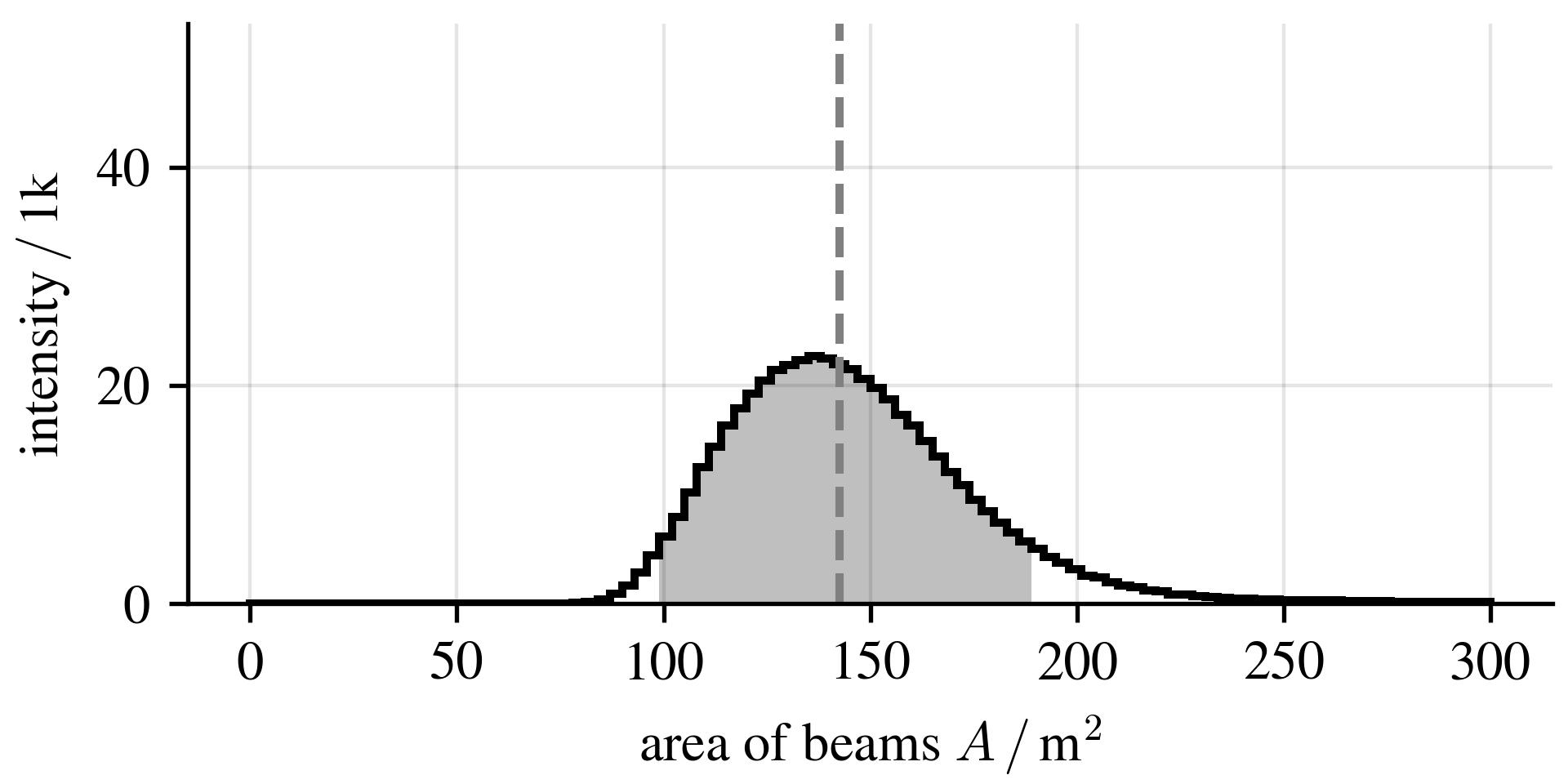
                }
                \includegraphics[width=1.0\columnwidth]{
                    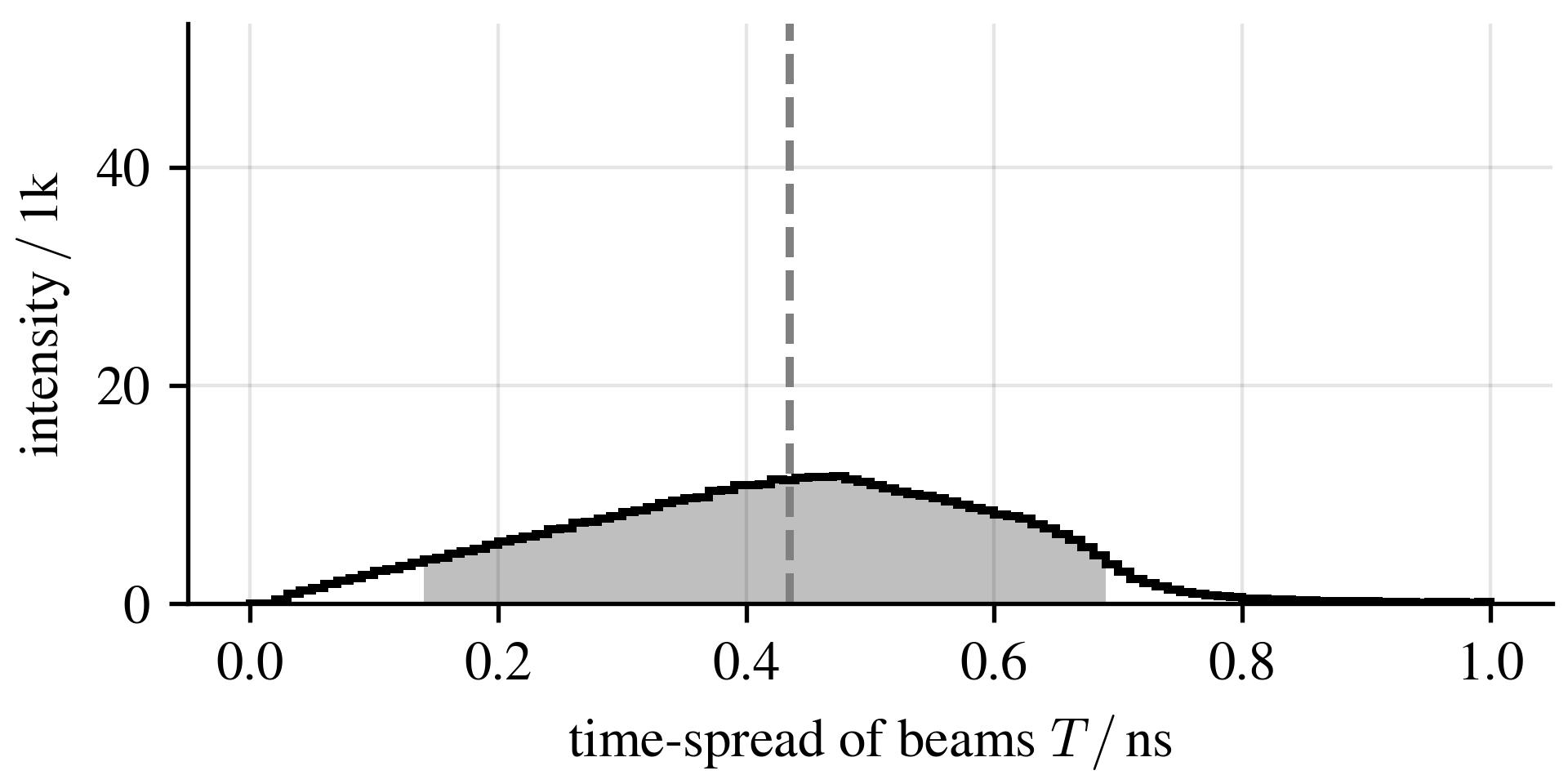
                }
                \includegraphics[width=1.0\columnwidth]{
                    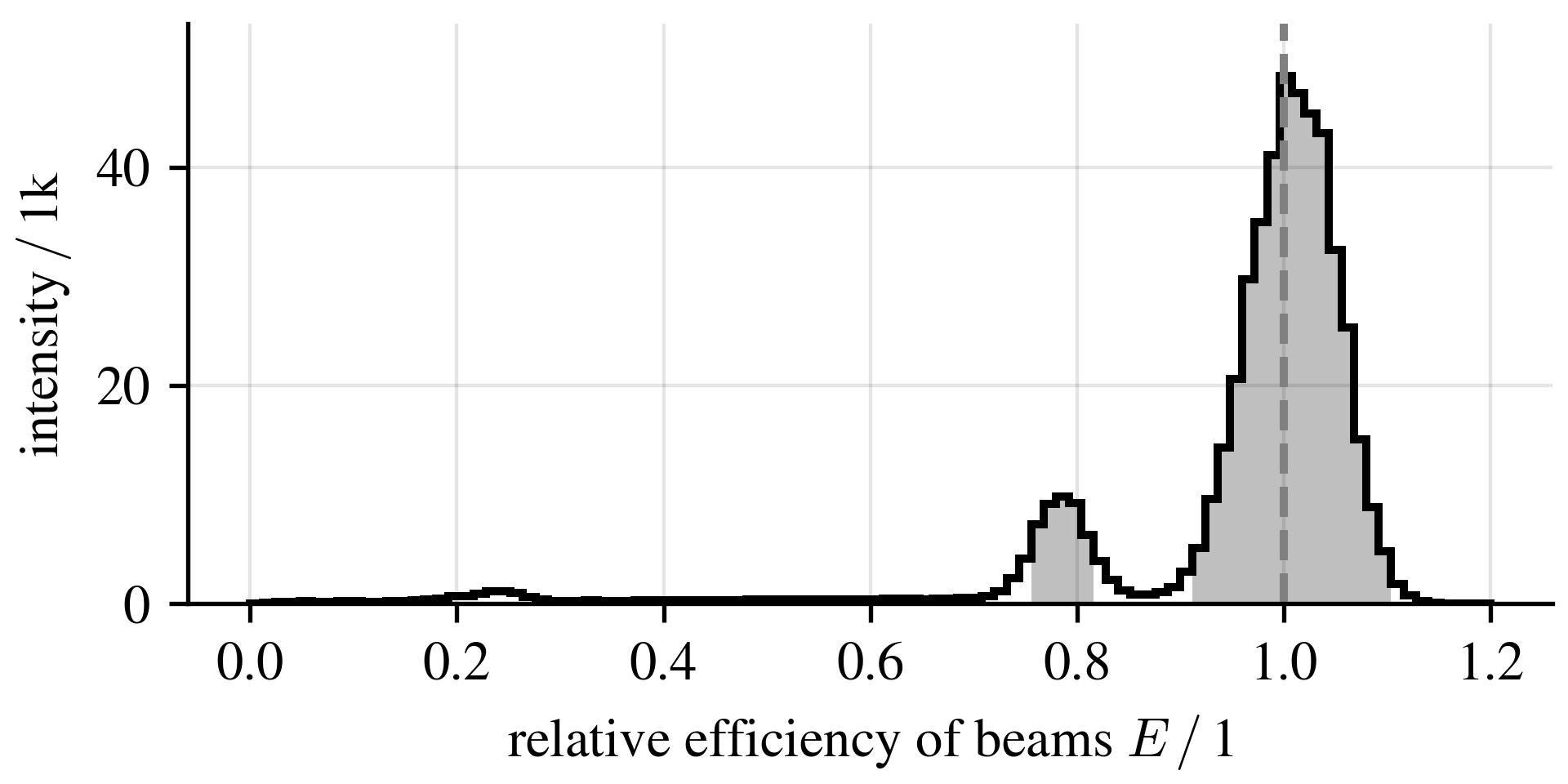
                }
                \caption{
                    Statistics of \NameAcp{}'s $K$ beams gathered from its light-field~calibration $\LightFieldGeometry{}_\text{default}$ when \NameAcp{} is in its default geometry, i.e. \NameAcp{}'s mirror is not deformed and its light-field~camera is both centered on, and orientated perpendicular to the mirror's optical axis.
                    Dashed, vertical, gray line is the median.
                    Light gray areas show the $90\%$ percentile.
                    Statistics is $\approx1000$ optical paths per beam.
                    Compare \autoref{SecCharacterizingBeam}, \autoref{EqBeamSolidAngle} to \autoref{EqBeamEfficiency}.
                    State of \NameAcp{}: \StateMirrorCamera{\MirrorGood{}}{\CameraGood{}}.
                }
                \label{FigBeamStatistics}
            \end{figure}

            We will discuss \NameAcp{}'s point-spread-function for imaging in the following sections but because images are only projections of light~fields we first want to look at the beams which sample this light~field.
            A simple characterization of the $k$-th beam $\Beam{}_k$ is its spread in solid angle in the sky $\BeamSolidAngle{}_k$, its spread in area on the aperture's principal plane $\BeamArea{}_k$, its spread in time that can not be resolved $\BeamTimeSpread{}_k$, and its optical efficiency $\BeamEfficiency{}_k$.
            \autoref{FigBeamStatistics} shows the distributions of these spreads for \NameAcp{}, and \autoref{SecCharacterizingBeam} shows how we define those.
            In summary: \NameAcp{}'s beams resolve solid angles of 1.8\,$\mu$sr, (a cone with an half-angle of $0.043^{\circ}$), they resolve areas of $160\,$m$^2$ (a disk of $14.3$\,m in diameter) on the aperture's principal plane, they induce an unresolvable spread in time of $0.4\,$ns, and their optical efficiencies are rather similar.
            The clusters in the beams efficiencies in \autoref{FigBeamStatistics} change with the camera's alignment and the mirror's shape and are caused by the camera shadowing the center of the mirror.

\begin{figure}
                \centering
                \includegraphics[width=1\columnwidth]{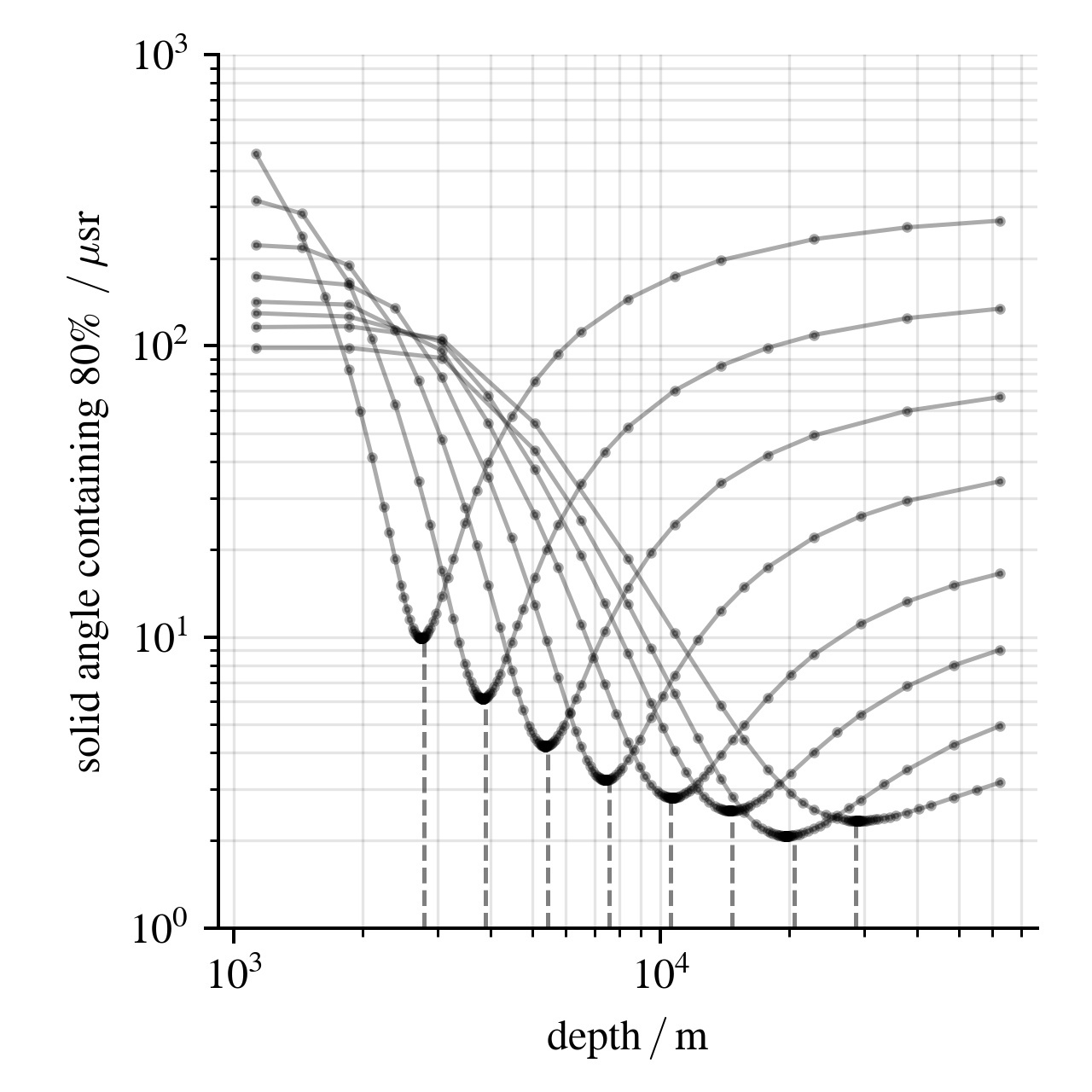}
                \caption[]{
                    \NameAcp{} observes eight isolated point like sources in eight different depths.
                    Each solid line is one observation and the dots indicate projections of the observed light~field onto an image focusing to a specific depth.
                    The dashed vertical lines indicate the true depth of the point like sources.
                    State of \NameAcp{}: \StateMirrorCamera{\MirrorGood{}}{\CameraGood{}}.
                }
                \label{FigRefocusRunsP61}
            \end{figure}

\begin{figure}{}
                \centering
                \includegraphics[width=1.0\columnwidth]{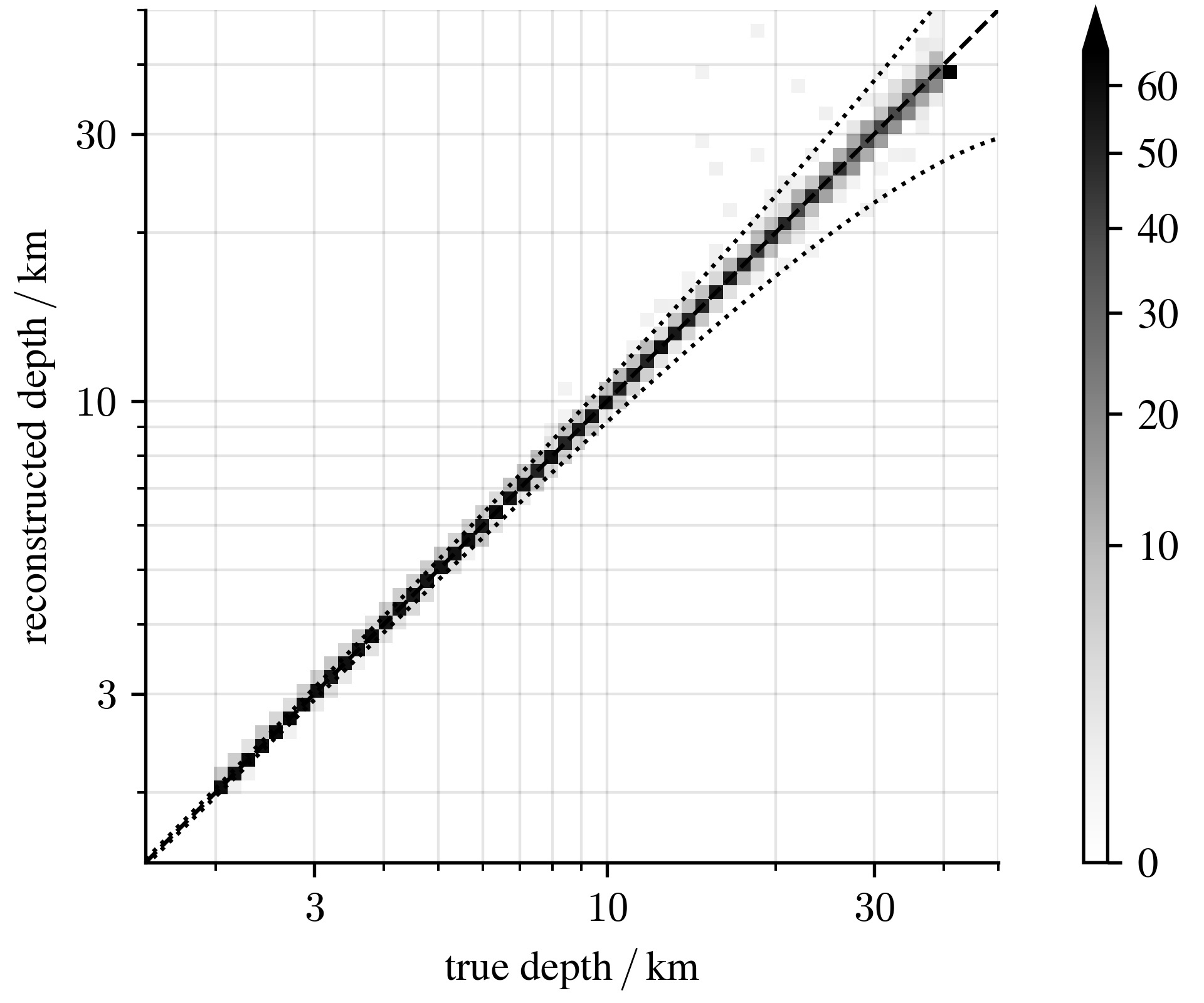
                }
                \caption{
                    \NameAcp{}'s confusion in depth for isolated point like sources randomly scattered in depth and directions (across \NameAcp{}'s field-of-view).
                    Dotted lines show \NameAcp{}'s $g_{-}$ and $g_{+}$ according to \autoref{EqDepthOfField}.
                    State of \NameAcp{}: \StateMirrorCamera{\MirrorGood{}}{\CameraGood{}}.
                }
                \label{FigResolvingDepth}
            \end{figure}

            The plenoscope's performance to perceive depth illustrates the overlap of its beams.
            \autoref{FigRefocusRunsP61} shows how the blur in an image of a point like source changes when the focus of the image changes.
            Indeed, \autoref{FigRefocusRunsP61} shows this for eight isolated sources which where observed independently.
            One finds that the focus with the minimal spread is a good estimator for the true depth of the source.
            \autoref{FigResolvingDepth} histograms the confusion of many of such estimates where a source is observed and has its depth reconstructed by \NameAcp{} minimizing the spread in its refocused image.
            The sources are randomly placed in a depth ranging from $2\,$km to $40\,$km and are uniformly distributed in \NameAcp{}'s field-of-view.
            In the histogram one finds \NameAcp{}'s confusion to show more at further depths, as one would expect from any stereoscopic observation with a finite baseline, and as the range $(g_{+} - g_{-})$ in \autoref{EqDepthOfField} predicts it.
            Still, the range in depth with good resolution comfortably covers a typical shower induced by a cosmic gamma~ray.

            But keep in mind that \autoref{FigRefocusRunsP61} and \autoref{FigResolvingDepth} show the case of observing a bright point like source what gives better performance than what is to be expected from the reconstruction of an actual shower, initiated by a 1\,GeV gamma~ray, which emits less Cherenkov~photons and is more spread out in space.
            Similar to the telescope's point-spread-function, which is also showing a bright point like source, it might be good to interpret \autoref{FigRefocusRunsP61} and \autoref{FigResolvingDepth} as the plenoscope's `depth-spread-function'.
        \subsection{Comparing to a Telescope, and to a Variant}
            \label{SecNameAcpsVariants}
            To compare the \NameAcp{} Cherenkov~plenoscope to a telescope, we introduce a Cherenkov~telescope named `\Pone{}'.
            The difference between \Pone{} and \NameAcp{} is in how well they can resolve the position where a photon is reflected on their mirrors.
            The \Pone{} telescope naturally resolves its mirror in only 1\,bin, hence the `1' in \Pone{}.
            The \NameAcp{} plenoscope resolves its mirror in 61\,bins because its \Eye{}s have 61\,photosensors each.
            One might refer to \NameAcp{} as `\PsixtyOne{}' in this scheme.
            To further demonstrate how this resolution impacts plenoptic perception we also introduce a plenoscope `\Pseven{}' which is also identical to \NameAcp{} except for it only has 7\,bins to resolve its mirror.
            To make a fair comparison, we estimate the light-field~calibration for each instrument, including \Pone{}, and apply it to compute images.
            This way, the few effects that a telescope can compensate (distortion, vignetting, specific misalignments) are taken care of equally for every instrument.

            In addition, one wants to adjust the focus of the telescope \Pone{}, such that its images of stars are sharp.
            This is because when we will compare the variants (\Pone{}, \Pseven{}, and \PsixtyOne{}), we will use star light.
            Since \NameAcp{} can compensate translations of its camera relative to its mirror, this adjustment of the focus is not important for the \NameAcp{} (\PsixtyOne{}) plenoscope.
            But for the telescope \Pone{}, adjusting the focus is important \citep{trichard2015enhanced,hofmann2001focus}.
            When searching for the highest concentration of star light from \NameAcp{}'s mirror, one finds that the focal-point, which one estimates this way, is $1.66$\,m closer to the base of the mirror's parabola than the parabola's theoretical focal-point.
            After all, segmented mirrors with facets on a parabola are not the same as parabolic mirrors.
            To account for this, one translates the cameras screens of all variants (\Pone{}, \Pseven{}, and \PsixtyOne{}) into this estimated focal-point, and orientates the camera's screen to be perpendicular to the mirror's optical axis.
            For reference, we mark this using the expression \mbox{`camera: \textit{\CameraGood{}}'} wherever it is applicable, see e.g. \autoref{FigPsfDefaultDefault}.

            With the estimated focal-point, one also found the aperture's principal plane of \NameAcp{}'s mirror which is located $1.66$\,m below the mirror's parabola which mounts the facets.
            When one estimates the light-field calibration for the variants (\Pone{}, \Pseven{}, and \PsixtyOne{}), one uses this estimated principal plane together with the optical axis as the frame of reference, compare \autoref{SecDefiningTheLightFieldCalibration}.
    \section{Compensating Aberrations}
        \label{SecCompensatingAberrations}
\begin{figure}
            \centering
            Telescope \Pone{}\\
            \begin{minipage}{.39\columnwidth}
                \includegraphics[width=1.0\columnwidth]{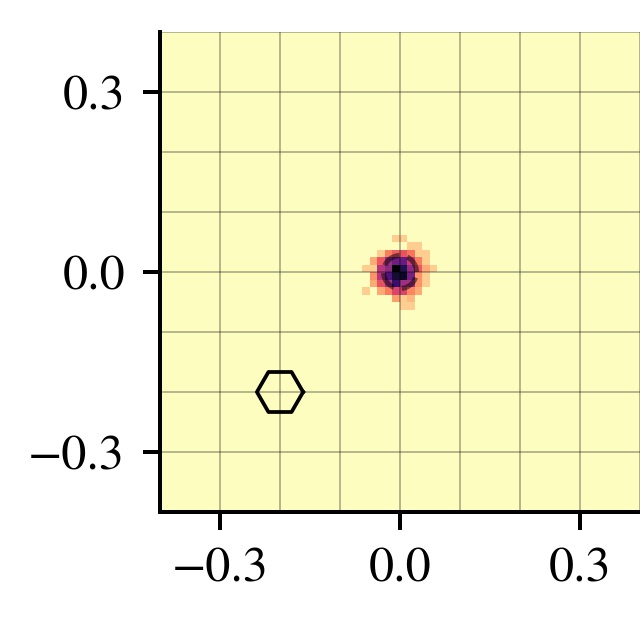}
            \end{minipage}
            \begin{minipage}{.29\columnwidth}
                \includegraphics[width=1.0\columnwidth]{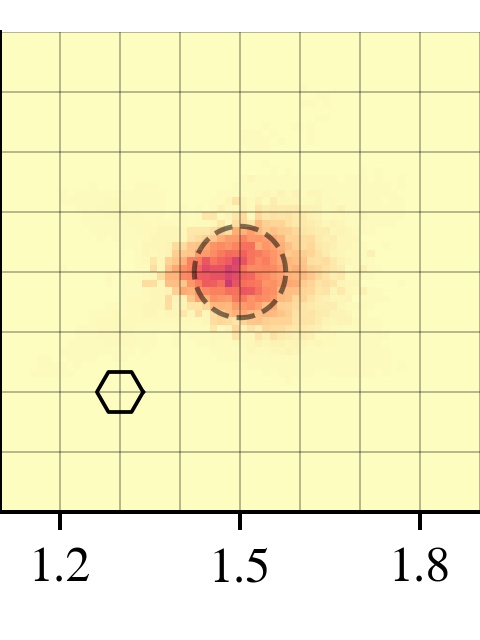}
            \end{minipage}
            \begin{minipage}{.29\columnwidth}
                \includegraphics[width=1.0\columnwidth]{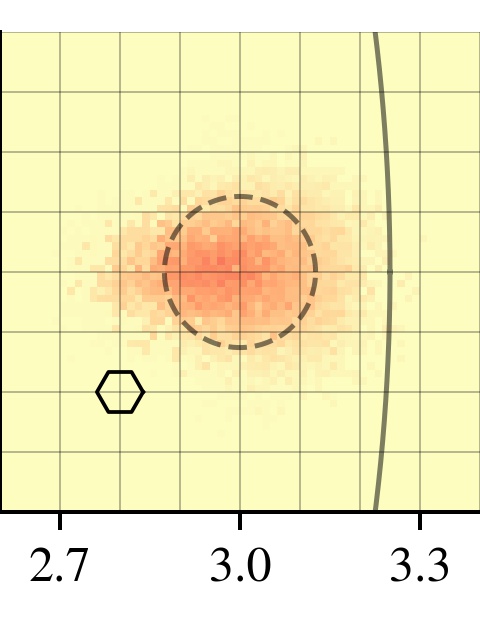}
            \end{minipage}
            Plenoscope \Pseven{}\\
            \begin{minipage}{.39\columnwidth}
                \includegraphics[width=1.0\columnwidth]{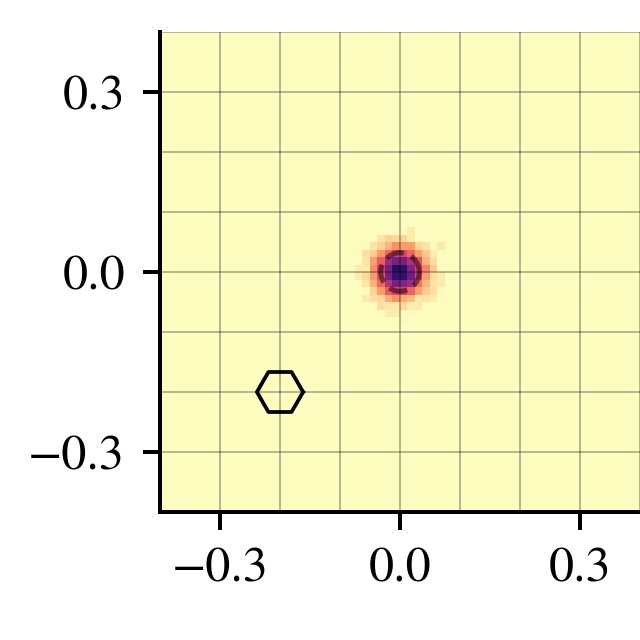}
            \end{minipage}
            \begin{minipage}{.29\columnwidth}
                \includegraphics[width=1.0\columnwidth]{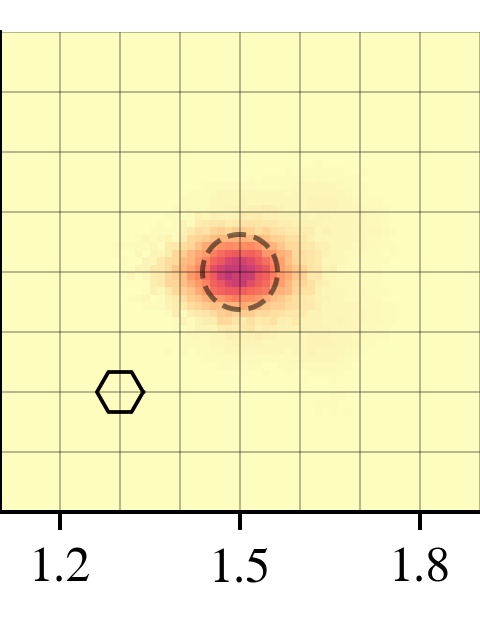}
            \end{minipage}
            \begin{minipage}{.29\columnwidth}
                \includegraphics[width=1.0\columnwidth]{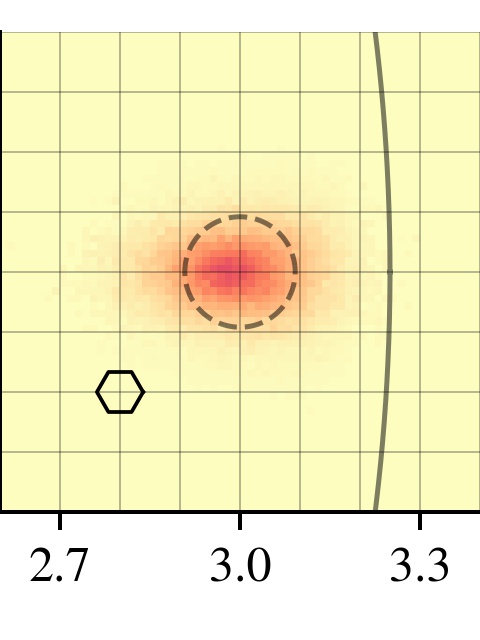}
            \end{minipage}
            \NameAcp{} Plenoscope \PsixtyOne{}\\
            \begin{minipage}{.39\columnwidth}
                \includegraphics[width=1.0\columnwidth]{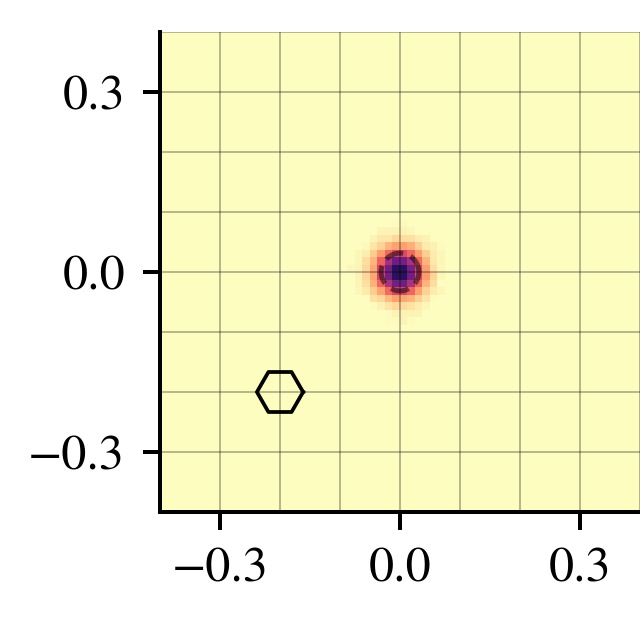}
            \end{minipage}
            \begin{minipage}{.29\columnwidth}
                \includegraphics[width=1.0\columnwidth]{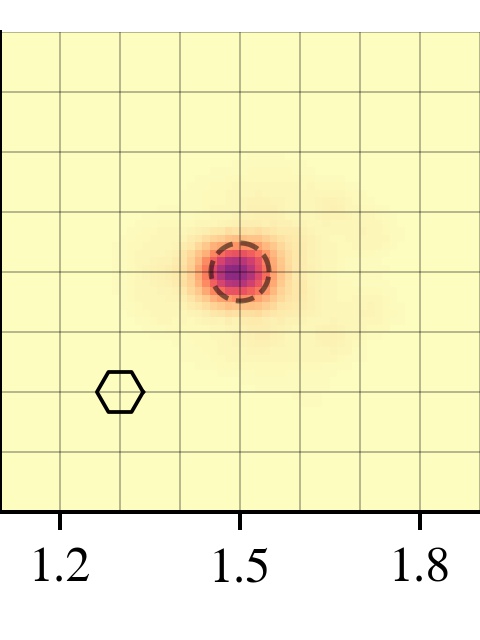}
            \end{minipage}
            \begin{minipage}{.29\columnwidth}
                \includegraphics[width=1.0\columnwidth]{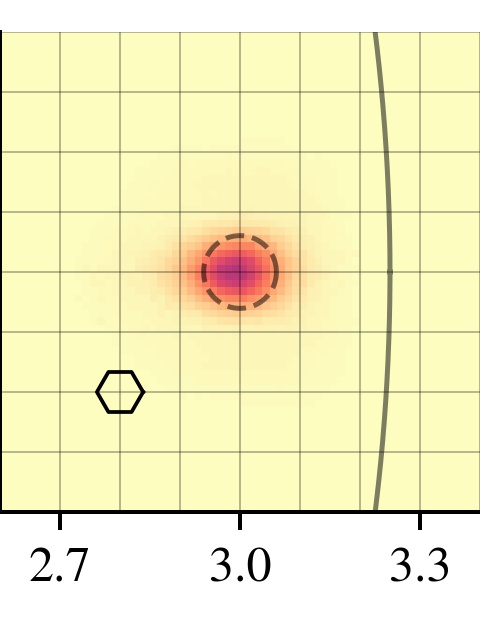}
            \end{minipage}
            \includegraphics[width=1.0\columnwidth]{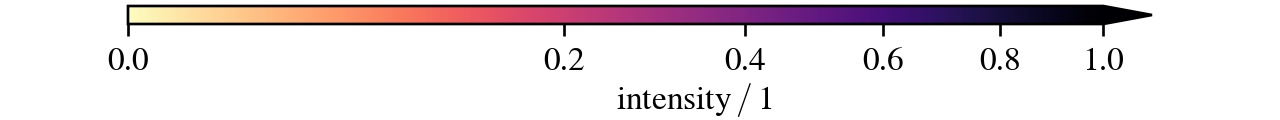}
            \caption[]{
                \FigPsfTextIntro{} while all optics have their targeted geometry: \StateMirrorCamera{\MirrorGood{}}{\CameraGood{}}.
                \FigPsfTextAutoRef{FigPsfVsOffAxisDefaultDefault}\\
                Legend:
                The three columns show the star being 0.0$^{\circ}$, 1.5$^{\circ}$, and 3.0$^{\circ}$ off the mirror's optical axis.
                Axes are in units of $1^{\circ}$.
                The little hexagon shows the field-of-view of a telescope's photosensor and plenoscope's \Eye{} respectively.
                The dashed ring shows the containment of 80\% of the star's light.
                The big ring in the most right panels shows the limit of the instrumented field-of-view.
            }
            \label{FigPsfDefaultDefault}
        \end{figure}

\begin{figure}
            \centering
            \includegraphics[width=1.0\columnwidth]{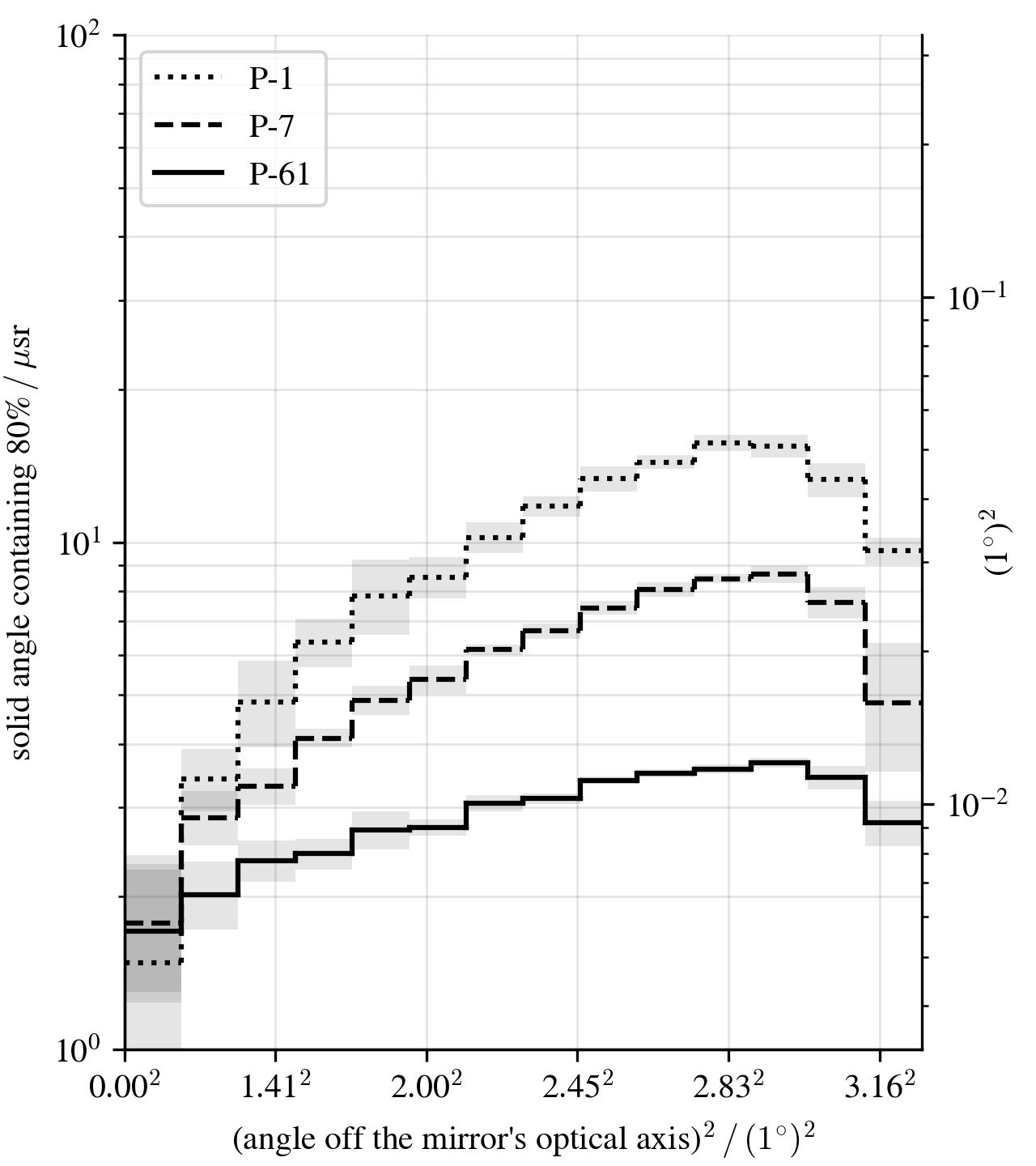}
            \caption[]{
                \FigPsfVsOffAxisTextIntro{} while all optics have their targeted geometry: \StateMirrorCamera{\MirrorGood{}}{\CameraGood{}}.
                \FigPsfVsOffAxisTextAutoRef{FigPsfDefaultDefault}\\
                Legend:
                Stars are randomly placed in the sky.
                The stars are binned according to their angles off the optical axis.
                The bins cover equally large solid angles.
                The lines show the average spread of the stars images within a bin.
            }
            \label{FigPsfVsOffAxisDefaultDefault}
        \end{figure}

\begin{figure}{}
            \centering
            \subfloat[][An overview of \NameAcp{}'s plane of photosensors.]{
                \includegraphics[width=1\columnwidth]{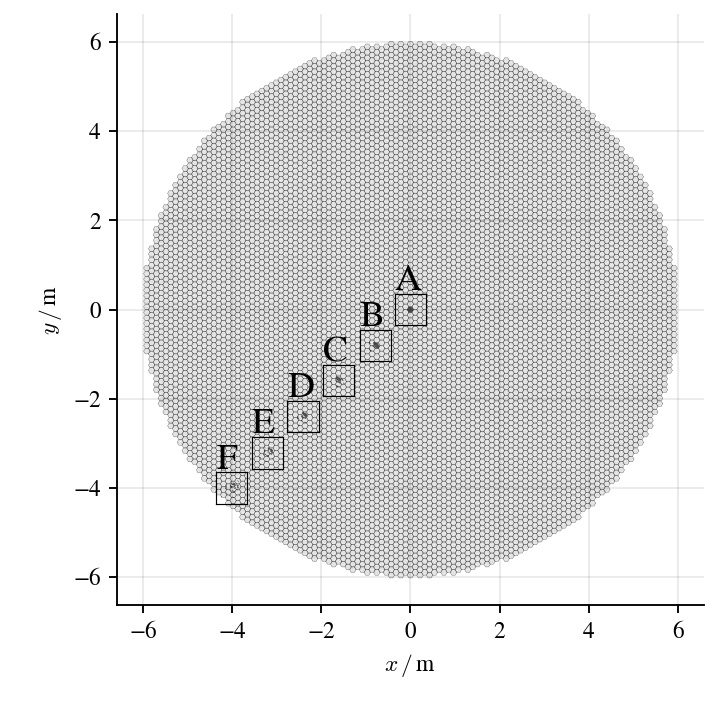}
                \label{FigAberrationsNameAcpOverview}
            }
            \\
            \subfloat[][Close-ups on \NameAcp{}'s plane of photosensors.]{
                \begin{tabular}[b]{ccc}
                    \begin{minipage}{.32\columnwidth}
                        \includegraphics[width=\columnwidth]{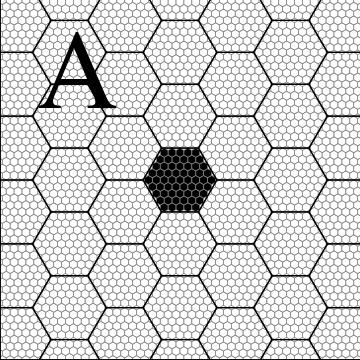}
                    \end{minipage}
                    \hspace{.003\columnwidth}
                    \begin{minipage}{.32\columnwidth}
                        \includegraphics[width=\columnwidth]{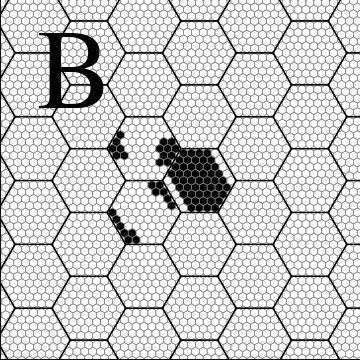}
                    \end{minipage}
                    \hspace{.003\columnwidth}
                    \begin{minipage}{.32\columnwidth}
                        \includegraphics[width=\columnwidth]{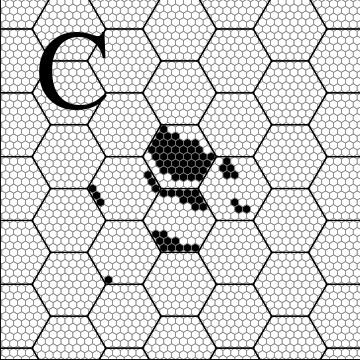}
                    \end{minipage}
                    \vspace{2mm}
                    \\
                    \begin{minipage}{.32\columnwidth}
                        \includegraphics[width=\columnwidth]{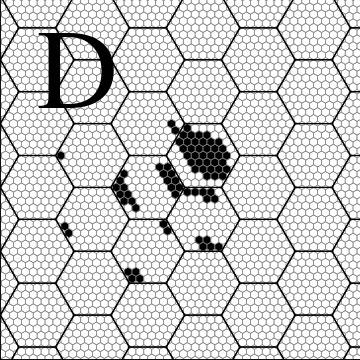}
                    \end{minipage}
                    \hspace{.003\columnwidth}
                    \begin{minipage}{.32\columnwidth}
                        \includegraphics[width=\columnwidth]{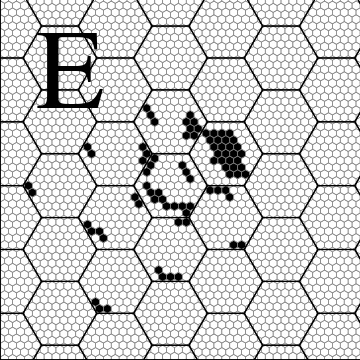}
                    \end{minipage}
                    \hspace{.003\columnwidth}
                    \begin{minipage}{.32\columnwidth}
                        \includegraphics[width=\columnwidth]{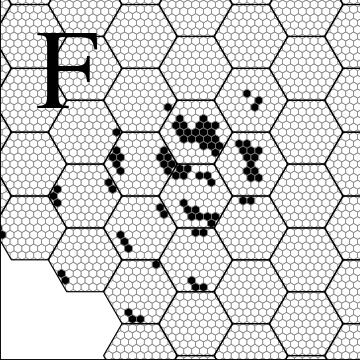}
                    \end{minipage}
                    \label{FigAberrationsNameAcpCloseUp}
                \end{tabular}
            }
            \caption{
                The pixels angles off the mirror's optical axis are: A: 0.0$^\circ$, B: 0.6$^\circ$, C: 1.2$^\circ$, D: 1.8$^\circ$, E: 2.4$^\circ$, and F: 3.0$^\circ$.
                Dark photosensors show the non-zero elements in the rows of the imaging matrix $U(g)$ in \autoref{EqRefocusedImagingMatrix}.
                State of \NameAcp{}: \StateMirrorCamera{\MirrorGood{}}{\CameraGood{}}.
            }
        \end{figure}

        Real mirrors induce aberrations in images.
        In Cherenkov~telescopes, the mirror's aberrations limit the field-of-view as its blurring gets stronger with larger angles off the mirror's optical axis.
        A significant enlargement of the field-of-view is currently only possible when concatenating aspheric, and non-flat optical surfaces \citep{vassiliev2007schwarzschild}.
        \autoref{FigPsfDefaultDefault} shows the reconstructed images of a star observed by the \NameAcp{} plenoscope and the telescope \Pone{} for three different angles off the mirror's optical axis.
        The images are computed according to \autoref{SecProjectingTheLightFieldOntoImages}.
        Both telescope and plenoscope offer a small point-spread-function in the center of the field-of-view, but for larger angles off the mirror's optical axis, the blurring in the telescope's image increases rapidly.
        On the other hand, the \NameAcp{} plenoscope's point-spread-function is hardly effected by the light's direction.
        \autoref{FigPsfVsOffAxisDefaultDefault} shows how the plenoscope's and the telescope's point-spread-functions scale with the light's angle off the mirror's optical axis.
        On average across the field-of-view, the telescope \Pone{} contains 80\% of the light from a star in a cone with a solid angle of 12\,$\mu$sr (half-angle 0.11$^{\circ}$).
        The \NameAcp{} plenoscope contains the same light in a quarter of the solid angle of 3.2\,$\mu$sr (half-angle 0.058$^{\circ}$).
        By means of the atmospheric Cherenkov~method, \NameAcp{}'s images are very sharp and very `flat' i.e. very independent of the direction of the light.
        To visualize how \NameAcp{} compensates the aberrations one can again color the elements of the imaging matrix $U$ for certain pixels, see. \autoref{FigAberrationsNameAcpOverview}, and \autoref{FigAberrationsNameAcpCloseUp}.
        One finds that the aberrations of \NameAcp{}'s mirror are small when close to the optical axis (A), but become stronger the further the light is off the optical axis (F).
        The aberration's spread becoming stronger with larger angles corresponds to the growing distances between the colored photosensors.
        Just like in \autoref{FigPixelRefocusedLinearCombinationSensorPlaneColored}, in the special case of the pixel being on the mirror's optical axis the summation of photosensors simply runs over the central \Eye{}, see panel (A) in \autoref{FigAberrationsNameAcpCloseUp}.
    \section{Compensating Deformations}
        \label{SubSecCherenkovPlenoscopeDeformations}
\begin{figure}
            \centering
            \small{Surface's height along the optical axis.}
            \includegraphics[width=1\columnwidth]{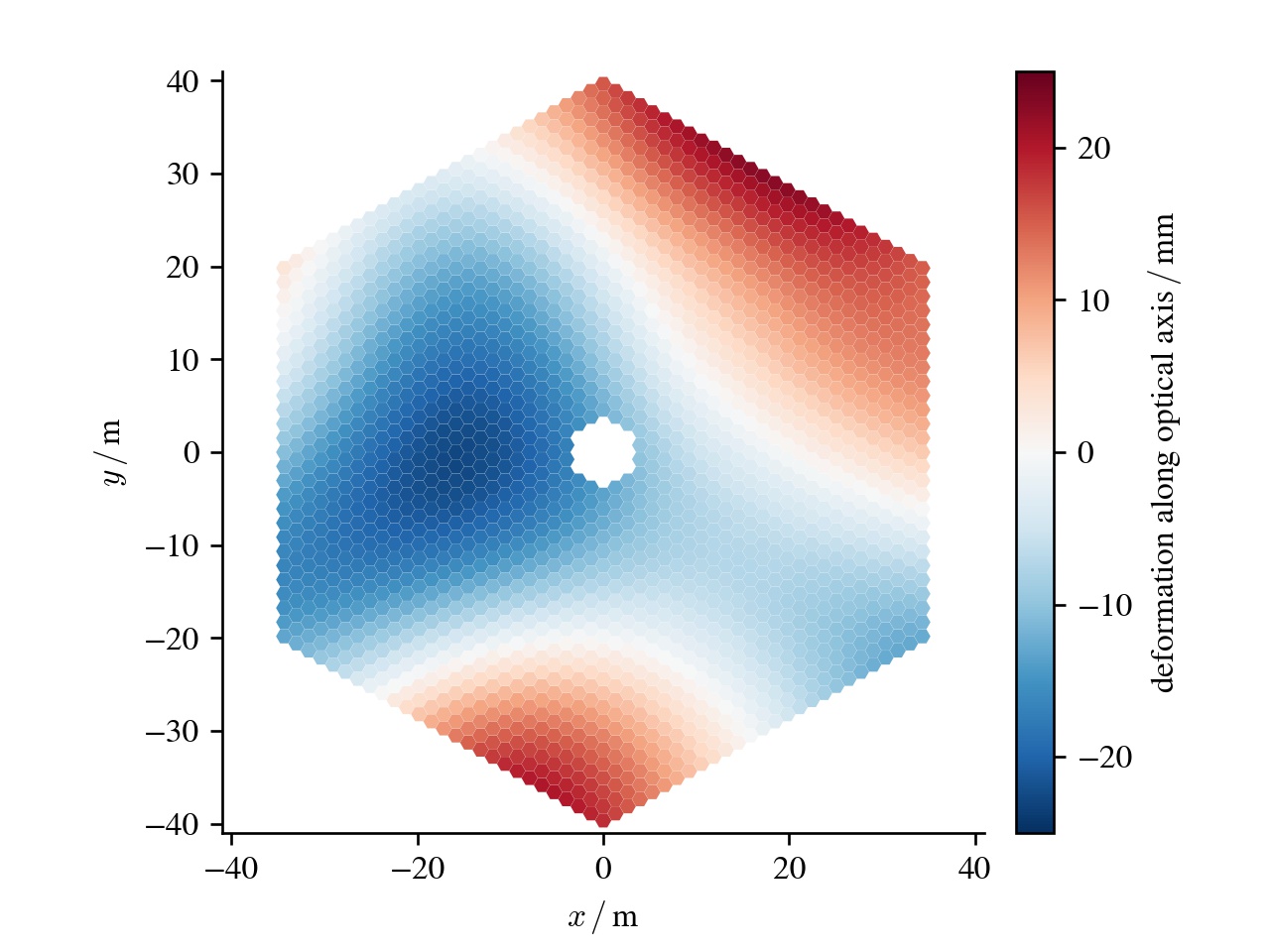}
            \small{Surface's normal vector}.
            \includegraphics[width=1\columnwidth]{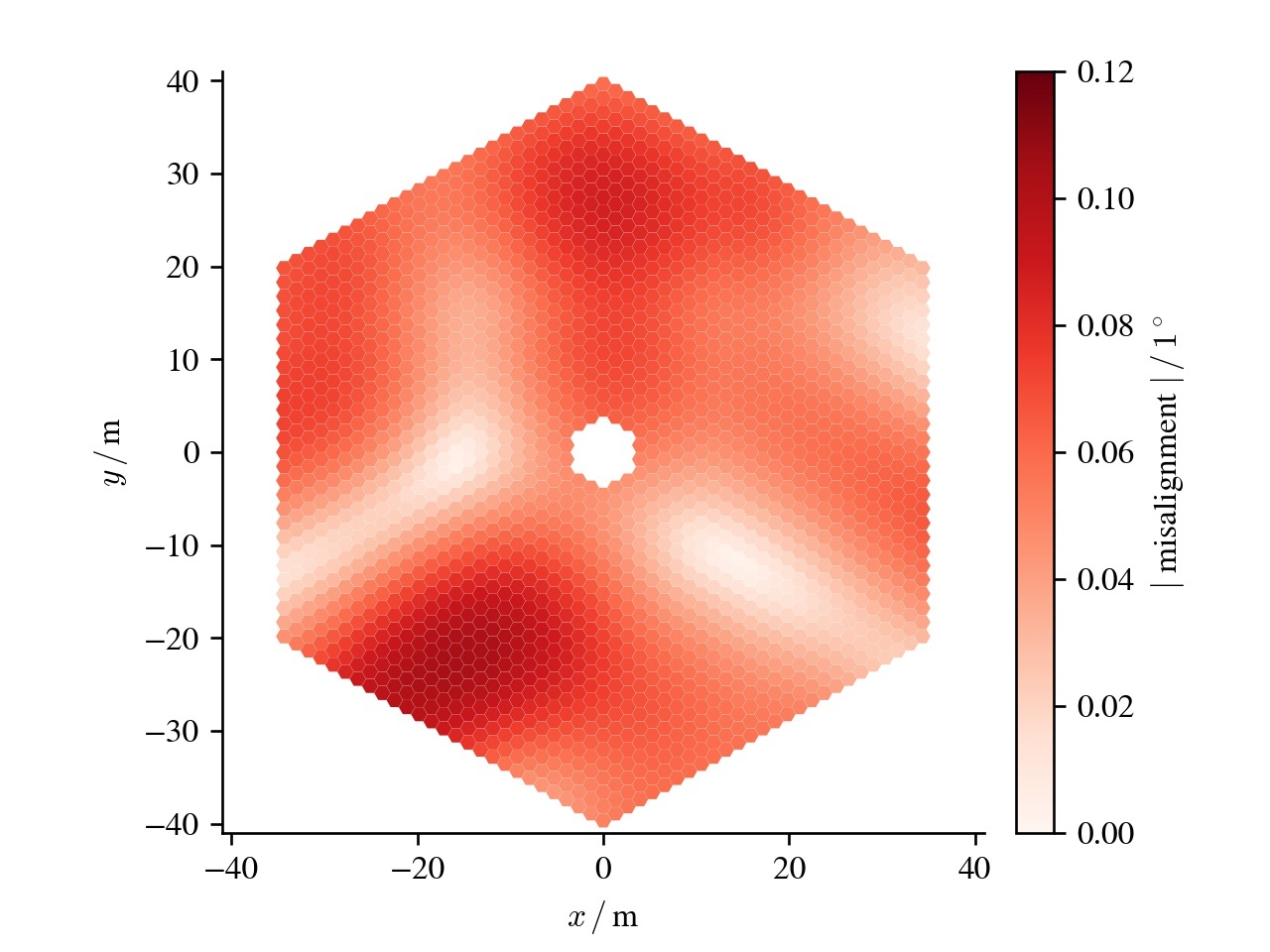}
            \caption[]{
                Deformations in the mirror effect its surface's normal vector.
            }
            \label{FigMirrorDeformation}
        \end{figure}

\begin{figure}
            \centering
            Telescope \Pone{}\\
            \begin{minipage}{.39\columnwidth}
                \includegraphics[width=1.0\columnwidth]{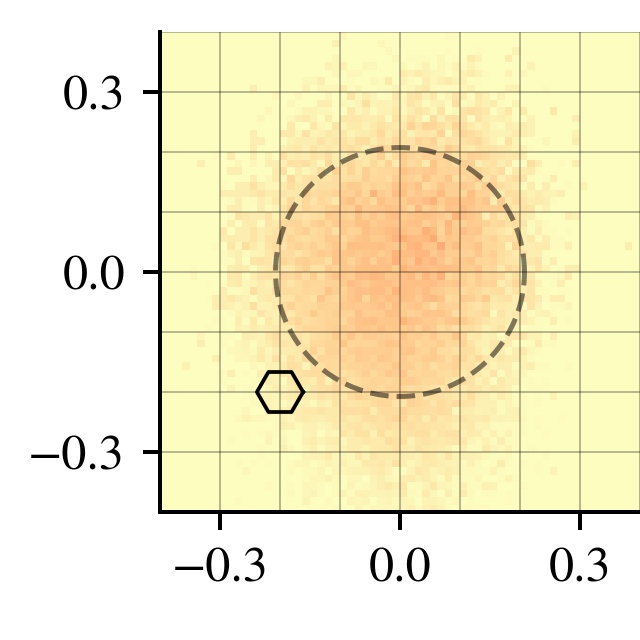}
            \end{minipage}
            \begin{minipage}{.29\columnwidth}
                \includegraphics[width=1.0\columnwidth]{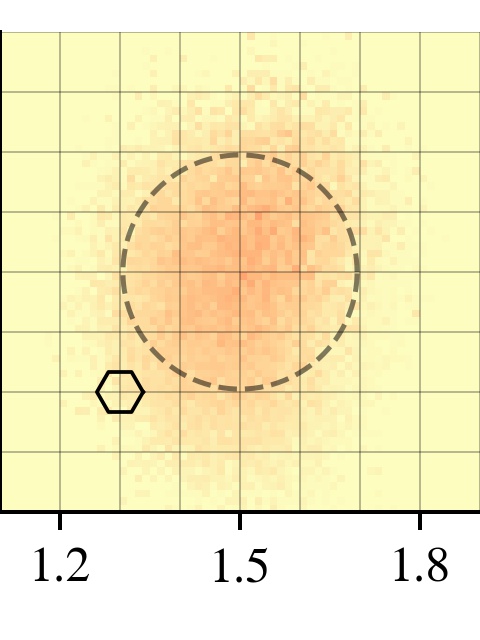}
            \end{minipage}
            \begin{minipage}{.29\columnwidth}
                \includegraphics[width=1.0\columnwidth]{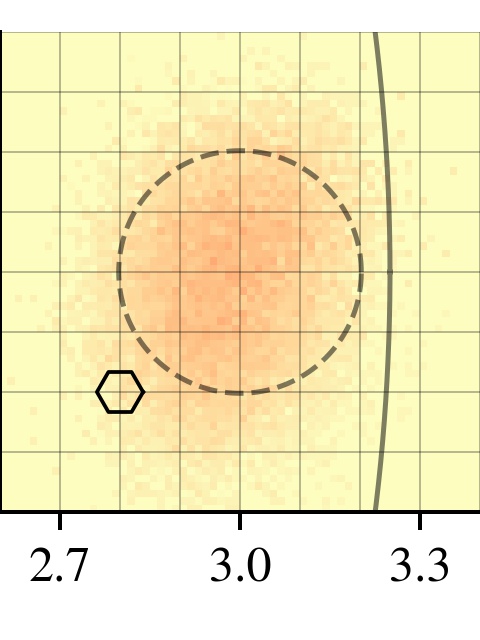}
            \end{minipage}
            Plenoscope \Pseven{}\\
            \begin{minipage}{.39\columnwidth}
                \includegraphics[width=1.0\columnwidth]{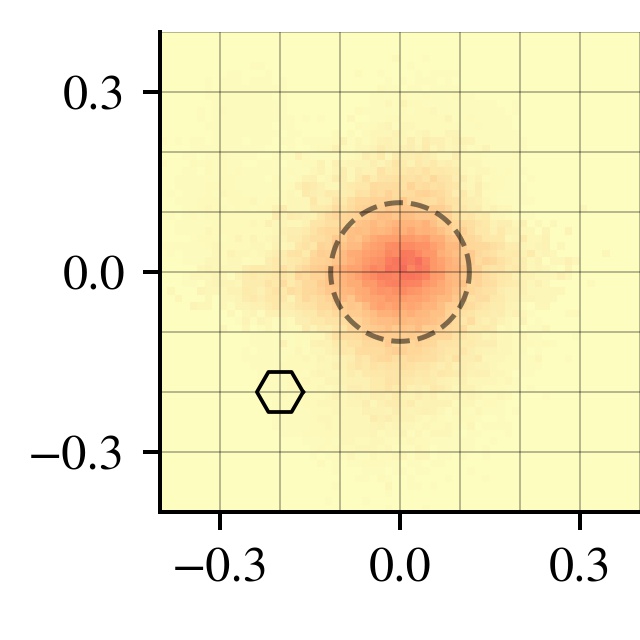}
            \end{minipage}
            \begin{minipage}{.29\columnwidth}
                \includegraphics[width=1.0\columnwidth]{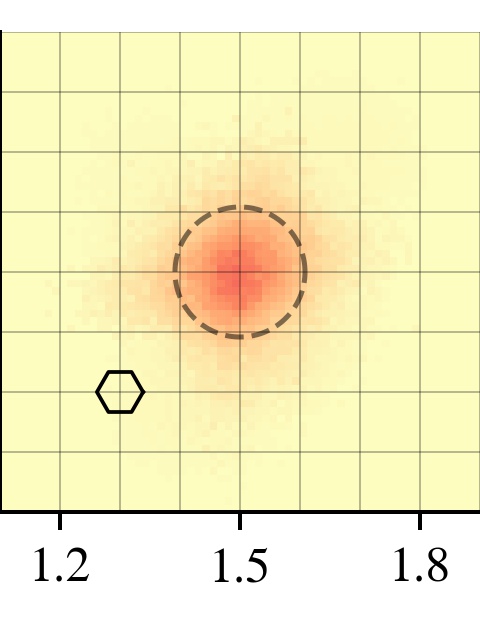}
            \end{minipage}
            \begin{minipage}{.29\columnwidth}
                \includegraphics[width=1.0\columnwidth]{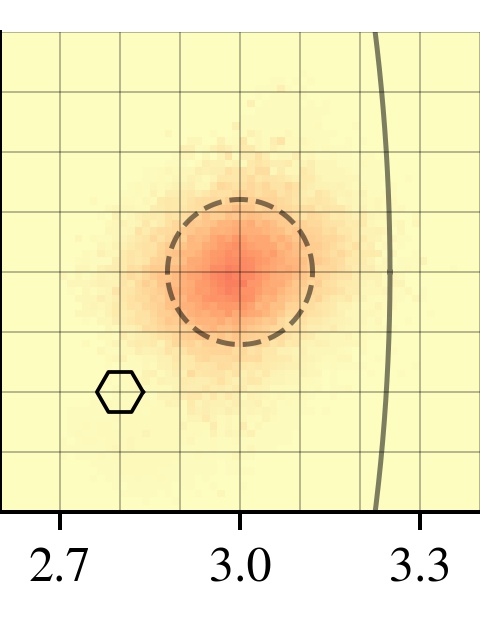}
            \end{minipage}
            \NameAcp{} Plenoscope \PsixtyOne{}\\
            \begin{minipage}{.39\columnwidth}
                \includegraphics[width=1.0\columnwidth]{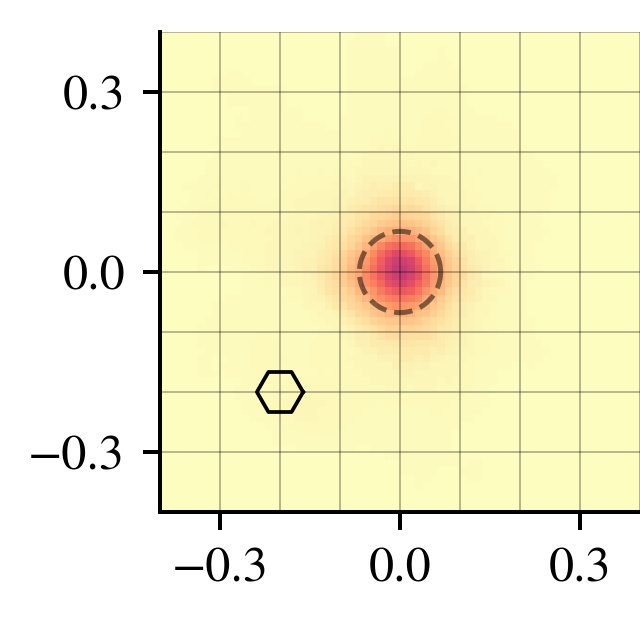}
            \end{minipage}
            \begin{minipage}{.29\columnwidth}
                \includegraphics[width=1.0\columnwidth]{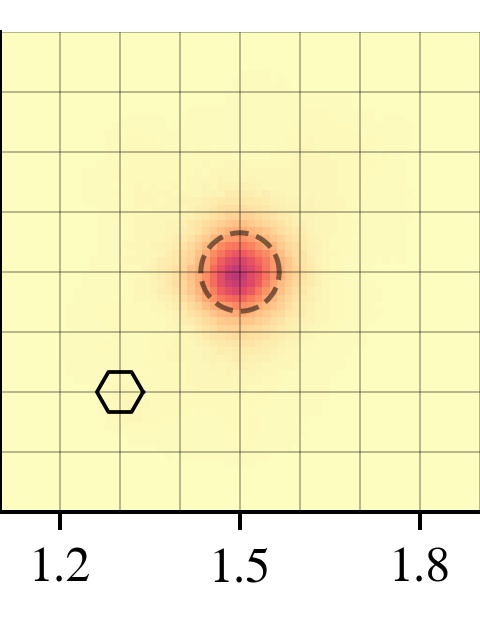}
            \end{minipage}
            \begin{minipage}{.29\columnwidth}
                \includegraphics[width=1.0\columnwidth]{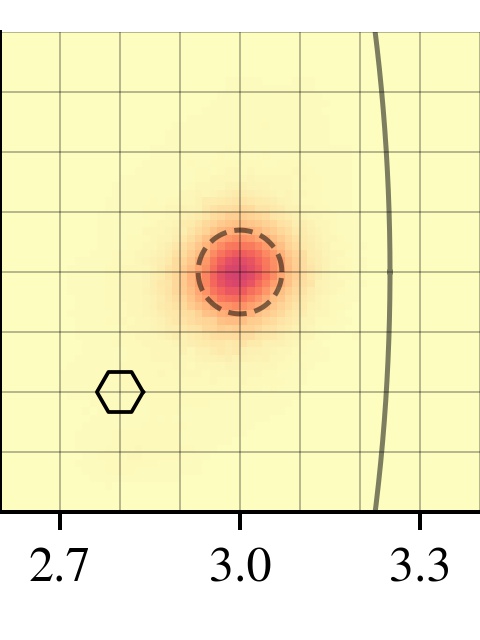}
            \end{minipage}
            \includegraphics[width=1.0\columnwidth]{figures-2023-05-23_optics-plots-guide_stars-magma_r-cmap_magma_r.jpg}
            \caption[]{
                \FigPsfTextIntro{} while their mirrors are deformed: \StateMirrorCamera{\MirrorBad{}}{\CameraGood{}}.
                \FigPsfTextAutoRef{FigPsfVsOffAxisDeformedDefault}
                \FigPsfTextSeeLegend{FigPsfDefaultDefault}
            }
            \label{FigPsfDeformedDefault}
        \end{figure}

\begin{figure}
            \centering
            \includegraphics[width=1.0\columnwidth]{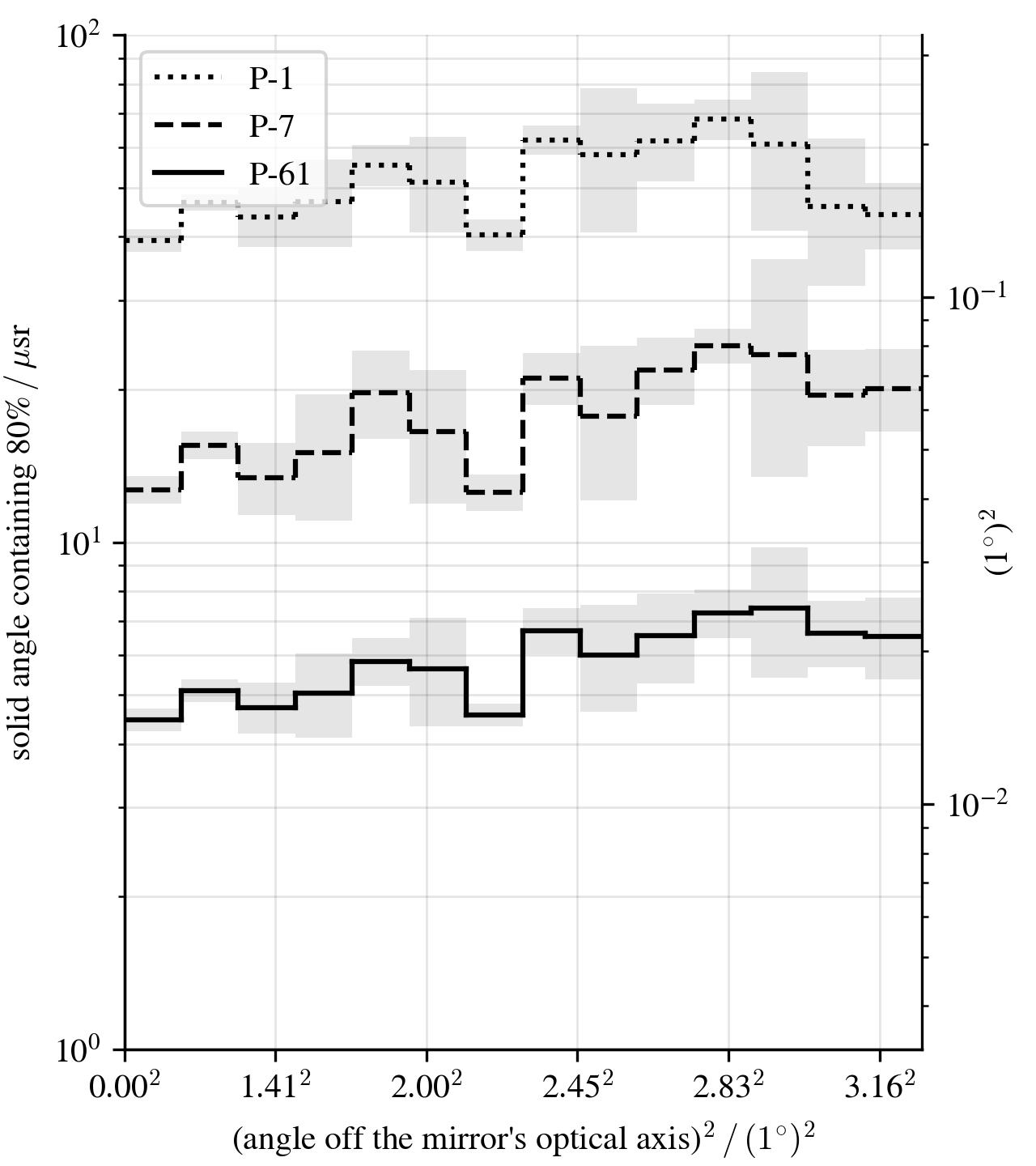}
            \caption[]{
                \FigPsfVsOffAxisTextIntro{} while their mirrors are deformed: \StateMirrorCamera{\MirrorBad{}}{\CameraGood{}}. 
                \FigPsfVsOffAxisTextAutoRef{FigPsfDeformedDefault}
                \FigPsfTextSeeLegend{FigPsfVsOffAxisDefaultDefault}
            }
            \label{FigPsfVsOffAxisDeformedDefault}
        \end{figure}

        Deformations in the mirror's supporting structure effect the normal vector of the mirror's surface.
        Deformations induce aberrations which blur the image.
        However, a plenoscope can compensate such deformations when the deformation's spatial frequencies across the mirror are low enough.
        Deformations with spatial frequencies lower then the spatial resolution a plenoscope has to resolve its mirror do only re-orientate the plenoscope's beams but they do not widen the plenoscope's beams.
        And as long as the beams are not widened, the plenoscope's performance is hardly reduced.
        To demonstrate plenoptic's capabilities to compensate such deformations we compare \NameAcp{}'s images to the images taken by the telescope \Pone{}.
        All instruments use identical mirrors with the deformation shown in \autoref{FigMirrorDeformation}.
        We deform the mirror randomly using \citet{perlin1985image} and set the magnitude such that some of the mirror's facets deviate from their targeted orientations by $0.12^{^\circ}$.
        For a telescope, where the field-of-view of a pixel is $0.067^\circ$, such a deformation is significant as it potentially spills light from a star over several neighboring photosensors in the telescope's image~camera.
        \autoref{FigPsfDeformedDefault} shows the images of a star taken by the different instruments for three different angles, and \autoref{FigPsfVsOffAxisDeformedDefault} shows how the spread of the images changes with the star's angle off the optical axis.
        As expected, the telescope's image is significantly washed out and it takes a solid angle of $\approx 53\,\mu$sr (half-angle: 0.23\,$^\circ$) to contain 80\% of the light.
        In contrast, the \NameAcp{} Cherenkov~plenoscope reduces the 80\%-cone's solid angle by almost one order of magnitude down to only $\approx 6.1\,\mu$sr (half-angle: 0.08\,$^\circ$).

        Since the Cherenkov~plenoscope can compensate deformations after it has recorded a shower, it can compensate deformations as rapidly as it can measure them.
        Rapid deformations might be caused by gusts of wind or by a fast pointing in order to hunt a transient source of gamma~rays.
        On a structure with the size of \NameAcp{}, distance- and angle-sensors can sample with rates in a regime of several $10$\,s$^{-1}$ what allows to compensate changes in the mirror's deformation with frequencies as high as several $1$\,Hz.
        The plenoscope's rapid compensation of deformations is a powerful augmentation to the precise, but gentle, re-orientation of facets in a Cherenkov~telescope's active mirror~control\,\citep{biland2007active} which in the end is also subject to mechanical wear and tear.
        Combining the two approaches covers a wide range of scenarios and makes the \NameAcp{} Cherenkov~plenoscope promising to choose a more forgiving, more elastic, and thus less expensive design for its mirror.
        Although the plenoscope's light-field~calibration will allow the compensation of a broad range of deformations in the mirror, \NameAcp{} still has an active mirror~control because it eases the initial alignment of the many facets, and because it can reduce the risk of having concentrated sunlight, see \autoref{SecSunLight}.
    \section{Compensating Misalignments}
        \label{SubSecCompensatingMisalignments}
\begin{figure}
            \centering
            Telescope \Pone{}\\
            \begin{minipage}{.39\columnwidth}
                \includegraphics[width=1.0\columnwidth]{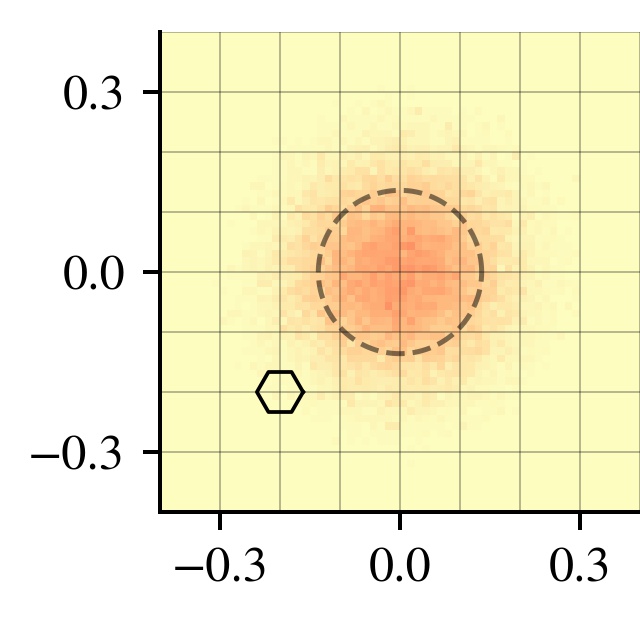}
            \end{minipage}
            \begin{minipage}{.29\columnwidth}
                \includegraphics[width=1.0\columnwidth]{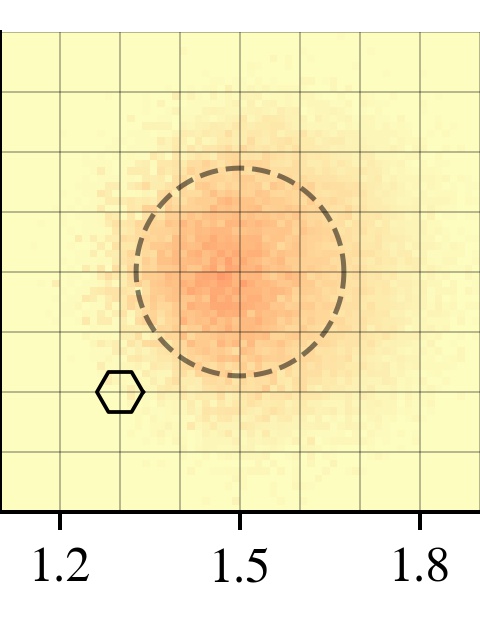}
            \end{minipage}
            \begin{minipage}{.29\columnwidth}
                \includegraphics[width=1.0\columnwidth]{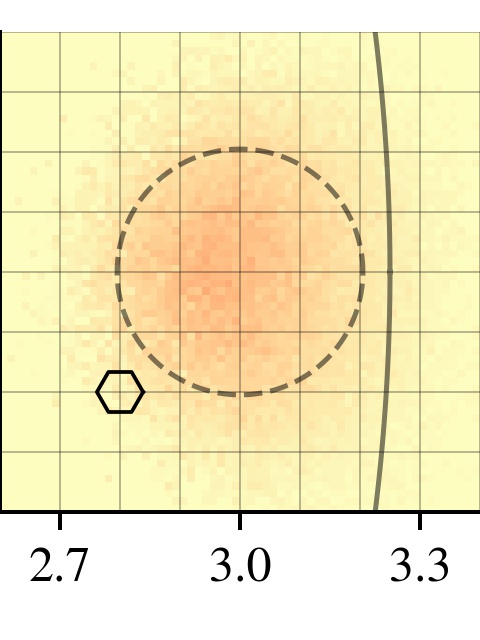}
            \end{minipage}
            Plenoscope \Pseven{}\\
            \begin{minipage}{.39\columnwidth}
                \includegraphics[width=1.0\columnwidth]{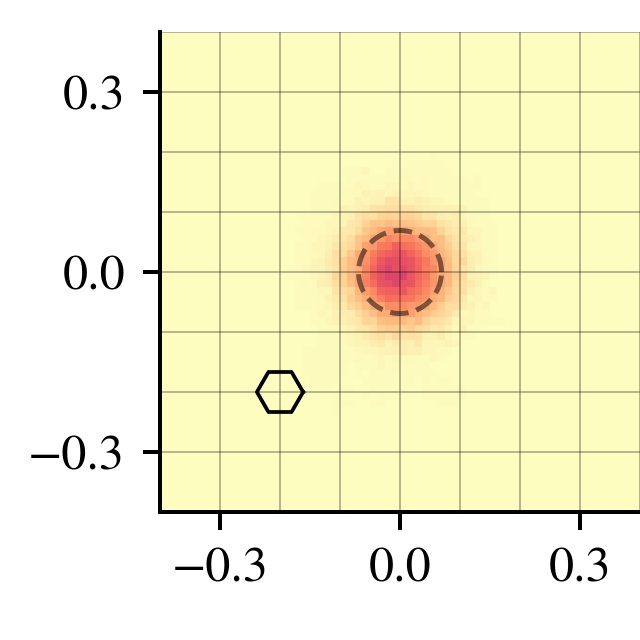}
            \end{minipage}
            \begin{minipage}{.29\columnwidth}
                \includegraphics[width=1.0\columnwidth]{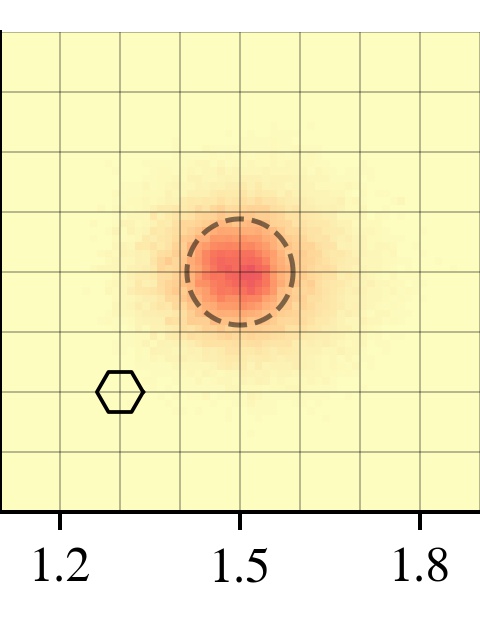}
            \end{minipage}
            \begin{minipage}{.29\columnwidth}
                \includegraphics[width=1.0\columnwidth]{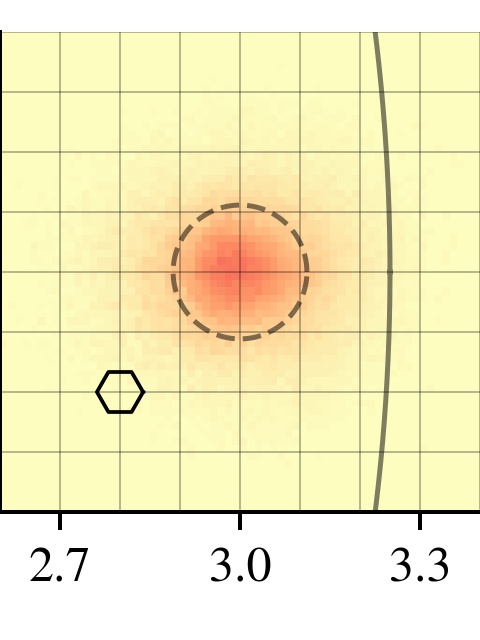}
            \end{minipage}
            \NameAcp{} Plenoscope \PsixtyOne{}\\
            \begin{minipage}{.39\columnwidth}
                \includegraphics[width=1.0\columnwidth]{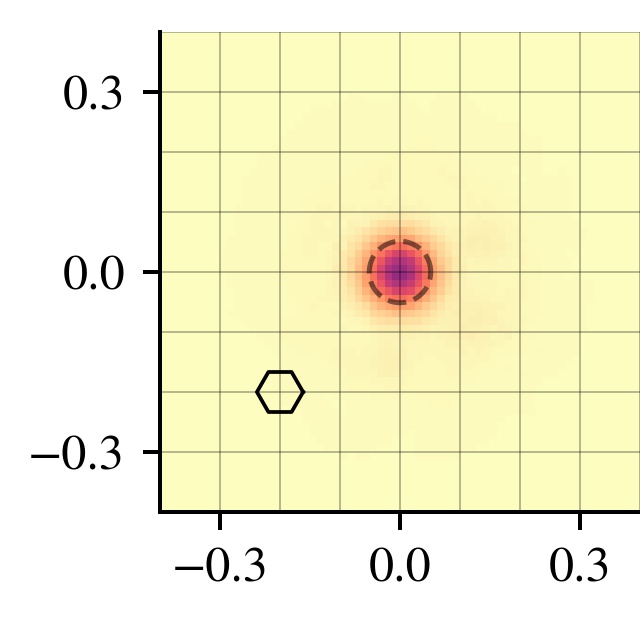}
            \end{minipage}
            \begin{minipage}{.29\columnwidth}
                \includegraphics[width=1.0\columnwidth]{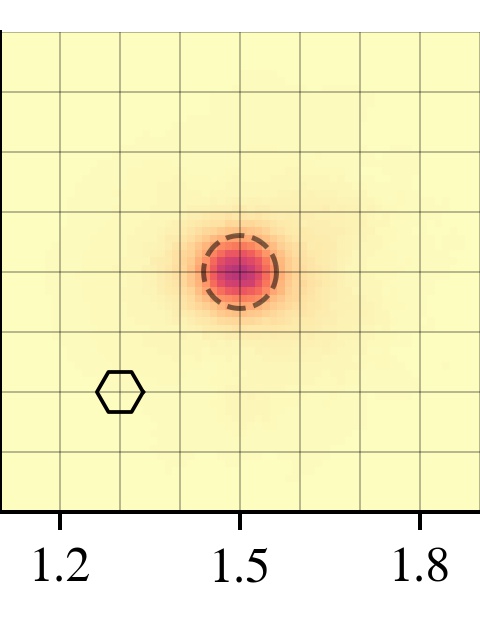}
            \end{minipage}
            \begin{minipage}{.29\columnwidth}
                \includegraphics[width=1.0\columnwidth]{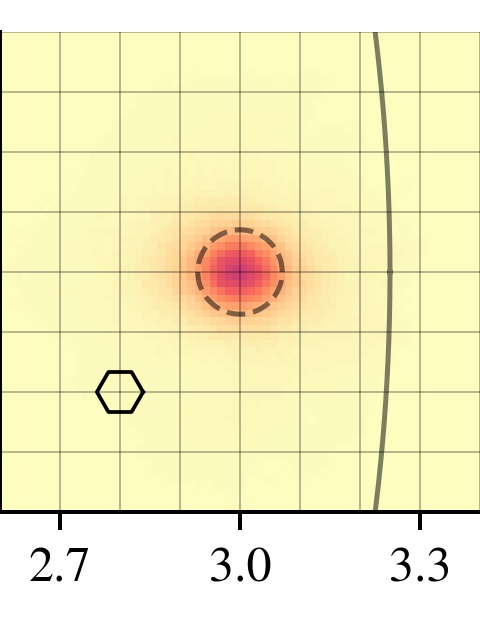}
            \end{minipage}
            \includegraphics[width=1.0\columnwidth]{figures-2023-05-23_optics-plots-guide_stars-magma_r-cmap_magma_r.jpg}
            \caption[]{
                \FigPsfTextIntro{} while their cameras are misaligned: \StateMirrorCamera{\MirrorGood{}}{\CameraBad{}}.
                \FigPsfTextAutoRef{FigPsfVsOffAxisDefaultGentle}
                \FigPsfTextSeeLegend{FigPsfDefaultDefault}
            }
            \label{FigPsfDefaultDefaultGentle}
        \end{figure}

\begin{figure}
            \centering
            \includegraphics[width=1.0\columnwidth]{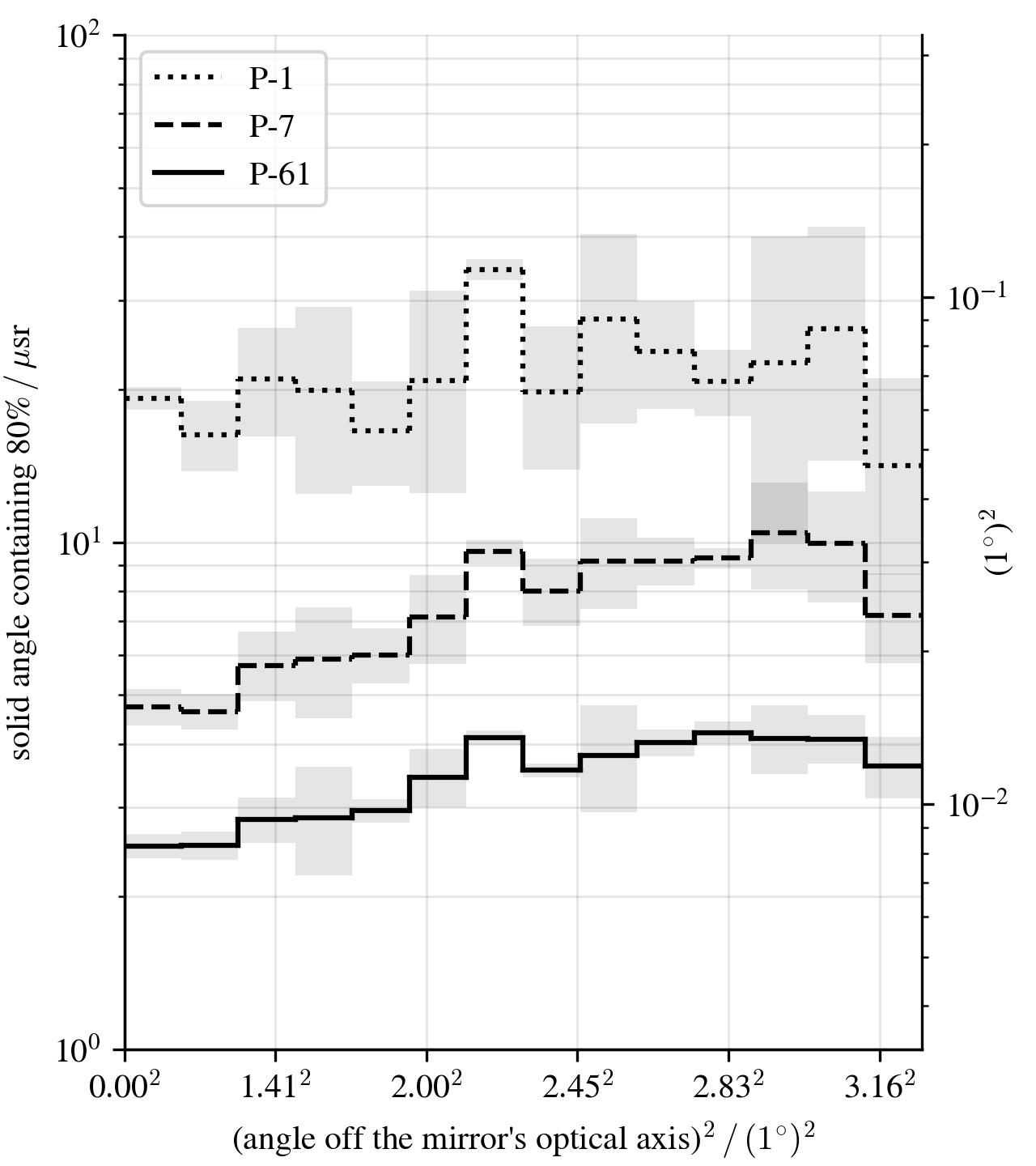}
            \caption[]{
                \FigPsfVsOffAxisTextIntro{} while their cameras are misaligned: \StateMirrorCamera{\MirrorGood{}}{\CameraBad{}}. 
                \FigPsfVsOffAxisTextAutoRef{FigPsfDefaultDefaultGentle}
                \FigPsfTextSeeLegend{FigPsfVsOffAxisDefaultDefault}
            }
            \label{FigPsfVsOffAxisDefaultGentle}
        \end{figure}

        The prevention of misalignments between a telescope's mirror and its camera rapidly complicates the construction of larger telescopes.
        While a telescope can compensate certain components of misalignments when these are known, others blur its image irrecoverably.
        \autoref{TabMisalignmentComponents} shows that the telescope \Pone{} can tolerate absolute misalignments in the regime of $10^{-3}\,f$ before its image degrades significantly.
        \autoref{SecEstimatingLimitsOfMisalignment} shows how we estimate these tolerances.
        The components of misalignment which the telescope can not compensate make independent mountings of the mirror and the camera impractical.
        A rigid mechanical support is required between the mirror and the camera to prevent misalignments in the first place.
        The plenoscope relaxes this limitation as it can compensate all components of a misalignment.
        To demonstrate the plenoscope's power to compensate misalignments we significantly misalign the camera on purpose, see \autoref{TabMisalignmentAmplitude}.
        \autoref{FigPsfDefaultDefaultGentle} and \autoref{FigPsfVsOffAxisDefaultGentle} show the images of stars taken with the camera being misaligned.
        Since we apply the light-field~calibration also to the images of the telescope \Pone{}, the components of misalignment which the telescope \Pone{} can compensate for, are compensated for.
        This is why the average of the star's light is reconstructed with its correct direction in the images of \Pone{}.
        But the large spread of the stars in \Pone{}'s images shows that we pushed the telescope \Pone{} beyond its limits.
        In contrast, the misaligned \NameAcp{} plenoscope provides images which still outperform the images of the telescope \Pone{} when \Pone{}'s camera is perfectly aligned, compare \autoref{FigPsfVsOffAxisDefaultDefault} and \autoref{FigPsfVsOffAxisDefaultGentle}.
        \begin{table}[]
            \begin{center}
                \begin{footnotesize}
                    \begin{tabular}{llcc}
                        &
                        tolerance when not
                        & \multicolumn{2}{c}{can be compensated}\\
                        & compensated for
                        & telescope
                        & plenoscope
                        \\
                        \toprule
                            \multirow{2}{.5cm}{$\TransPara{}$} & Translating parallel & \\
                             & $\Delta_\parallel \approx \pm 96\,$mm $\approx 10^{-3}\,f$ & No & Yes \vspace{1mm} \\
                        \hline
                            \multirow{2}{.5cm}{$\TransPerp{}$} & Translating perpendicular & \\
                             & $\Delta_\perp \approx \pm 63\,$mm $\approx 10^{-3}\,f$&
                            Yes$^{\star}$ &
                            Yes \vspace{1mm} \\
                        \hline
                            \multirow{2}{.5cm}{$\RotPara{}$} & Rotating parallel & \\
                             & $ \Phi_\parallel \approx \pm 0.59^\circ$ &
                            Yes &
                            Yes \vspace{1mm} \\
                        \hline
                            \multirow{2}{.5cm}{$\RotPerp{}$} & Rotating perpendicular & \\
                             & $ \Phi_\perp \approx \pm 0.83^\circ$ &
                            No$^{\star{}\star{}}$ &
                            Yes \vspace{1mm} \\
                    \end{tabular}
                \end{footnotesize}
            \end{center}
            \caption[]{
                The four components of misalignments between a camera and a mirror.
                The tolerances are estimated for the telescope \Pone{}.
                ($^{\star}$ Aberrations increase with the angle off the mirror's optical axis.
                $^{\star{}\star{}}$ The axis of the rotation forms a line across the camera's screen which remains in focus and thus will not be blurred. But the rest of the image will be.
                ).
            }
            \label{TabMisalignmentComponents}
        \end{table}
        \begin{table}[]
            \begin{center}
                \begin{footnotesize}
                    \begin{tabular}{llcc}
                        & component & absolute & relative \\
                        \toprule
                            \multirow{2}{.5cm}{$\TransPara{}$} & Translating parallel & \\
                             & & -532.5\,mm & 5.5\,$\Delta_\parallel$ \\
                        \hline
                            \multirow{2}{.5cm}{$\TransPerp{}$} & Translating perpendicular & \\
                             & & 223.6\,mm & 3.5\,$\Delta_\perp$ \\
                        \hline
                            \multirow{2}{.5cm}{$\RotPara{}$} & Rotating parallel & \\
                             & & 5.0\,$^{\circ{}}$ & 8.5\,$\Phi_\parallel$ \\
                        \hline
                            \multirow{2}{.5cm}{$\RotPerp{}$} & Rotating perpendicular & \\
                             & & 3.2\,$^{\circ{}}$ & 3.9\,$\Phi_\perp$ \\
                    \end{tabular}
                \end{footnotesize}
            \end{center}
            \caption[]{
                The misalignment of \NameAcp{}'s camera in our demonstration.
                Explicit:
                Translation-$xyz$: (-100, 200, -532.5)\,mm, rotation tait-bryan-$xyz$: (1.0,\,3.0,\,5.0)$^\circ$.
            }
            \label{TabMisalignmentAmplitude}
        \end{table}
    \section{Observing a Phantom~Source}
        \label{SecObservingAPhantomSource}
\begin{figure}
            \centering
            \includegraphics[width=0.8\columnwidth]{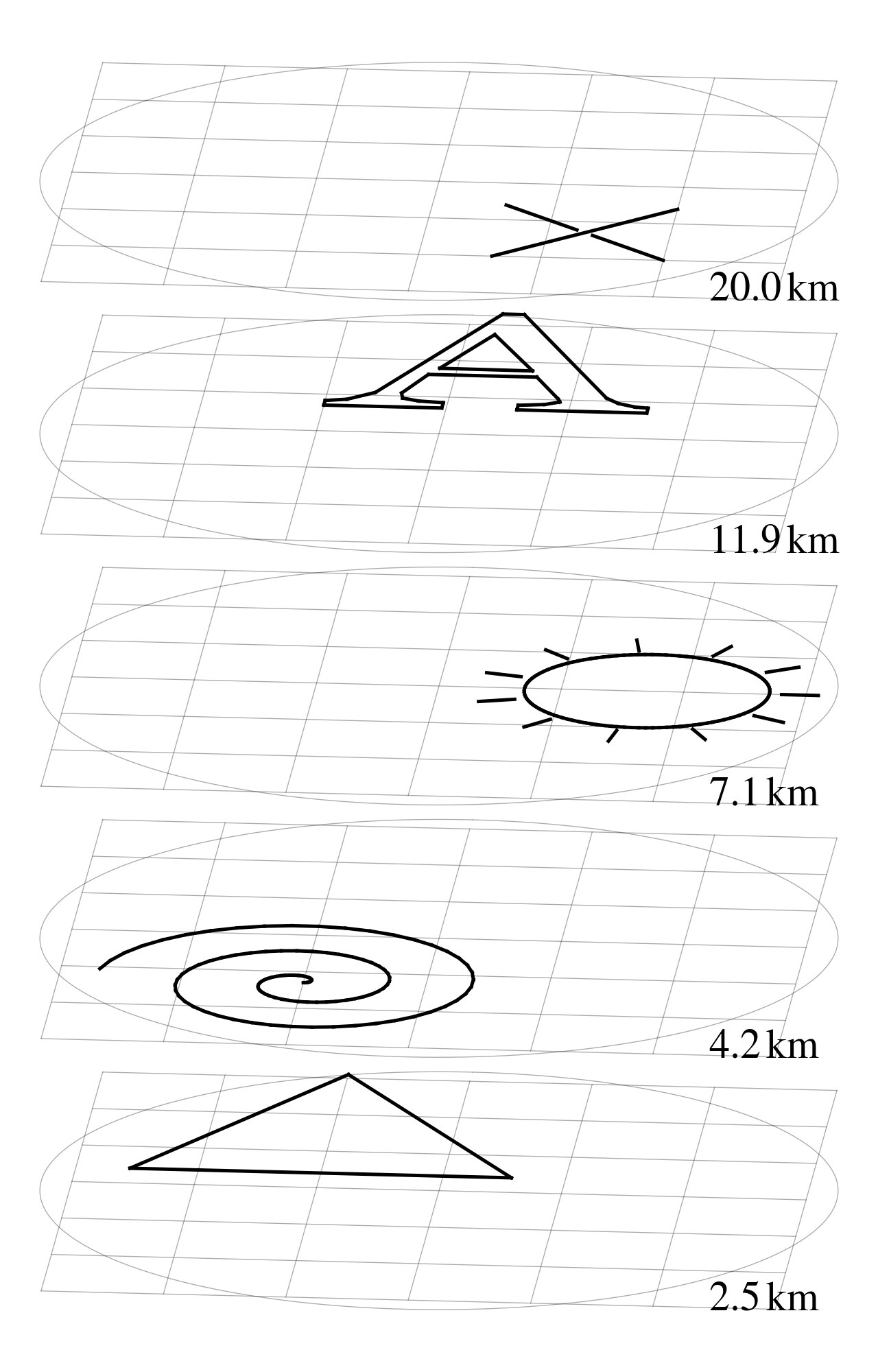}
            \caption[]{
                A phantom~source in the atmosphere with five unique light emitting structures in five different depths.
                Grid has a spacing of $1^{\circ}$.
                Ring indicates the instrument's field-of-view with a half-angle of $3.25^{\circ}$.
            }
            \label{FigPhantomSource}
        \end{figure}

\begin{figure}
            \centering
            Telescope \Pone{}\\
            \begin{minipage}{.49\columnwidth}
                \centering
                \includegraphics[width=1.0\columnwidth]{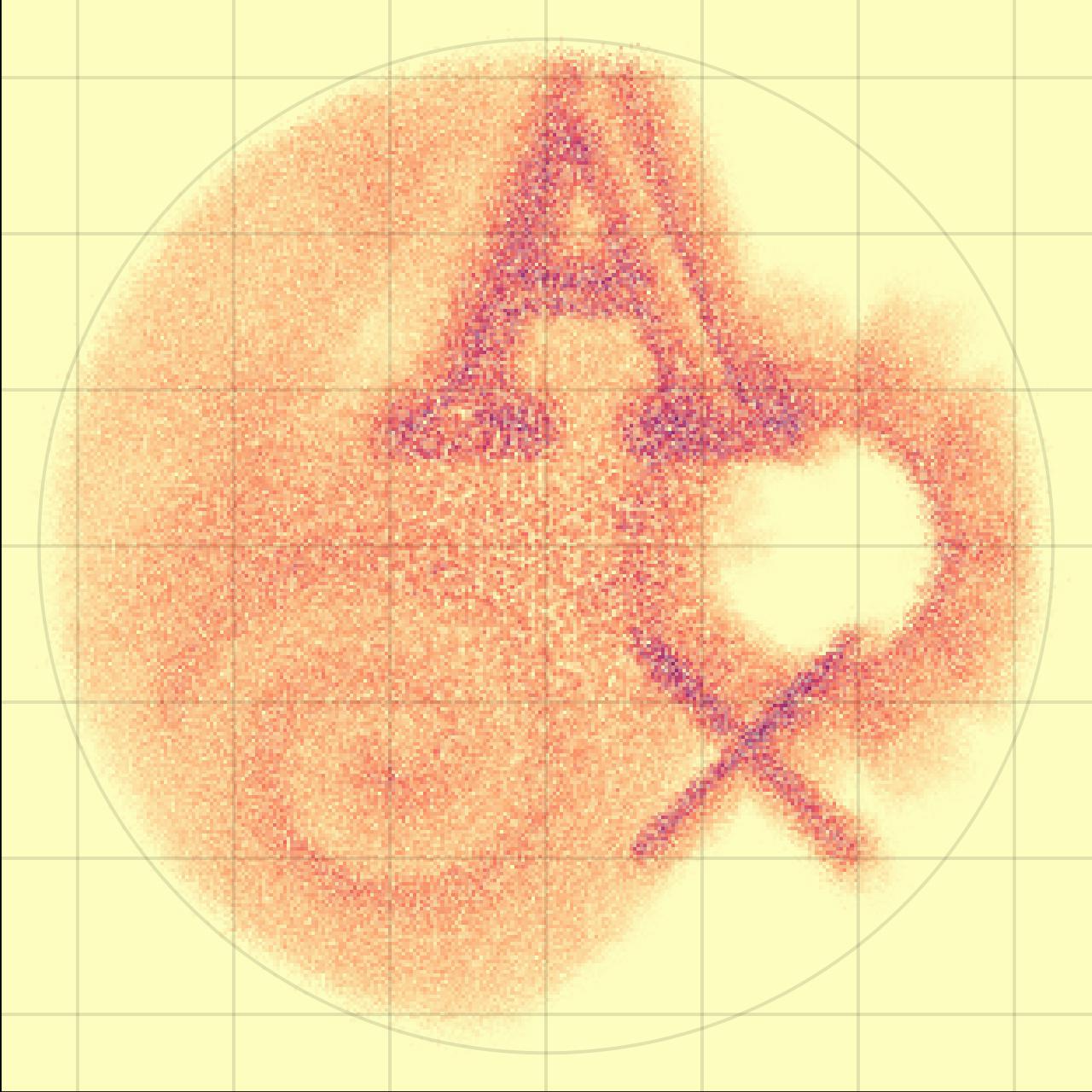}
            \end{minipage}
            \begin{minipage}{.49\columnwidth}
                \centering
                \includegraphics[width=1.0\columnwidth]{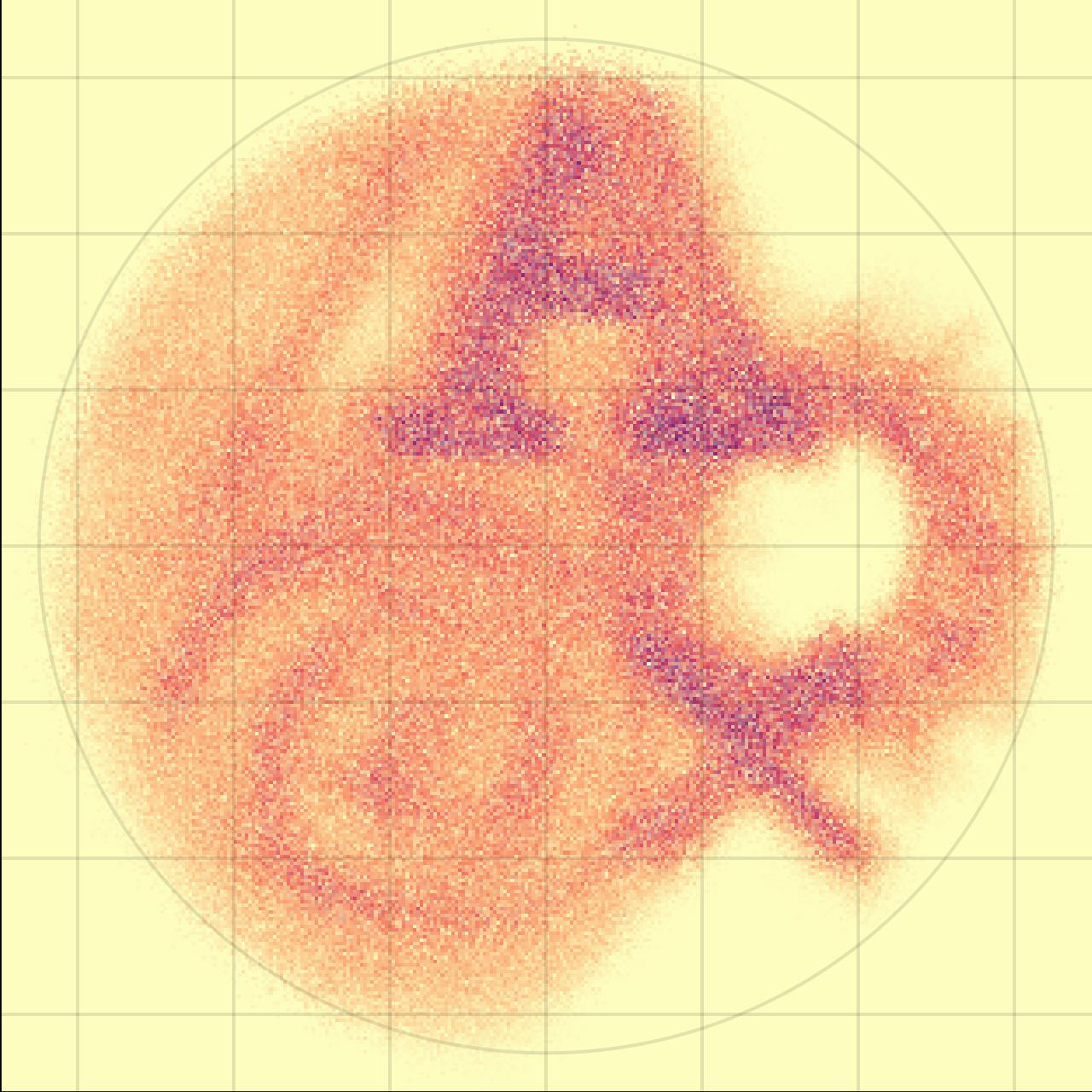}
            \end{minipage}
            \vspace{3mm}\\
            \includegraphics[width=1.0\columnwidth]{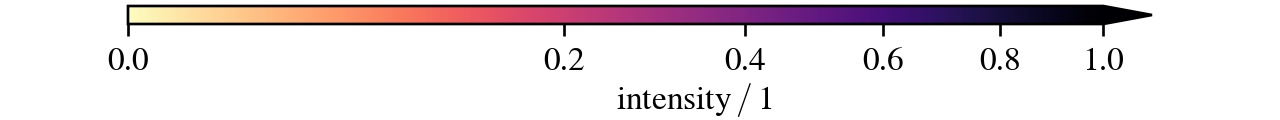}
            \caption[]{
                Observing the phantom~source shown in \autoref{FigPhantomSource} with the \Pone{} Cherenkov~telescope.
                In the left column \Pone{}'s state is \StateMirrorCamera{\MirrorGood{}}{\CameraGood{}}, while in the right column its state is \StateMirrorCamera{\MirrorBad{}}{\CameraBad{}}.
            }
            \label{FigObservingPhantomT1}
        \end{figure}

        The observation of a phantom~source in the atmosphere can demonstrate how efficiently the Cherenkov~plenoscope can compensate aberrations, deformations in the mirror and misalignments of the camera.
        \autoref{FigPhantomSource} shows the phantom~source.
        Instead of a shower induced by a cosmic particle this phantom~source is composed of structures which are easy to identify in an image, and which allow the viewer to formulate a well defined expectation of what the images should look like.
        The phantom~source ranges from $2.5\,$km to $20.0\,$km in depth what matches the depth where most showers reach their maximum emission of Cherenkov~light.
        Both the telescope \Pone{} and the \NameAcp{} plenoscope observe the source.
        The geometries of both the telescope and the plenoscope do not change during the observation, and thus neither the telescope nor the plenoscope can physically adjust its focus, just as if they observe a flash of Cherenkov~light.
        \autoref{FigObservingPhantomT1} shows the observations of \Pone{}.
        One might argue that \Pone{}'s physical focus is not set properly to any of the structures in the phantom~source what prevents \Pone{} from producing a sharp image of at least one structure in the phantom~source.
        But this is just what a narrow depth-of-field is all about:
        A large telescope like \Pone{} is unlikely to have a structure of interest in its narrow focus.
        In the images made by \Pone{} the triangle~structure is blurred beyond recognition.
        By chance the cross~structure and the letter-`A' structure are closest to the focus.
        Yet despite the blurring caused by the narrow depth-of-field aberrations can be found as the structures are not completely flat, i.e. their sharpness depends on where they are in the image.

\begin{figure}
            \centering
            \NameAcp{} Plenoscope (\PsixtyOne{})\\
            \begin{minipage}{.49\columnwidth}
                \centering
                \includegraphics[width=1.0\columnwidth]{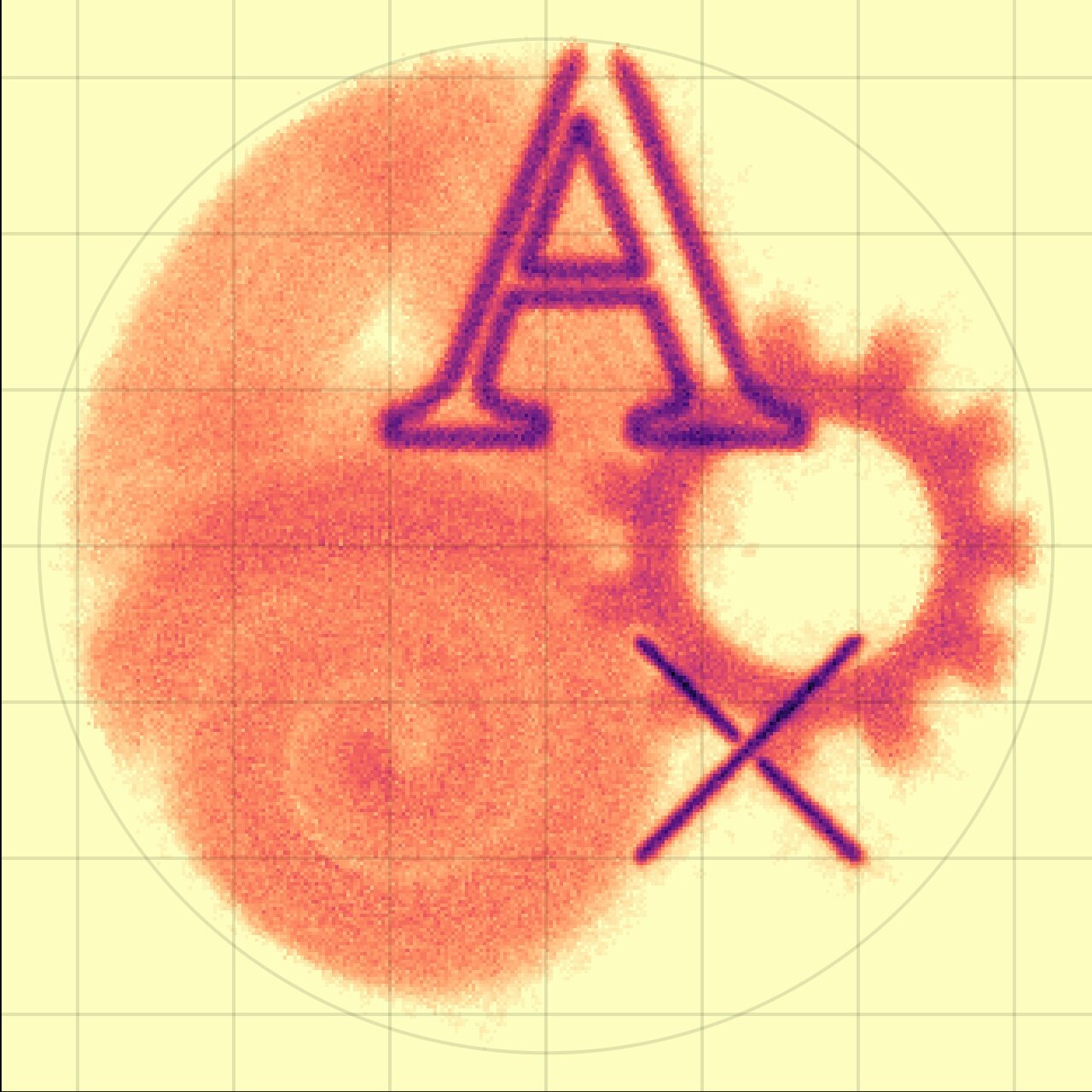}
                \includegraphics[width=1.0\columnwidth]{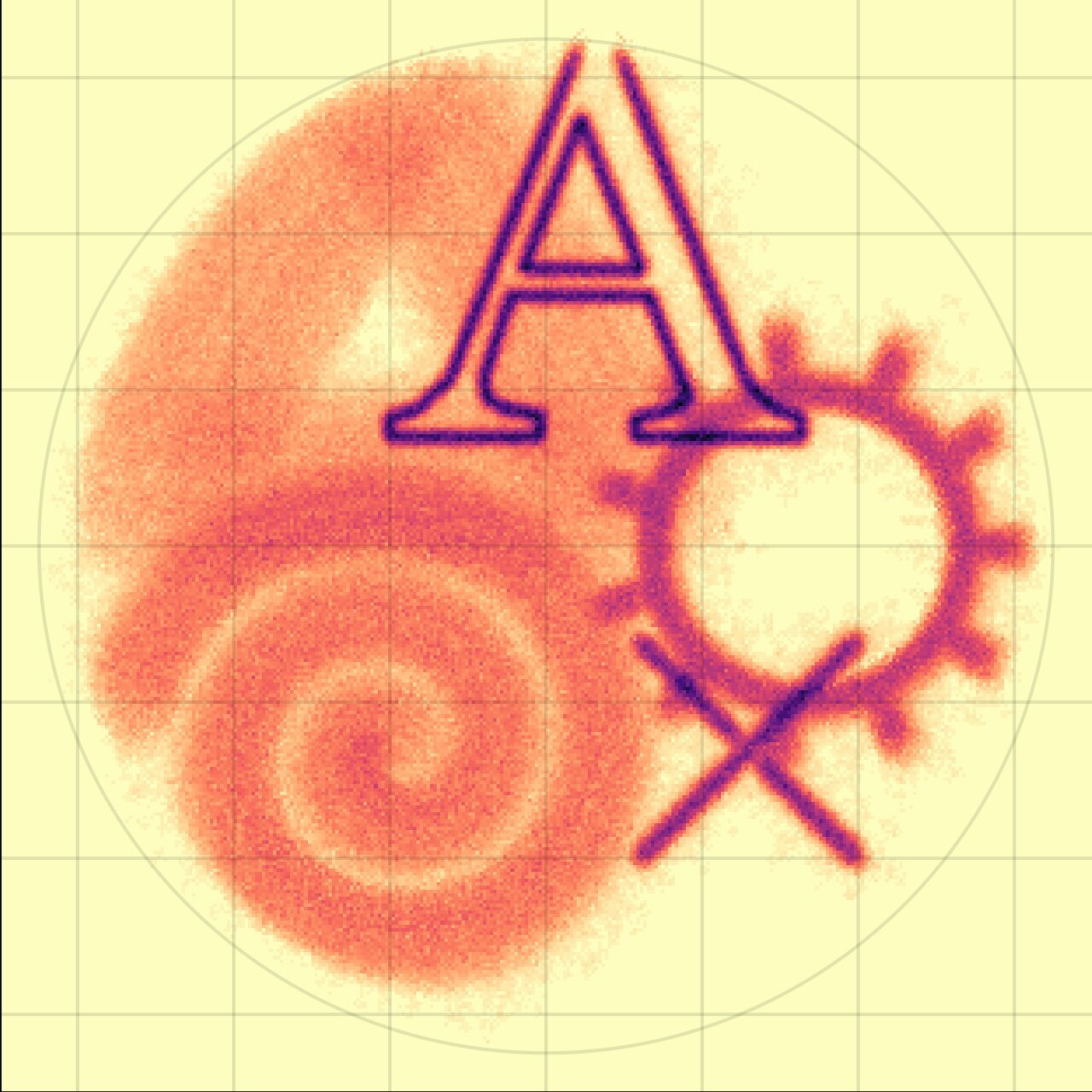}
                \includegraphics[width=1.0\columnwidth]{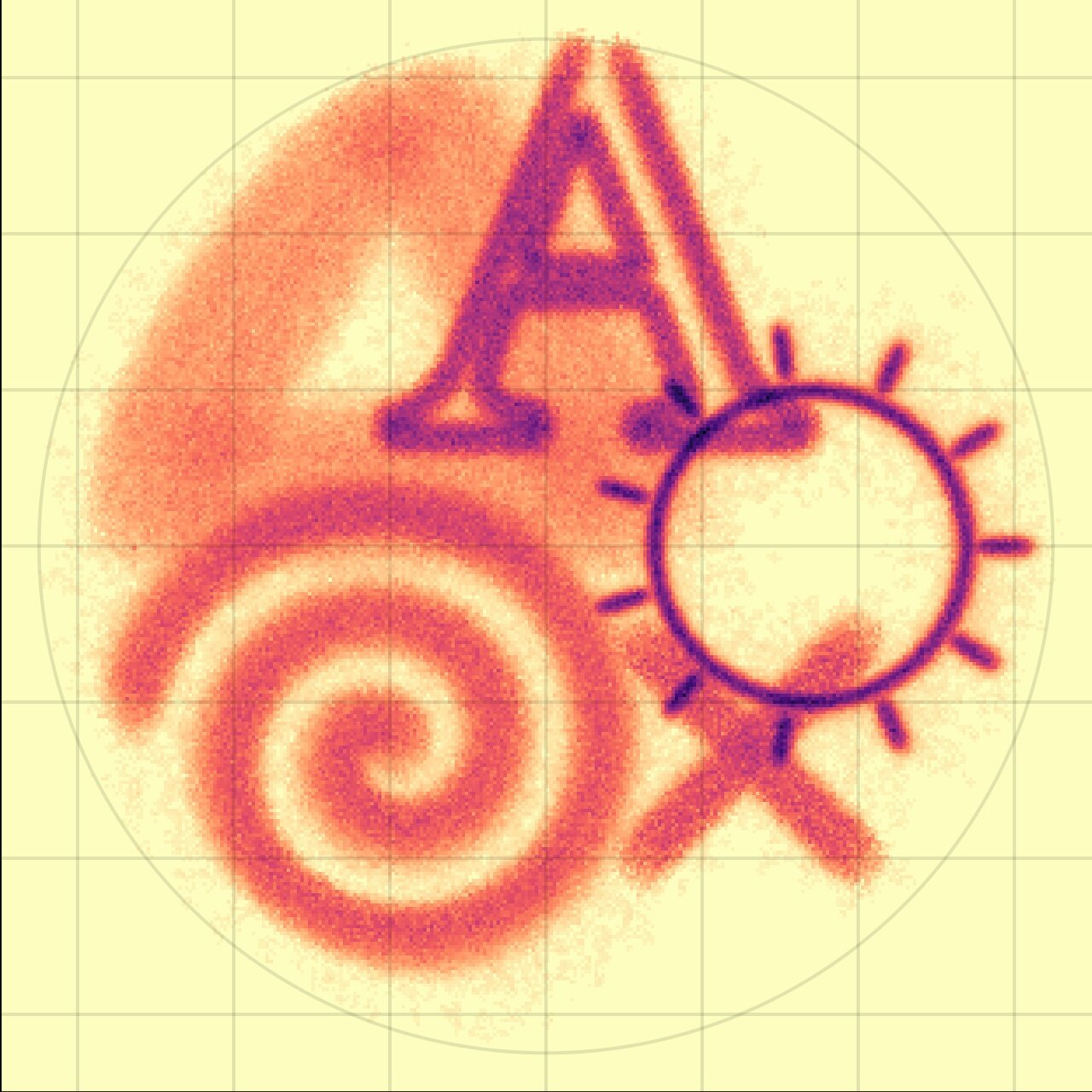}
                \includegraphics[width=1.0\columnwidth]{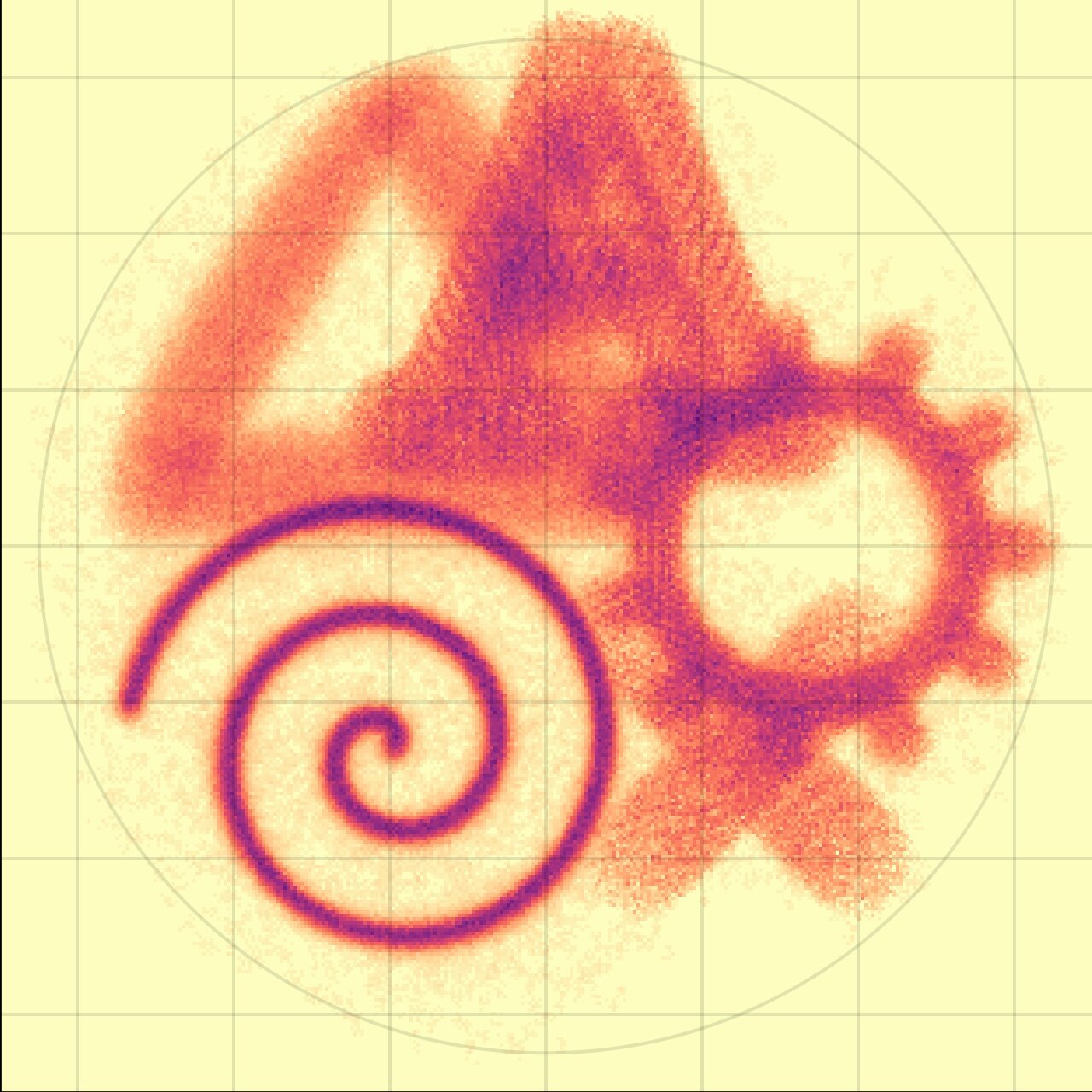}
                \includegraphics[width=1.0\columnwidth]{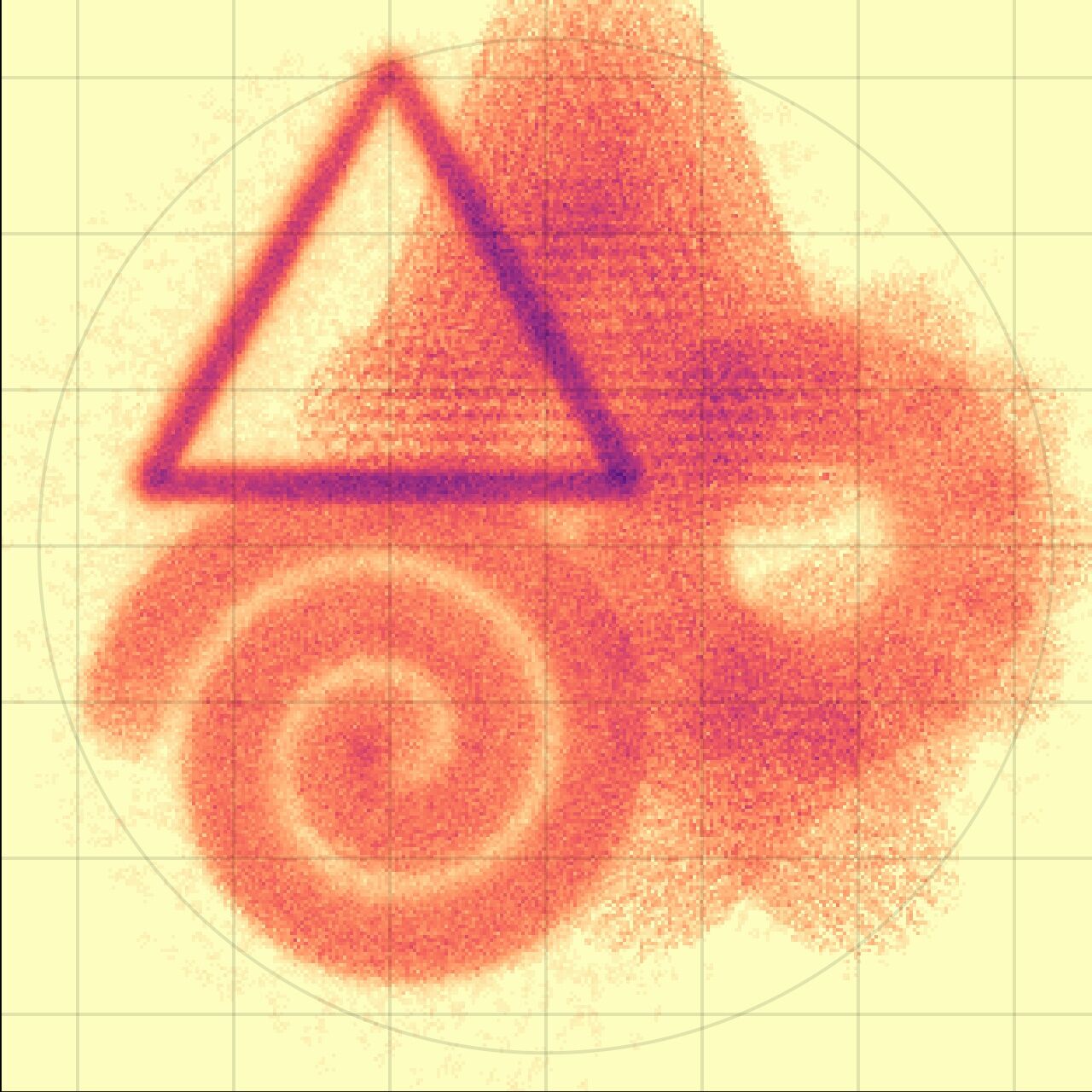}
            \end{minipage}
            \begin{minipage}{.49\columnwidth}
                \centering
                \includegraphics[width=1.0\columnwidth]{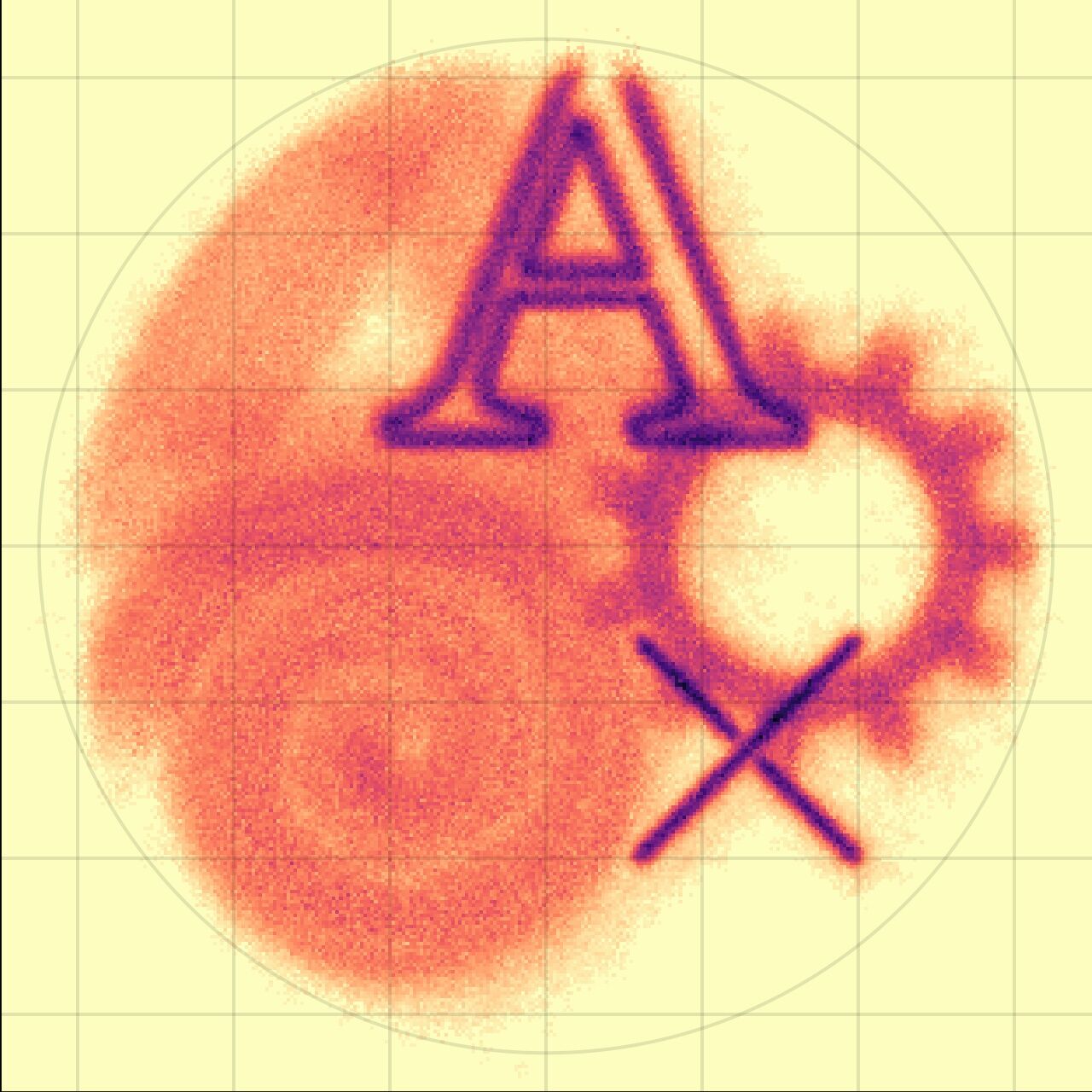}
                \includegraphics[width=1.0\columnwidth]{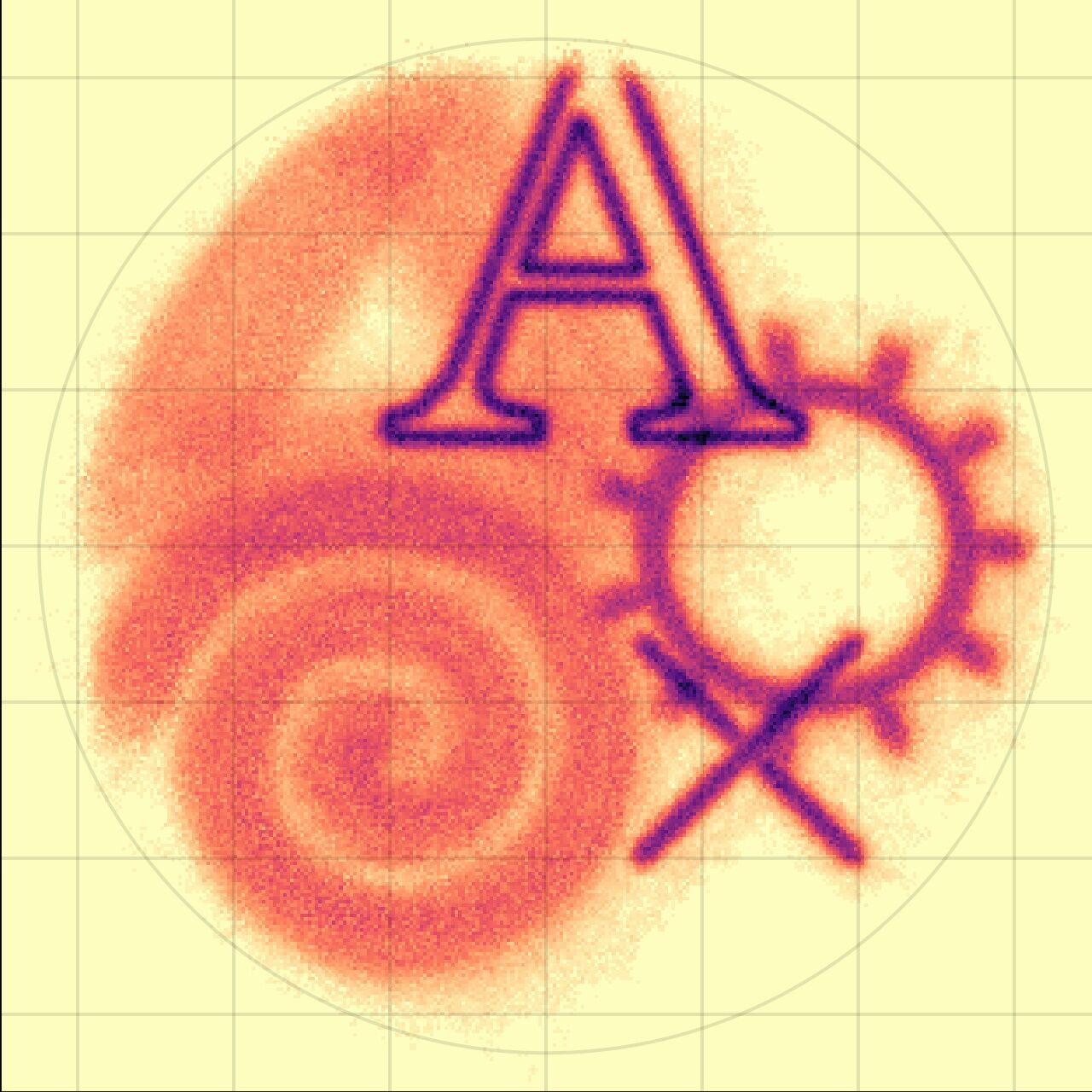}
                \includegraphics[width=1.0\columnwidth]{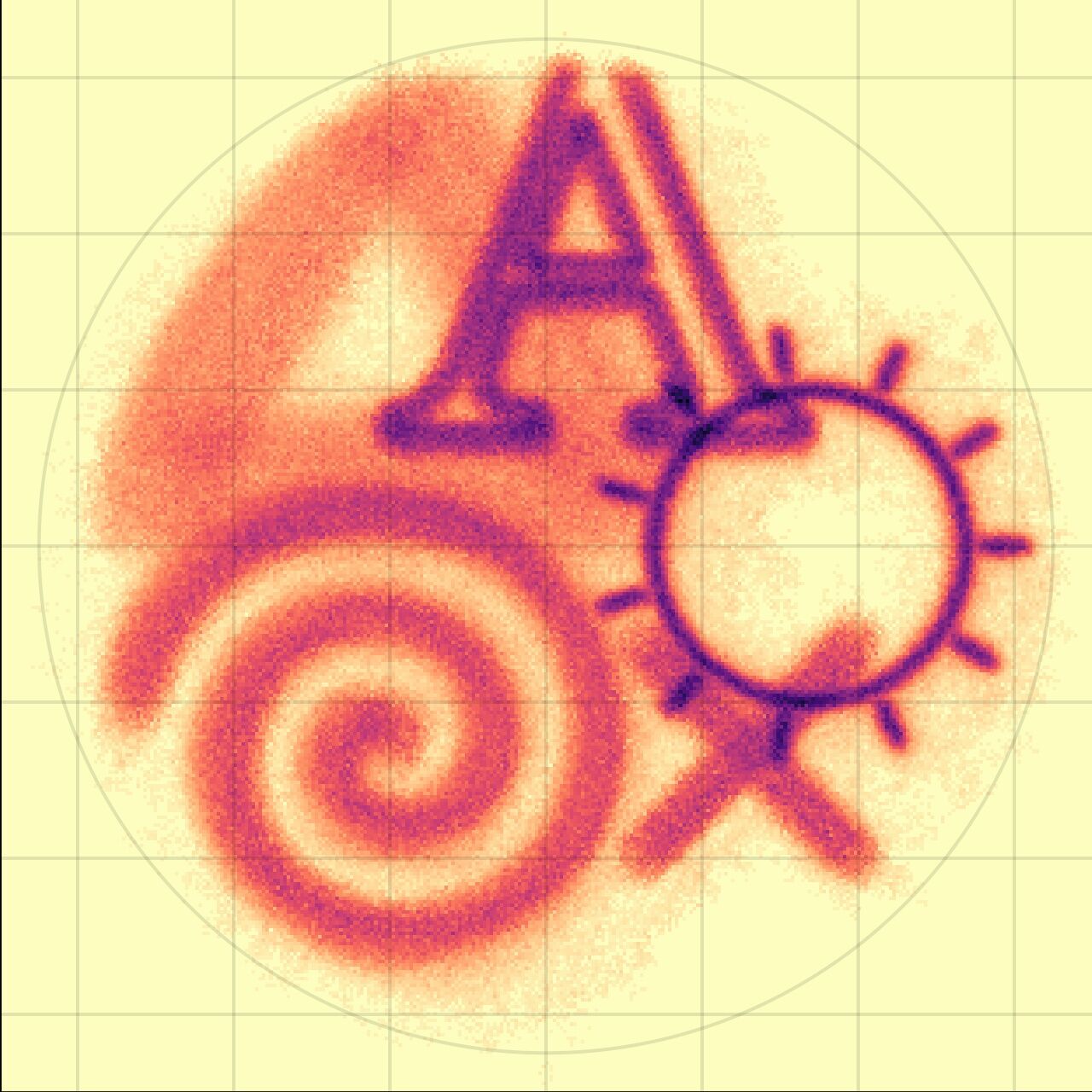}
                \includegraphics[width=1.0\columnwidth]{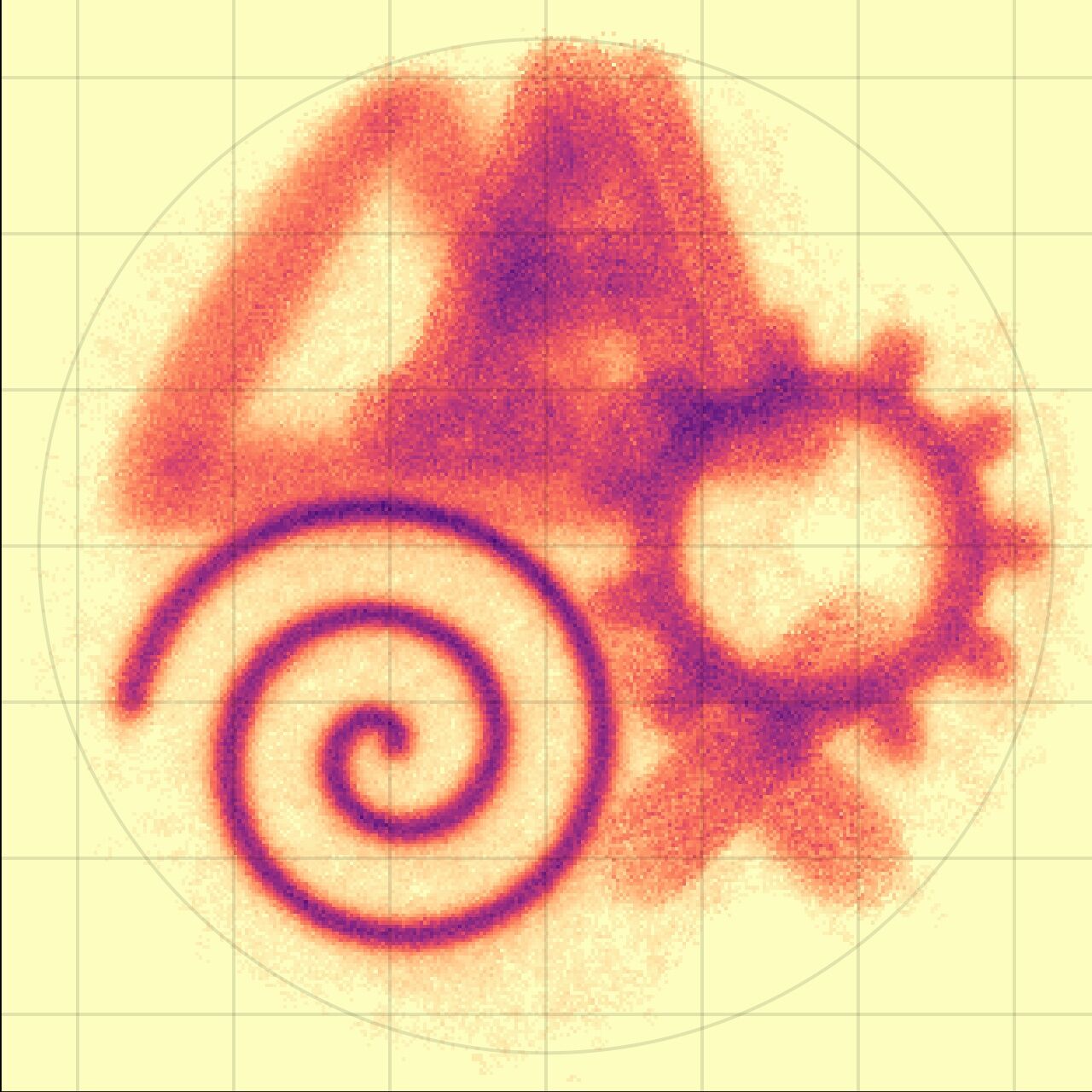}
                \includegraphics[width=1.0\columnwidth]{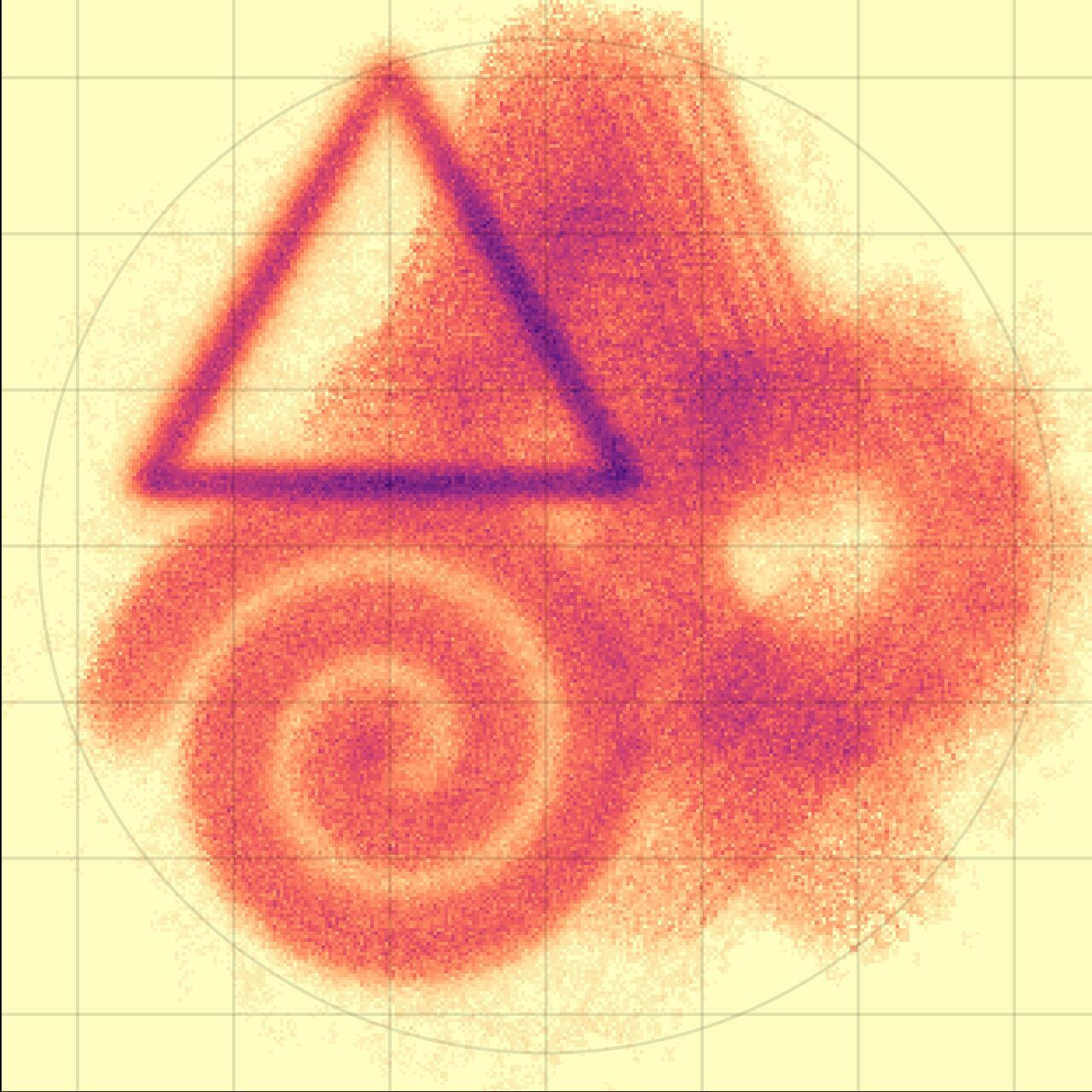}
            \end{minipage}
            \caption[]{
                Observing the phantom~source shown in \autoref{FigPhantomSource} with the \NameAcp{} Cherenkov~plenoscope (\PsixtyOne{}).
                In the left column \NameAcp{}'s state is \StateMirrorCamera{\MirrorGood{}}{\CameraGood{}}, while in the right column its state is \StateMirrorCamera{\MirrorBad{}}{\CameraBad{}}.
                See \autoref{FigObservingPhantomT1} for color~bar.
            }
            \label{FigObservingPhantomP61}
        \end{figure}

        \autoref{FigObservingPhantomP61} shows the observations of the \NameAcp{} Cherenkov~plenoscope (\PsixtyOne{}).
        \NameAcp{} projects its recorded light~field onto images with different foci.
        The five images minimize the spread of individual structures of the phantom~source.
        This way \NameAcp{} estimates the depth of these structures, compare \autoref{FigRefocusRunsP61}.
        The images show that despite deformations and despite misalignments \NameAcp{}'s images are very flat.
        For most showers induced by cosmic particles, this is enough range in depth to refocus a shower from start, to maximum, to finish.
        When comparing the phantom~source's true geometry in \autoref{FigPhantomSource} to the image taken by \Pone{} in \autoref{FigObservingPhantomT1}, and to the light~field taken by \NameAcp{} in \autoref{FigObservingPhantomP61} one finds a striking difference in the amount of perceived information and quality.
        While \Pone{} can not perceive the structures depths, \NameAcp{} can perceive the depth and further can reveal a multitude of small features, such as the sun~structure having 11 flares or the existence of the triangle~structure.
        And on top of this, this comparison does not yet even show all the information in \NameAcp{}'s light~field but only shows projections of it onto images.
        To take full advantage, a reconstruction has to use the full light~field at once as it is proposed by tomographic reconstructions by \citet{ng2006lightfieldmicroscopy} and \citet{engels2017master}.
    \section{Further Potential}
        \label{SecFurtherBenefits}
        \subsection{Widening the Field-of-View}
            \autoref{SecCompensatingAberrations} demonstrates plenoptic's potential to widen the field-of-view of imaging Cherenkov~instruments.
            Already with \NameAcp{}, which is not meant to push the boundaries of larger field-of-views, we find a comfortable 6.5\,$^\circ$ of exceptional flat field-of-view.
            First simulations indicate the potential for a wide field-of-view, single mirror, all spherical surfaces plenoscope.
            We propose a design similar to \NameAcp{} but with the screen of the light-field~camera being curved to make its \Eye{}s point towards the center of the mirror.
            A field-of-view being significantly larger than 10$^{\circ{}}$ seems reasonable.
        \subsection{Detecting Gamma~Rays with Energies $>25$\,GeV}
            For the detection of cosmic gamma~rays with energies above $25$\,GeV, arrays of Cherenkov~telescopes are great.
            But plenoptics can improve Cherenkov~telescopes, even existing ones, to profit from reduced aberrations, wider field-of-views, and more cost effective optics with only single mirrors and solely spherical surfaces.
            Arrays of Cherenkov~telescopes do not need all of plenoptic's features to profit.
            On telescopes with mirrors $<20\,$m there is not much depth to be perceived anyhow and arrays of such telescopes can combine their images after the fact to perceive depth.
            Also mirrors $<10\,$m will probably not profit too much from plenoptic's adaptive compensation of deformations and misalignments.
            But to compensate aberrations one can build a simplified light-field~camera where the imaging~matrix $U(g)$ from \autoref{SecProjectingTheLightFieldOntoImages} is fix and implemented into the analog summation of the photosensors.
            This way one only needs one analog-to-digital~converter for each pixel in the final image.
            Such simplified light-field~cameras might be a worthy upgrade for Cherenkov~telescopes to reduce aberrations, widen the field-of-view, and potentially improve the directional reconstruction of the cosmic gamma~rays.
        \subsection{Pushing the Stellar Intensity Interferometer}
            Beside observing the sky of gamma~rays, the \NameAcp{} Cherenkov~plenoscope might at the same time image bright stars with angular resolutions approaching $\lambda/D = 500\,$nm$/71\,$m$\,\approx 2\times10^{-3}$\,arc-seconds.
            Its plenoptic perception offers a unique opportunity for a stellar intensity~interferometer.
            A stellar intensity~interferometer needs ample bandwidths to correlate the signals of photosensors observing optical beams with same directions ($c_x$,\,$c_y$) but different impacts ($x$,\,$y$).
            Currently, in arrays of Cherenkov~telescopes the beams to be correlated are scattered throughout the entire array and need adjustable delays in time in order to compensate the elongation or shortening of optical paths for different pointings.
            Therefore interferometers proposed for arrays of Cherenkov~telescopes \citep{dravins2012stellar} are held back by the limited bandwidth that can leave the individual telescope, effectively limiting either their field-of-view or exposure~time.
            Now, the plenoscope on the other hand might continuously run an intensity~interferometer across its entire field-of-view because here the beams to be correlated have their photosensors physically close to each other, inside the same housing, and do not need adjustable delays in time.
        \subsection{Perceiving Depth in Showers and Limits of the Atmospheric Cherenkov Method}
            \label{SecPerceivingDepthInAtmosphericShowers}
            After \NameAcp{} has triggered, one can extract the Cherenkov photons from the pool of random photons coming from the night sky by means of clustering.
            Simulations indicate that the extracted Cherenkov-light field has a size of $\approx$100 photons for gamma~rays with energies in the regime of $1\,$GeV to $2\,$GeV.
            With such sizes, the perception of depth becomes feasible.
            Eventually, the perception of depth is based on the statistics of multiple photons, as the individual photon does not have an observable which correlates to the depth of its emission.
            As images are a good start to discuss the statistics of multiple photons, one can project the recorded Cherenkov-light field onto many (several 1000) images, each with its focus set to a different depth $g$ in a range from e.g. 2.5\,km to 25\,km.
            In this stack of images, one finds a prominent cluster of Cherenkov light which grows and shrinks in solid angle when the stack of images is traversed along its dimension in depth.
            Also, the Cherenkov cluster deforms and slightly moves in the images when the stack is traversed along its dimension in depth.

            First, one can find the depth along the stack of refocused images that minimizes the Cherenkov cluster's spread in solid angle, compare \autoref{FigRefocusRunsP61}.
            We find that this depth correlates with the depth where the shower has its maximum.

            Second, one can characterize how quickly the Cherenkov cluster grows and shrinks in solid angle when the stack is traversed along its dimension in depth.
            If the Cherenkov cluster's solid angle changes quickly with depth, the observed region where light was emitted is probably narrow in depth.
            On the other hand, if the Cherenkov cluster's solid angle changes slowly with depth, the observed region where light was emitted is probably extended in depth.

            Third, one can investigate how the Cherenkov cluster's most dense region moves in the images when the stack of images is traversed along its dimension in depth.
            This hints to the orientation of the shower's trajectory relative to the aperture's principal plane and can potentially participate in the reconstruction of the cosmic gamma~ray's direction.

            The Cherenkov cluster discussed here is less predictable than the Cherenkov ellipse which \citet{hillas1985cerenkov} discusses based on simulations of gamma~rays with energies in the regime of $1\,$TeV.
            Gamma~rays with energies as low as 1\,GeV, often induce showers which create no more than 10 light emitting particles.
            Here, fluctuations in the shower's development have a noticeable effect on the statistics of the shower's Cherenkov pool.
            Probably, these noticeable fluctuations will become the main challenge, and at some point the limit, of any atmospheric Cherenkov instrument which aims to observe cosmic gamma~rays with energies as low as $1\,$GeV.
    \section{Outlook}
        \label{SecOutlook}
        We are looking forward to compile a series of papers discussing the all new Cherenkov~plenoscope.
        This part discusses the optics.
        In the following parts we will evaluate the route towards an explorer for the timing in the sky of gamma~rays which is based on the \NameAcp{} Cherenkov~plenoscope shown here.
        We are aiming for an energetic threshold of one giga electronvolt for cosmic gamma~rays.
        We will evaluate \NameAcp{}'s response~functions, discuss the background from cosmic~rays, and evaluate \NameAcp{}'s performance to explore sudden cosmic events in the rapid sky of gamma~rays.
    \section{Conclusion}
        \label{SecConclusion}
        The Cherenkov~plenoscope has the potential to push the atmospheric Cherenkov~method beyond the limits of the Cherenkov~telescope.
        The concept of modeling beams of light and computing images after the fact has a great potential to widen the field-of-view, to compensate deformations in the mirror, to compensate misalignments of the camera, and to effectively overcome astronomy's arch nemesis: Aberrations.
        By significantly reducing the demand for structural rigidity, the plenoscope has the potential to be built larger and more cost effective.
        By completely overcoming the telescope's narrow depth-of-field and turning it into the perception of depth the plenoscope has the potential to be large enough to significantly lower the energetic threshold for cosmic gamma~rays.
        As the Cherenkov~plenoscope can have all these advantages at the same time, we might have the opportunity to implement the long anticipated and physically motivated vision\citep{aharonian2005next} of a \mbox{gamma~ray~timing~explorer}.
    \section*{Acknowledgments}
        This work would not exist without the early backing provided by Felicitas Pauss (ETH-Zurich), and the current backing provided by Jim Hinton (MPIK-Heidelberg).
        We acknowledge the discussions with Georgios Zinas and Mario Fontana (ETH-Zurich) on civil~engineering, the discussions with Johannes Stoll and Marc Fabritius (Fraunhofer IPA) on cable~robots,
        and the discussions with Felix Goehring (DLR Solar\,towers\,Juelich) on the protection of solar concentrators.
        We acknowledge the fruitful exchanges with Felix Aharonian, Jie-Shuang Wang, and Heinz Voelk (MPIK-Heidelberg) on possible observations with \NameAcp{}.
        S.A.M. fondly recalls when Wolfgang Rhode (TU-Dortmund) introduced him to the fascinating optics of Cherenkov telescopes.
    \bibliographystyle{elsarticle-harv}
    \bibliography{sebastians_references}
    \section*{Appendix}
    \appendix
        \section{Calculating Image~Rays}
            \label{SecCalculatingImageRays}
            For the image~ray's direction $\vec{\delta}$ one first calculates the point
            \begin{footnotesize}
            \begin{eqnarray}
                \label{EqThinLensPointOfLightOnFocalPlane}
                \vec{b} &=& f \left( - \frac{c_x}{\sqrt{1 - c_x^2 - c_y^2}},\,\,- \frac{c_y}{\sqrt{1 - c_x^2 - c_y^2}},\,\,1 \right)^T
            \end{eqnarray}
            \end{footnotesize}
            where the image~ray intersects with the plane in focal-length $f$ according to the thin lens,
            and second subtracts the support from this point
            \begin{eqnarray}
                \vec{\delta} &=& \frac{\vec{b} - \vec{s}}{\vert \vec{b} - \vec{s} \vert}.
            \end{eqnarray}
            Because \autoref{EqThinLensPointOfLightOnFocalPlane} has no dependency of the optical path's support $\vec{s}$ on the aperture, this image is free of aberrations.
        \section{Calculating the Imaging Matrix}
            \label{SecCalculatingImagingMatrix}
            One calculates the elements $u_{n,\,k}(g)$ using the $k$-th image~ray, and the $n$-th pixel's position on a virtual screen which focuses on depth $g$.
            First, one calculates the distance of the image
            \begin{eqnarray}
                b &=& \frac{1}{\frac{1}{f} - \frac{1}{g}},
                \label{EqVirtualSensorPlaneDistance}
            \end{eqnarray}
            based on one's desired depth and puts the virtual screen right into this distance
            \begin{eqnarray}
                d &=& b
                \label{EqVirtualSensorPlaneDistanceEqualsImageDistance}
            \end{eqnarray}
            to make it focus on $g$, see \autoref{FigThinLens}.
            Second, one calculates how far
            \begin{eqnarray}
                \chi_g &=& \frac{d}{(0,0,1) \cdot \vec{\delta}_k}
                \label{EqChiToReachVirtualSensorPlane}
            \end{eqnarray}
            one has to travel along the image~ray $\vec{\rho}(\chi)_k$ to intersect with the virtual screen.
            Third, one calculates the intersection
            \begin{eqnarray}
                \vec{i}_{k,g} &=& \vec{\rho}(\chi_g)_k
                \label{EqIntersectionVirtualSensorPlane}
            \end{eqnarray}
            of the image~ray and virtual screen.
            Fourth, one looks up the position
            \begin{eqnarray}
                \vec{p}_{n} &=& \left(x_n,\,y_n,\,d\right)^T
                \label{EqVirtualSensorPlanePixelPosition}
            \end{eqnarray}
            of the area on the virtual screen that feeds into the $n$-th pixel.
            These arbitrarily shaped areas, which feed into pixels, are defined in advance and describe how the image is binned.
            In most cases, one would probably choose this binning to be a regular grid of squares or hexagons.
            Fifth, one computes the distance
            \begin{eqnarray}
                o_{n,k,g} &=& \vert \vec{i}_{k,g} - \vec{p}_{n} \vert
                \label{EqDistanceBetweenPixelAndImageBeamIntersection}
            \end{eqnarray}
            between the pixel's area $\vec{p}_{n}$ and the image~beam's intersection $\vec{i}_{k,g}$ in the virtual screen.
            Finally, one can set the matrix's elements
            \begin{eqnarray}
                u_{n,\,k}(g) &=&
                \left\{
                \begin{array}{ll}
                1 & \text{if}\,\,o_{n,k,g} \leq r_n\\
                0 & \text{else} \\
                \end{array}
                \right.{}
                \label{EqPostRefocusedImagingPixelAssignmentGeneral}
            \end{eqnarray}
            by comparing the distance $o_{n,k,g}$ to a threshold $r_n$ which can be e.g. the radius of the area feeding into the $n$-th pixel.
        \section{Characterizing an Optical Beam}
            \label{SecCharacterizingBeam}
            This is a simple definition for the spread of a beam which uses the instrument's light-field~calibration to gather statistics.
            In the light-field~calibration, the $k$-th beam $\Beam{}_k$ is represented by a list of $P_k$ optical paths, see \autoref{EqOpticalPathDefinition}, and \autoref{EqBeamDefinition}.
            We define the beam's spread in solid angle to be the solid angle of an elliptical cone.
            The two half-angles of this elliptical cone are the standard deviations of the directions of the optical paths rays.
            For simplicity, we directly estimate these standard deviations from the components ${c_x}_{k,p}$, and ${c_y}_{k,p}$ and thus neglect potential smaller radii along the distribution's principal axes.
            We apply the same method to estimate the beam's spread in area, see \autoref{EqBeamSolidAngle}, and \autoref{EqBeamArea}.
            \begin{equation}
                \label{EqBeamSolidAngle}
                \begin{aligned}
                    \BeamSolidAngle{}_k =& (4\pi\,\std{}([{c_x}_{k,1}, \dots {c_x}_{k,P_k}])\\
                    & \,\std{}([{c_y}_{k,1}, \dots {c_y}_{k,P_k}]))
                \end{aligned}
            \end{equation}
            \begin{equation}
                \label{EqBeamArea}
                \begin{aligned}
                    \BeamArea{}_k =& (4\pi\,\std{}([{x}_{k,1}, \dots {x}_{k,P_k}])\\
                    & \,\std{}([{y}_{k,1}, \dots {y}_{k,P_k}])
                \end{aligned}
            \end{equation}
            For a beam's spread in time we just compute the standard deviation of its delays ${\tau}_{k,p}$
            \begin{eqnarray}
                \label{EqBeamTimeSpread}
                \BeamTimeSpread{}_k &=& \std{}([{\tau}_{k,1}, \dots {\tau}_{k,P}]),
            \end{eqnarray}
            and for the beam's efficiency we compute the median of its optical paths efficiencies
            \begin{eqnarray}
                \label{EqBeamEfficiency}
                \BeamEfficiency{}_k &=& \median{}([{\eta}_{k,1}, \dots {\eta}_{k,P}]).
            \end{eqnarray}
            In \autoref{FigBeamStatistics} we show the beams efficiencies relative to the median of all the $K$ beams efficiencies to not be spoiled by the particular coefficient of reflection of the mirror.
        \section{Estimating Limits for Misalignments on a Telescope}
            \label{SecEstimatingLimitsOfMisalignment}
            These limits are motivated by \autoref{EqDepthOfField} which apparently defines a change in an image to be significant when a considerable fraction ($\approx$ 100\%) of the light expected to be assigned to one pixel ends up being assigned to a different pixel.
            \subsection*{Translating Parallel $\TransPara{}$}
                Translating the camera's screen parallel to the mirror's optical axis in the range
                \begin{eqnarray}
                    \Delta_\parallel &=& d - d_\pm
                \end{eqnarray}
                will not significantly change the image of a telescope.
                The start and stop distances of the camera's screen
                \begin{eqnarray}
                    d_\pm &=& \frac{1}{\frac{1}{f} - \frac{1}{g_\pm}}
                    \label{EqToleranceTranslationParallel}
                \end{eqnarray}
                are estimated using \autoref{EqThinLens} and \autoref{EqDepthOfField}.
            \subsection*{Translating Perpendicular $\TransPerp{}$}
                Translating the camera's screen perpendicular to the mirror's optical axis up to about half the extent of a photosensor
                \begin{eqnarray}
                    \Delta_\perp &\approx& p/2
                \end{eqnarray}
                will not significantly effect the image.
            \subsection*{Rotating Parallel $\RotPara{}$}
                Rotating the camera's screen mostly translates its outer edge which is in radial distance
                \begin{eqnarray}
                    r_\text{camera} &=& f \tan{(\Theta/2)}
                \end{eqnarray}
                from its optical axis.
                Here $\Theta/2$ is the half-angle of the instrumented field-of-view.
                Thus to not effect the image, a parallel rotation of the camera's screen
                \begin{eqnarray}
                    \Phi_\parallel &\approx& \frac{\Delta_\perp}{r_\text{camera}}
                    \label{EqToleranceRotationParallel}
                \end{eqnarray}
                must not translate a screen's photosensor perpendicular by more than $\Delta_\perp$.
            \subsection*{Rotating Perpendicular $\RotPerp{}$}
                Similar, rotating the camera's screen perpendicular to the mirror's optical axis
                \begin{eqnarray}
                    \Phi_\perp &=& \frac{\Delta_\parallel}{r_\text{camera}}
                    \label{EqToleranceRotationPerpendicular}
                \end{eqnarray}
                must not translate a photosensor parallel by more than $\Delta_\parallel$.
    \section*{Author Contributions}
        S.A.M. leads the investigations, proposed the Cherenkov~plenoscope, writes, runs, and analyses the simulations, and writes this manuscript.
        S.D. investigates \NameAcp{}'s cable-robot~mount in his master~thesis in civil~engineering with A.E. and E.C. reviewing his findings.
        A.B. supported the early investigations, reviews the feasibility of the optics, and supported A.A.E. to evaluate plenoptic reconstructions of atmospheric showers in A.A.E.'s master~thesis.
        A.A.E. helps formulating the mindset of a light-field~calibration.
        W.H. supports the investigations now and provides critical reviews on optics, statistics and language.
        M.L.A. and D.N. helped with early estimates of the plenoscope's performance with M.L.A. further helping to find collaborators and motivating S.A.M. to commit and go `all in' on the quest for a `gamma~ray~timing~explorer'.
    \section*{Additional Information}
        Computer~simulations are public on \url{https://github.com/cherenkov-plenoscope}.
        Correspondence and requests for materials should be addressed to S.A.M.
    \section*{Competing Financial Interests}
        The authors declare no competing financial interests.
\end{document}